\documentclass[nofootinbib,
 amsmath,amssymb,
prb,
reprint
]{revtex4-2}

\usepackage{graphicx}
\usepackage{dcolumn}
\usepackage{bm}
\usepackage{amsmath}
\usepackage{amssymb}
\usepackage{physics}
\usepackage{xcolor}
\usepackage{hyperref}
\usepackage{cleveref}

\begin{document}

\title{A bosonic matrix product state description of Read-Rezayi states
\\
and its application to quasi-hole spins}

\author{Alexander Fagerlund}
\email{alexander.fagerlund@fysik.su.se}
\author{Eddy Ardonne}%
 \email{ardonne@fysik.su.se}
\affiliation{%
 Department of Physics, Stockholm University, AlbaNova University Center, SE-106 91 Stockholm, Sweden
}%

\date{\today}

\begin{abstract}
We study the $k=3$ Read-Rezayi quantum Hall state by means of a purely bosonic matrix product state formulation, which is described in detail. We calculate the density profiles in the presence of bulk quasi-holes of six different types: one for each $\mathbb{Z}_3$ parafermion sector. From the density profiles, we calculate the (local) spins of these quasi-holes. By employing a spin-statistics relation, we obtain the exchange statistics parameters. Our results, which are entirely based on \textit{local} properties of the quasi-holes, corroborate previous results obtained by explicitly braiding quasi-holes, showing that the exchange statistics can be read off from the monodromy properties of the wave functions, i.e., that the associated Berry phase vanishes. We also discuss the entanglement spectrum, to show that our bosonic matrix product state formulation correctly captures the $\mathbb{Z}_3$ parafermionic structure of the $k=3$ Read-Rezayi states.
\end{abstract}

\maketitle

\section{Introduction}
\label{sec:introduction}
The fractional quantum Hall (FQH) effect is of great physical interest: since its discovery \cite{Tsui1982FQHE}, it has provided many important examples of topologically ordered states \cite{Wen1995Order}, as well as being the only setting where anyons have been experimentally observed so far \cite{nakamura2020braiding}. An important class of FQH states are the Read-Rezayi states \cite{Read1999Parafermions}. These states are constructed by subdividing the particles into $k$ subsets, symmetrizing over all possible ways to divide the particles into such subsets and multiplying by an overall Jastrow factor. This gives the state
\begin{equation}\label{Read-Rezayi}
    \psi(\{z\}) = \mathcal{S}\bigg[\prod_{l=1}^k\prod_{i<j}\big(z_i^{(l)}-z_j^{(l)}\big)^2\bigg]\prod_{r<s}(z_r-z_s)^M \ ,
\end{equation}
where the filling fraction is given by $\nu = k/(k M+2)$.
For $k=1$, one obtains the famous Laughlin state \cite{Laughlin_1983}, while the case $k=2$ corresponds to the Moore-Read state \cite{Moore1991Nonabelions}, which is a candidate for the $\nu=\frac{5}{2}$ quantum Hall state \cite{Willett1987Observation,Greiter1992paired}.

In this paper, we are solely concerned with the $k=3$ Read-Rezayi states. 
The (particle-hole conjugate of the) $k=3$ Read-Rezayi state with $M=1$ is a candidate for describing the $\nu=\frac{12}{5}$ quantum Hall state, which was
observed experimentally \cite{Xia2004Correlation}.
In the present paper, we perform numerical calculations for the wave function \cref{Read-Rezayi} in the presence of quasi-hole excitations.

To perform numerics for quantum Hall states, a useful method is that of matrix product states (MPS) \cite{Orus2014mps,Schollwock2011mps} which was first used to study quantum Hall systems in \cite{Zaletel2012MPS}.
The advantage of MPS-based approaches for the Read-Rezayi states is best seen when comparing to the calculation of $k=3$ Read-Rezayi observables using Monte Carlo methods. 
Finding the density of the quantum Hall fluid with (say) the Metropolis algorithm \cite{metropolis1953equation} would involve the repeated evaluation of $|\psi|^2$, and hence of the wave function, to estimate the contributions to the density $\rho=\int |\psi(\mathbf{r})|^2d^{2(N-1)}r$. Since the $k=3$ Read-Rezayi wave function \cref{Read-Rezayi} involves an explicit symmetrization over all ways of grouping the electron coordinates $z_1,\ldots, z_N$ into three sets, it quickly becomes demanding to evaluate as the number of electrons $N$ increases. Having to evaluate it frequently for many electrons makes the Monte Carlo approach impractically slow.
We remark that for the Laughlin ($k=1$) and Moore-Read ($k=2$) cases, where the latter can be written in terms of a Pfaffian,  
\begin{equation} \psi(\{z\})=\mathrm{Pf}\bigg(\frac{1}{z_i-z_j}\bigg)\prod_{k<l}(z_k-z_l)^{M+1},
\end{equation}
it is perfectly possible to use a Monte Carlo scheme because the Pfaffian obeys $\mathrm{Pf}(A)^2=\mathrm{det}(A)$, and techniques exist for rapidly evaluating determinants. %
For the $k=3$ state, however, another approach is needed. The technique we use is based on MPS states, but deviates from the MPS approach of \cite{Estienne_2013,Wu2014Braiding,Wu2015MPS,estienne2015entropies} (see also \cite{Herviou_2024}) by exclusively using free boson fields instead of also including a $\mathbb{Z}_3$ parafermion theory \cite{Fateev-Zamolodchikov1985}. Hence, our method works as an alternative MPS setup for the $k=3$ Read-Rezayi state and can be fruitfully compared to the $\mathbb{Z}_3$ approach. It is based on the way to write the Read-Rezayi states introduced in \cite{Cappelli2001Parafermion}, and uses three free chiral boson fields. MPS studies of the Halperin~\cite{Halperin_1983,Halperin_1984} and Haldane-Rezayi~\cite{Haldane_1988} states, using two chiral boson fields, were performed in~\cite{Crepel_2018,Crepel_2019}.

One can also write explicit wave functions for the Read-Rezayi states in the presence of quasi-holes; we refer to section \ref{sec:wavefcns} for explicit examples. The resulting wave functions are obtained by evaluating expectation values of vertex operators in a conformal field theory (CFT) (see section \ref{sec:cft}), where the vertex operators correspond to both electron operators and operators describing the quasi-hole(s). By considering the operator product expansion (OPE) of the quasi-hole operators in a certain `minimal' description%
\footnote{In the case of the Read-Rezayi states, the minimal description is in terms of the $\mathbb{Z}_k$ parafermion CFT, not the free boson approach we take.}, the braiding phase in a given fusion channel can simply be read off: if the quasi-hole operators are at positions $w_1$ and $w_2$, their OPE in a given fusion channel is proportional to $(w_1-w_2)^{\gamma}$, for some number $\gamma$. Details are provided in section \ref{sec:spin}. Under the highly nontrivial assumption that the Berry phase \cite{Berry1984phase} associated with moving quasi-holes around in the $k=3$ Read-Rezayi state vanishes, the number $\gamma$ then describes the mutual statistics of the quasi-holes. This assumption holds for the Laughlin and Moore-Read states \cite{Laughlin_1983,Belavin1984conformal,Bonderson2011Plasma,Herland2012Screening}, and has been numerically demonstrated to hold also for the $k=3$ Read-Rezayi state by performing explicit braiding of quasi-holes \cite{Wu2014Braiding}. However, one can obtain the mutual statistics of the quasi-holes based on purely \textit{local} quantities. In particular, we will see that a recently derived spin-statistics theorem for quasi-holes in quantum Hall systems \cite{Nardin2023SSR} can be used to calculate the mutual statistics from the local spins of the quasi-holes. These results agree with the results obtained by calculating the monodromy properties of the quasi-holes from the minimal CFT description. This shows that the Berry phases of the states derived from the minimal CFT description indeed vanish.
We arrive at this conclusion by considering purely local quantities: the quasi-hole spins.

The remainder of the paper is structured as follows. In section \ref{sec:wavefcns}, we describe how to obtain the $k=3$ Read-Rezayi wave functions when there is a quasi-hole in the system. This is done for the $\sigma_1,\sigma_2,\mathbf{1},\epsilon,\psi_1$ and $\psi_2$ quasi-holes, where the labels are those of the $\mathbb{Z}_3$ CFT description. Some generalities about how to express wave functions as CFT correlation functions are given in section \ref{sec:cft}, before showing in section 
\ref{sec:operators} which operators we use to represent electrons and quasi-holes. The link to  the $\mathbb{Z}_3$ description is made explicit in section \ref{sec:ident}. After that, we focus on the MPS description. We describe the finite cylinder geometry of the MPS implementation and introduce the various quantum numbers of our auxiliary Hilbert spaces (section \ref{sec:Hilbert}), and give the matrix elements for unoccupied orbitals, for orbitals occupied by electrons, and for the quasi-hole operators (section \ref{sec:matrices}). In the latter section, we also discuss the imaginary time evolution used to put operators in the right positions along the cylinder, and give an expression for a factor resulting from this time evolution. This factor makes terms decay exponentially in the quantum numbers of the auxiliary spaces. Hence, there is a natural cutoff for system configurations that lead to extreme values of the quantum numbers. This cutoff is described further in section \ref{sec:implementation}, together with other implementation details, which are relevant for e.g. section \ref{sec:density}, where we describe how the MPS technique was utilized to compute the quasi-hole density profiles. The density profiles are shown in section 
\ref{sec:density-profiles}, and the density information is used to compute the charges of the quasi-holes
in section \ref{sec:compute charge}. After explaining in section \ref{sec:spin} how -- under the nontrivial assumption that the Berry phase vanishes -- one can predict the spins of the different quasi-holes, we also use the MPS-based density profiles to compute the quasi-hole spins in section \ref{sec:compute spin}. This direct computation requires no assumption about the value of the Berry phase. The numerical results are found to agree well with the predictions, and to converge to the predicted values as the dimension of the auxiliary space in the MPS computation increases. The final results shown are about the entanglement spectrum, and can be found in section \ref{sec:entanglement}, followed by a discussion about how our results depend on the cylinder circumference, the cutoff parameter, and the integration scheme used when computing observables in section \ref{sec:error}. Our conclusions follow in section \ref{sec:concl}. Finally, we give a more detailed derivation of the factor from the imaginary time evolution of section \ref{sec:matrices} in appendix \ref{app:exponent}.

Although the reader is welcome to read all of the sections outlined above, parts may be omitted based on interest and prior knowledge. In particular, readers familiar with the Read-Rezayi states and their quasi-holes may skip section \ref{sec:wavefcns}, whereas those with a working knowledge of the free boson CFT need not read section \ref{sec:cft}. Those who mainly want the results about the quasi-hole profiles, charges and spins will find them in sections \ref{sec:density-profiles}, \ref{sec:compute charge} and \ref{sec:compute spin}, together with a discussion of how the spin relates to the braiding phase in section \ref{sec:spin}. Readers more interested in the particularities of our MPS setup are instead invited to read sections \ref{sec:operators}, \ref{sec:Hilbert}, \ref{sec:matrices}, \ref{sec:implementation} and \ref{sec:error}, as well as the identification in section \ref{sec:ident} between our MPS operators and those from earlier work based on the $\mathbb{Z}_3$ parafermion theory. Those interested specifically in the entanglement spectrum may wish to read section \ref{sec:entanglement}. Finally, for readers more interested in results than in methodological details, we also refer to the related paper \cite{Fagerlund2024nonAbelian}. This paper defines an edge spin for quantum Hall droplets with bulk quasi-holes, and demonstrates how this edge spin takes fractional values due to the fractional spin of the quasi-hole in the bulk.

\section{Quasi-hole wave functions}\label{sec:wavefcns}

From now on, we only consider the $k=3$ Read-Rezayi states, at filling $\nu = 3/(3M+2)$. We often
state results for arbitrary $M$, but for the numerical calculations below, we only consider the fermionic state
with $M=1$.

To describe the form of the quasi-hole wave functions that we analyze using the MPS formulation,
we start with the ground state in the presence of three $\sigma_1$ quasi-holes, at the locations
$w_1, w_2, w_3$, with the number of electrons being a multiple of three. For details, we refer to \cite{Read1999Parafermions,ARDONNE2007Registers}.
We label the quasi-holes using their $\mathbb{Z}_k$ parafermion field and their charge. In this notation, the minimal
quasi-hole is denoted as $(\sigma_1, \frac{1}{3M+2})$.

Generically, we divide the electrons into three groups $S_1, S_2, S_3$, whose sizes can
vary, depending on the different quasi-holes, and sum over the different ways of
dividing the electrons over the groups. We associate a Laughlin factor with each
group as follows
\begin{equation}
    \Psi_{S_a}^2 (\{z\}) = \Psi_{\rm L}^M \prod_{\substack{i<j\\i,j \in S_a}} (z_i - z_j )^2 \ ,
\end{equation}
where $\Psi_{\rm L}^M = \prod_{i<j} (z_i-z_j)^M$, with the product being over all particle coordinates.
We note that in this section, we drop the (geometry dependent) gaussian factors.

The $k=3$ Read-Rezayi ground state wave function can then be written as
\begin{equation}\label{RR}
    \Psi_{\rm RR}^{k=3} (\{z\})= \Psi_{\rm L}^M \sum_{S_1, S_2, S_3} \Psi_{S_1}^2 \Psi_{S_2}^2 \Psi_{S_3}^2 \ ,
\end{equation}
where the sum is over all ways to divide the particles into three groups of equal size.
This particular form of the Read-Rezayi wave function was first considered in \cite{Cappelli2001Parafermion}, which differs from the one used in the original paper \cite{Read1999Parafermions}.
The wave function with three $(\sigma_1,\frac{1}{3M+2})$ quasi-holes is given by
\begin{align}
    &\Psi_{\rm RR,3 \sigma_1}^{k=3} (\{z\})= \Psi_{\rm L}^M \sum_{S_1, S_2, S_3}
    \Bigl[
    \Psi_{S_1}^2
    \Psi_{S_2}^2 
    \Psi_{S_3}^2
    \nonumber\\
    &
    \prod_{i_1 \in S_1} (z_{i_1}-w_1)
    \prod_{i_2 \in S_2} (z_{i_2}-w_2)
    \prod_{i_3 \in S_3} (z_{i_3}-w_3)
    \Bigr] \ .
\end{align}

From this wave function, one obtains the form for a single $(\sigma_1,\frac{1}{3M+2})$ quasi-hole by sending the other two
quasi-holes to the edge of the system. This means that we only consider the part of the wave function
that is proportional to $w_2^{N_e/3} w_3^{N_e/3}$, with $N_e$ the
number of electrons, and we use the coordinate $w = w_1$.
Explicitly, one finds, for the $(\sigma_1,\frac{1}{3M+2})$ quasi-hole
\begin{align}
\label{wavefcn sigma1}
    &\Psi_{\rm RR,\sigma_1}^{k=3} (\{z\},w)=
    \Psi_{\rm L}^M \times \\ \nonumber
    &\sum_{S_1, S_2, S_3}
    \Bigl[
    \Psi_{S_1}^2
    \Psi_{S_2}^2 
    \Psi_{S_3}^2
    \prod_{i_1 \in S_1} (z_{i_1}-w)
    \Bigr] \ .
\end{align}
Similarly, the $(\sigma_2,\frac{2}{3M+2})$ quasi-hole is obtained by setting $w_1,w_2 \rightarrow w$ and
taking the part of the wave function that is proportional to $w_3^{N_e/3}$.
Explicitly, one finds
\begin{align}
\label{wavefcn sigma2}
    &\Psi_{\rm RR,\sigma_2}^{k=3} (\{z\},w)=
    \Psi_{\rm L}^M \times \\ \nonumber
    &\sum_{S_1, S_2, S_3}
    \Bigl[
    \Psi_{S_1}^2
    \Psi_{S_2}^2 
    \Psi_{S_3}^2
    \prod_{i_1 \in S_1} (z_{i_1}-w)
    \prod_{i_2 \in S_2} (z_{i_2}-w)
    \Bigr] \ .
\end{align}
The Laughlin quasi-hole, $(\mathbf{1},\frac{3}{3M+2})$, is obtained by setting $w_1,w_2,w_3 \rightarrow w$, giving
\begin{equation}\label{wavefcn Laughlin}
    \Psi_{\rm RR,\mathbf{1}}^{k=3} (\{z\},w)=
    \Psi_{\rm L}^M
    \prod_{i} (z_i - w)
    \sum_{S_1, S_2, S_3}
    \Bigl[
    \Psi_{S_1}^2
    \Psi_{S_2}^2 
    \Psi_{S_3}^2
    \Bigr] \ .
\end{equation}

We also consider the quasi-holes $(\psi_1, \frac{2}{3M+2})$, $(\epsilon, \frac{3}{3M+2})$ and
$(\psi_2, \frac{4}{3M+2})$. To obtain the wave function for a state with a $(\psi_1, \frac{2}{3M+2})$ quasi-hole,
we consider the ground state
wave function with $N_e$ a multiple of three, but consider one of the electron coordinates
as a quasi-hole coordinate, by changing the chiral vertex operator part of the electron operator and
denoting the coordinate by $w$ instead of $z$. Thus we take $N_e = 3 p +2$, with $p$ a non-negative
integer and we assume that $S_1$ and $S_2$ have $p+1$ elements, while $S_3$ has $p$ elements.
This results in
\begin{align}
\label{wavefcn psi1}
    &\Psi_{\rm RR,\psi_1}^{k=3} (\{z\},w)=
    \Psi_{\rm L}^M \times \\ \nonumber
    &\sum_{S_1, S_2, S_3}
    \Bigl[
    \Psi_{S_1}^2
    \Psi_{S_2}^2 
    \Psi_{S_3}^2
    \prod_{i_3 \in S_3} (z_{i_3} - w)^2
    \Bigr] \ .
\end{align}
To obtain the $(\psi_2, \frac{4}{3M+2})$ quasi-hole, we also consider the ground state
wave function with $N_e$ a multiple of three, but fuse two of the electron coordinates, to obtain
a $\psi_2$ and change the associated chiral vertex operator, to get the correct charge for the quasi-hole
(we also denote the coordinate by $w$). Thus we take $N_e = 3 p + 1$, with $p$ a non-negative
integer and we assume that $S_1$ has $p+1$ elements, while $S_2$ and $S_3$ have $p$ elements.
This results in
\begin{align}
\label{wavefcn psi2}
    &\Psi_{\rm RR,\psi_2}^{k=3} (\{z\},w)=
    \Psi_{\rm L}^M \times \\ \nonumber
    &\sum_{S_1, S_2, S_3}
    \Bigl[
    \Psi_{S_1}^2
    \Psi_{S_2}^2 
    \Psi_{S_3}^2
    \prod_{i_2 \in S_2} (z_{i_2} - w)^2
    \prod_{i_3 \in S_3} (z_{i_3} - w)^2
    \Bigr] \ .
\end{align}
Finally, to obtain the $(\epsilon, \frac{3}{3M+2})$ quasi-hole, we consider the wave function of the state with
the number of electrons a multiple of three, and three $\sigma_1$ quasi-holes.
We fuse one electron with one of the
$\sigma_1$ quasi-holes (say at $w_1$), to obtain an $\epsilon$ excitation, and modify the chiral
vertex operator, to ensure the correct charge. The remaining two quasi-holes are sent to the same
edge of the system in the same way as was done above%
\footnote{If we tried to obtain the more symmetric
version, by sending one quasi-hole to either end of the cylinder, we would have to symmetrize over
these quasi-hole locations before sending them to the different boundaries. It is hard to implement this in the
MPS formulation).}.
Thus, we take the number of electrons to be $N_e = 3p+2$; $S_1$ and $S_2$ have $p+1$ elements,
while $S_3$ has $p$ elements.
The resulting wave function reads
\begin{align}\label{wavefcn eps}
    &\Psi_{\rm RR,\epsilon}^{k=3} (\{z\},w)=
    \Psi_{\rm L}^M \times \\ \nonumber
    &\sum_{S_1, S_2, S_3}
    \Bigl[
    \Psi_{S_1}^2
    \Psi_{S_2}^2 
    \Psi_{S_3}^2
    \prod_{i_2 \in S_2} (z_{i_2} - w)
    \prod_{i_3 \in S_3} (z_{i_3} - w)^2
    \Bigr] \ .
\end{align}
The wave functions \cref{wavefcn sigma1,wavefcn sigma2,wavefcn psi1,wavefcn psi2,wavefcn Laughlin,wavefcn eps,RR} can be expressed in terms of a CFT of free bosons. A short summary of this technique is given in section \ref{sec:cft}. Readers familiar with the free boson CFT may wish to skip ahead to section \ref{sec:operators}, where the specific operators for our scheme are introduced.

\section{The chiral boson CFT}\label{sec:cft}
The aim is to express the RR wave functions with quasi-holes in \cref{wavefcn sigma1,wavefcn sigma2,wavefcn Laughlin,wavefcn psi1,wavefcn psi2,wavefcn eps} as expectation values in a CFT
 which in turn can be expressed as tensor contractions and treated with MPS methods. 
Generalizing \cite{Zaletel2012MPS} (but following the conventions of \cite{Kjall2018MPS}), we take three free boson fields $\phi,\chi_1$ and $\chi_2$, each compactified on an appropriate radius. The corresponding actions and OPEs are
\begin{align}\label{OPE}
S&= \frac{1}{8\pi}\int d^2x (\partial_\mu\phi)(\partial^{\mu}\phi)
\\\nonumber 
\langle\phi(z_1)\phi(z_2)\rangle &\sim -\ln(z_1-z_2)
\end{align}
and a mode expansion can be performed according to 
\begin{equation}\label{modexpbos}
    \phi(z) = \phi_0-i\pi_0\ln(z)+\sum_{n\neq 0}\frac{a_n}{n}z^{-n}
\end{equation}
with corresponding expressions for the $\chi_1$ and $\chi_2$ fields. Below, we focus on the $\phi$ boson field, with the understanding that the other fields can be treated analogously. When significant differences arise, these will be pointed out.

In light of the mode expansion \cref{modexpbos} for the boson field in terms of harmonic oscillator operators, the vertex operators of free boson fields obey the standard identity \cite{Francesco2012CFT}
\begin{equation}
    \langle :e^{A_1}::e^{A_2}: \cdots :e^{A_N}:\rangle=\exp\bigg(\sum_{i<j}\langle A_iA_j \rangle\bigg) \ ,
\end{equation}
where the $A_i$ are some linear combinations of annihilation and creation operators for harmonic oscillator modes
and the ``$:$" denote normal ordering.  
From the OPE for the boson fields, it follows that 
\begin{equation}\label{VEV}
    \langle :e^{i\alpha_1\phi(z_1)}::e^{i\alpha_2\phi(z_2)}: \cdots :e^{i\alpha_N\phi(z_N)}:\rangle = \prod_{i<j}(z_i-z_j)^{\alpha_i\alpha_j},
\end{equation}
which can be used to represent Jastrow factors and other polynomials relevant for the quantum Hall states as expectation values in a bosonic CFT. To render the expectation value invariant under constant translations of the fields $\phi\rightarrow \phi+c$, this presupposes the charge neutrality condition $\sum_i\alpha_i=0.$

From the properties outlined above, it is clear that the wave function without quasi-holes \eqref{RR} may be represented as a symmetrized linear combination of expectation values all on the schematic form
\begin{equation}\label{wavefcn}
    \langle V(z_1)V(z_2) \cdots V(z_N)\rangle,
\end{equation}
where $\{V(z_i)\}$ represent individual electrons. By similar methods, the polynomial factors involving the quasi-holes in \cref{wavefcn sigma1,wavefcn sigma2,wavefcn Laughlin,wavefcn psi1,wavefcn psi2,wavefcn eps}
can be reproduced, by inserting appropriate quasi-hole operators. The exact forms of the electron and quasi-hole operators are stated and motivated in section \ref{sec:operators}.

\section{The electron and quasi-hole operators}\label{sec:operators}
In this section, we obtain electron operators such that an expectation value analogous to \cref{wavefcn} can be used to reproduce the $k=3$ Read-Rezayi wave function, including the way electrons are grouped into three subsets. We continue by describing the quasi-hole operators whose OPEs with the electron operators give a factor $(z-w)^1$ or $(z-w)^0$ depending on whether or not the electron and quasi-hole operators belong to the same group.

We first point out that a generic electron operator in the Read-Rezayi states should obey
\begin{equation}\label{elec}
    V_e\propto \;:e^{i\sqrt{\frac{kM+2}{k}}\phi}: \ .
\end{equation}
Here and below, we set $k=3$ but keep $M$ general, although the state we will eventually show numerical results for is the $M=1$ state. 
The electron operator \eqref{elec} then obeys
\begin{equation}
    V_e\propto \;:e^{i\sqrt{\frac{3M+2}{3}}\phi}:.
\end{equation}
We introduce extra fields $\chi_1,\chi_2$, giving
\begin{equation}
    V_j= :e^{i\sqrt{\frac{3M+2}{3}}\phi}::e^{i\beta_j\chi_1}::e^{i\gamma_j\chi_2}:
\end{equation}
where $j\in\{a,b,c\}$ denotes which electron ``type" is meant: as we shall see, the way the electrons are grouped together in the Read-Rezayi wave function \cref{RR} is conveniently represented by introducing a different kind of electron operator for each electron group. Since the free boson OPE \cref{OPE} gives
\begin{equation}
    :e^{i\alpha\phi(z_r)}::e^{i\beta \phi(z_s)}:\sim (z_r-z_s)^{\alpha\beta},
\end{equation}
and as the wave function contains factors $(z_r-z_s)^{2+M}$ if $z_r,z_s$ are in the same subset, matching powers with the OPE of $V_i(z_r)V_i(z_s)$ gives
\begin{equation}
    \frac{3M+2}{3}+\beta_i^2+\gamma_i^2=2+M\implies \beta_i^2+\gamma_i^2=\frac{4}{3}.
\end{equation}
Meanwhile, if $z_r,z_s$ are in different groups, we should have a factor $(z_r-z_s)^M$. To obtain the same factor from the OPE of $V_i(z_r)V_j(z_s),\;i\neq j$, we require
\begin{equation}
    \frac{3M+2}{3}+\beta_i\beta_j+\gamma_i\gamma_j=M\implies \beta_i\beta_j+\gamma_i\gamma_j=-\frac{2}{3},\quad i\neq j.
\end{equation}
The last two equations can be solved by setting the vectors $(\beta_j,\gamma_j)$ to be at angles $2\pi/3$ relative to each other, with norm $\sqrt{\frac{4}{3}}$. We use, for simplicity,
\begin{equation}
\begin{cases}\label{e}
    (\beta_a,\gamma_a)=\sqrt{\frac{4}{3}}(1,0),\;\\ (\beta_b,\gamma_b)=\sqrt{\frac{4}{3}}(-1/2,\sqrt{3}/2),\; \\(\beta_c,\gamma_c)=\sqrt{\frac{4}{3}}(-1/2,-\sqrt{3}/2).
    \end{cases}
\end{equation}
Our electron vertex operators are thus
\begin{equation}
\begin{cases}    \label{electrons}
V_a= :e^{i\frac{3M+2}{\sqrt{q_0}}\phi}::e^{i\frac{4}{\sqrt{q_1}}\chi_1}:,\; \\V_b= :e^{i\frac{3M+2}{\sqrt{q_0}}\phi}::e^{-i\frac{2}{\sqrt{q_1}}\chi_1}::e^{i\frac{2}{\sqrt{q_2}}\chi_2}:,\; \\V_c= :e^{i\frac{3M+2}{\sqrt{q_0}}\phi}::e^{-i\frac{2}{\sqrt{q_1}}\chi_1}::e^{-i\frac{2}{\sqrt{q_2}}\chi_2}:,
\end{cases}
\end{equation}
where $q_0=3(3M+2)=15$ for $M=1$, $q_1=12$ and the $\chi_2$ exponents have been rewritten to make the compactification radius $\sqrt{q_2}=\sqrt{4}$. As we shall see, this is useful to allow for the smallest possible quasi-holes to be described in a unified way.
We emphasize that the physical electron charge is related to $q_0$.

In the Read-Rezayi state, only one subset of electrons should have zeros $(z_i-w_j)^{1}$ with the smallest possible quasi-hole, for which there are three different representations. Thus, the quasi-hole operators for the smallest quasi-hole must obey
\begin{equation}\label{anticomm}
    V_{j}(z)H_{l}(w)\sim(z-w)^{\delta_{j,l}},
\end{equation}
with $H_l$ (where $l\in\{a,b,c\}$) is a quasi-hole of type $l$. One can use
\begin{equation}\label{qhops}
   \begin{cases} H_{a}(w)=:e^{i\frac{1}{\sqrt{q_0}}\phi(w)}::e^{i\frac{2}{\sqrt{q_1}}\chi_1(w)}:,\\
   H_{b}(w)=:e^{i\frac{1}{\sqrt{q_0}}\phi(w)}::e^{-i\frac{1}{\sqrt{q_1}}\chi_1(w)}::e^{i\frac{1}{\sqrt{q_2}}\chi_2(w)}:,\\
   H_{c}(w)=:e^{i\frac{1}{\sqrt{q_0}}\phi(w)}::e^{-i\frac{1}{\sqrt{q_1}}\chi_1(w)}::e^{-i\frac{1}{\sqrt{q_2}}\chi_2(w)}:.
\end{cases}
\end{equation}
It is now clear why we rewrote the electron operators such that $q_2 = 4$. If not, the numerators in the vertex operators would not have been integers, as they must be.

We remark that the generalization of the charge neutrality condition $\sum_i\alpha_i=0$ for the VEV $\langle :e^{i\alpha_1\phi(z_1)}: \cdots :e^{i\alpha_n\phi(z_n)}:\rangle$ is that the exponents for each of the boson field zero modes $\phi_0,\chi_{1,0},\chi_{2,0}$ independently sum to $0$.  Therefore, the total charge due to the electrons in the system -- which is $q_eN_e$ for $N_e$ electrons of charge $q_e$ each -- has to be cancelled by an equally large background charge with the opposite sign for the VEV to produce a nonzero wave function. An additional, physical justification is that we expect the electron charge to be balanced by the positive charge of the underlying lattice. For MPS purposes, it is more convenient to spread out the background charge between the different orbitals instead of inserting all the compensating charge at one location. This procedure allows the matrices for the occupied orbitals to be orbital-independent. Additionally, it helps to keep the $Q_i$ quantum numbers (defined in section \ref{sec:Hilbert}) closer to zero and thus decreases the number of auxiliary states needed in the auxiliary Hilbert spaces.

We now calculate how much background charge should reside on each orbital. From the electron operators in \cref{electrons}, the charge of an electron in our units is $q_{e}=3M+2$, since the charges related to the $\chi_1$ and $\chi_2$ fields are unrelated to the physical electric charge. The filling fraction is $\nu=\frac{3}{3M+2}$. Hence, there are $3(3M+2)$ charge units across $3M+2$ orbitals. There must therefore be a cancelling background charge of $-3(3M+2)$ units over $3M+2$ orbitals, or $-3$ units per orbital. An appropriate background charge operator to insert on each orbital is consequently
\begin{equation}\label{Obg}
    \mathcal{O}_{bg}=e^{-i\frac{3}{\sqrt{q_0}}\phi_0}.
\end{equation}
This gives an operator $\mathcal{O}_{bg}$ for an empty orbital and $\mathcal{O}_{bg}V_a,\mathcal{O}_{bg}V_b,$ or $\mathcal{O}_{bg}V_c$ for an occupied orbital (ignoring quasi-holes).

There is a final complication regarding the charge neutrality condition. Our reasoning above assumes that the number of orbitals $N_{\phi}+1$ is related to the number of electrons $N_e$ as $N_{\phi}+1=\frac{1}{\nu}N_e.$ However, there is a constant deviation from this relation (known as the shift \cite{WenZee1992}), which can be seen as follows. Via the standard mapping from the plane to the cylinder, 
\begin{equation}\label{map}
    z_j\rightarrow e^{-i\frac{2\pi}{L}z_j},
\end{equation}
the number of orbitals on the cylinder corresponds to the maximum power of the $z_j$ in the Read-Rezayi state \cref{Read-Rezayi}, which is $N_{\phi}$. By examining (say) the term only containing factors of $z_1$, one sees that the maximum power is $N_{\phi}=2\big(\frac{N_e}{k}-1\big)+(N_e-1)M=\frac{kM+2}{k}N_e-(M+2).$ Identifying the filling fraction $\nu=\frac{k}{kM+2}$ shows that there are $M+1$ fewer orbitals compared to the value given above. Thus, $M+1$ copies of $\mathcal{O}_{bg}$ remain by the out state, giving
\begin{equation}
    \bra{Q_0} = \bra{0}e^{-i\frac{3(M+1)}{\sqrt{q_0}}\phi_0}=\bra{3(M+1)}.
\end{equation}
The ``out" charge $Q_0 = 3(M+1)$ fixes the total number of electrons. Charge neutrality for $Q_1$ and $Q_2$ then ensures that the number of electrons in each group is equal, in the absence of quasi-holes.

When the state contains a quasi-hole, the out state often needs to be modified. The reason is that the quasi-hole will carry $Q_0$ charge, and may carry $Q_1$ and/or $Q_2$ charge as well.
Also, the number of orbitals increases in the presence of a quasi-hole.
These issues are discussed in more detail in section \ref{sec:implementation}.

Finally, we remark that it is possible to identify combinations of the quasi-hole operators $H_{a},H_{b},H_{c}$ above with the
$\mathbb{Z}_3$ parafermion description \cite{Read1999Parafermions}, which was used in the MPS calculations described in \cite{Wu2014Braiding,Wu2015MPS}.
These identifications are given in \cref{table:ident} and motivated in the next section.
\begin{table}[bh]
    \centering
    \begin{tabular}{c|c}
       $\mathbb{Z}_3$ representation & Equivalent $3$-boson representation \\ \hline
         $\sigma_1e^{i\phi/\sqrt{q_0}}$ & $H_j$\\
         $\sigma_2e^{2i\phi/\sqrt{q_0}}$& $H_jH_k,\quad j\neq k$\\
         $\psi_1e^{2i\phi/\sqrt{q_0}}$& $H_j^2$\\
         $\mathbf{1}e^{3i\phi/\sqrt{q_0}}$& $H_aH_bH_c$\\
         $\epsilon e^{3i\phi/\sqrt{q_0}}$& $H_jH_k^2,\quad j\neq k$\\
         $\psi_2e^{4i\phi/\sqrt{q_0}}$& $H_j^2H_k^2,\quad j\neq k$ \\
    \end{tabular}
    \caption{Identifications between $\mathbb{Z}_3$ and $3$-boson representations of the quasi-holes. The indices $j,k$ are allowed to take any values $a,b,c$ as long as the constraints in the table are fulfilled.}
    \label{table:ident}
\end{table}

\section{Relating the bosonic and parafermionic descriptions of the quasi-holes}
\label{sec:ident}
In this section, we show how various quasi-holes allowed by the Read-Rezayi state can be written in terms of the quasi-hole operators $H_a,H_b$, and $H_c$, and how the parafermionic ``minimal description" discussed in \cite{Read1999Parafermions} and utilized in \cite{Wu2014Braiding,Wu2015MPS} is related to these operators, as shown in \cref{table:ident}.
Schematically, quasi-hole operators in the $\mathbb{Z}_3$ description are of the form
``$\mathbb{Z}_3\;\mathrm{parafermion}\times e^{i \frac{d}{\sqrt{q_0}}\phi}$",
with $d$ the quasi-hole charge in units for which the absolute value of the electron charge is $3M+2$, as can be seen from the electron operators in \eqref{e}. We focus on the quasi-holes for which the parafermion field is $\sigma_1$ ($d=1$), $\sigma_2$ or $\psi_1$ ($d=2$ for both), $\epsilon$ or $\mathbf{1}$ ($d=3$ for both, the latter corresponding to a Laughlin quasi-hole), and $\psi_2$ ($d=4$).

\indent Since the smallest-charge quasi-hole in the $\mathbb{Z}_3$ description is the $\sigma_1e^{i\phi/\sqrt{q_0}}$ quasi-hole, the $3$-boson equivalent of the $\sigma_1$ quasi-hole must be the one with the same electric charge, i.e. $H_a$, $H_b$ or $H_c$, with any choice being equally permissible. 

For the $d=2$ quasi-holes, we have that $\sigma_1\times\sigma_1 = \sigma_2+\psi_1$. Hence, we expect to be able to combine two $H_j$ operators ($j\in\{a,b,c\}$) to obtain either of these possibilities. Since $H_a,H_b,H_c$ all represent $\sigma_1$, the only distinction of importance is whether $\sigma_1\times \sigma_1\leftrightarrow H_jH_k$ has $j=k$ or $j\neq k$. For reasons which will become clear shortly, we must have $j\neq k$ for $\sigma_2$ and $j=k$ for $\psi_1$.

The Laughlin quasi-hole corresponds to $\mathbf{1}e^{i \frac{3}{\sqrt{q_0}}\phi(w)}$ in the $\mathbb{Z}_3$ theory. From the wave function \cref{wavefcn Laughlin}, there is a factor $(z_i-w)$ for each electron coordinate $z_i$, regardless of which particle group it belongs to. Therefore, the Laughlin quasi-hole must have a factor $H_j$ for each group $j=a,b,c$. Thus, we represent it as $H_aH_bH_c$. Indeed, one easily sees that
\begin{equation}
    H_{a}(w)H_{b}(w)H_{c}(w)=e^{i \frac{3}{\sqrt{q_0}}\phi(w)}=\mathbf{1}e^{i \frac{3}{\sqrt{q_0}}\phi(w)} \ .
\end{equation}
In other words, all quasi-hole operators have to be of different types. This is why we represent $\sigma_2e^{2i\phi/\sqrt{q_0}}$ using $H_jH_k,j\neq k$:
$\sigma_1\times \sigma_2$ can yield $\mathbf{1}$, which in the boson language is taken care of by choosing $H_j\leftrightarrow \sigma_1e^{i\phi/\sqrt{q_0}},H_kH_l\leftrightarrow \sigma_2e^{2i\phi/\sqrt{q_0}}$ such that $j,k,l$ are all different, giving $H_aH_bH_c$. The inability of $\sigma_1\times \psi_1$ to give $\mathbf{1}$ is enforced by representing $\psi_1e^{2i\phi/\sqrt{q_0}}$ as $H_j^2$, so that no choice of vertex operator $H_k$ for $\sigma_1e^{i\phi/\sqrt{q_0}}$ can make $H_kH_j^2$ equal $H_aH_bH_c$. 

It should be possible to use the fusion rule $\sigma_1\times\sigma_2=\mathbf{1}+\epsilon$  to obtain the $\epsilon e^{3i\phi/\sqrt{q_0}}$ quasi-hole. From our previous representations, it follows that $\epsilon e^{3i\phi/\sqrt{q_0}}\leftrightarrow H_jH_kH_l,$ with $k\neq l$ from the expression for the $\sigma_2$ quasi-hole operator. Here, $j$ must be coincident with either $k$ or $l$ not to give $j,k,l$ all different, which would give $H_aH_bH_c$, i.e. the Laughlin quasi-hole. We therefore demand that the $\epsilon$ quasi-hole is represented using $H_jH_k^2$ with $j\neq k$. 

Finally, the $\psi_2e^{4i\phi/\sqrt{q_0}}$ operator can be fused using $\psi_1\times \psi_1=\psi_2$, $\sigma_1\times \epsilon=\psi_2+\sigma_1$, or using $\sigma_2\times\sigma_2=\psi_2+\sigma_1$. The first fusion implies that the $\psi_2$ hole must be either $H_j^2H_k^2$, $j\neq k,$ or $H_j^4$.  The second fusion gives $H_jH_kH_l^2$ where $k\neq l$ but $j$ is indeterminate, i.e. we either have  $H_jH_kH_l^2$ (all indices different), $H_k^2H_l^2$ or $H_kH_l^3$. The third fusion gives $H_jH_kH_lH_m$ where $j\neq k$ and $l\neq m$. Because there are only three quasi-hole operators, not all indices can be different here, and we arrive at something of the form $H_jH_kH_l^2$ (all indices different) or $H_j^2H_k^2$ ($j\neq k$). Across our three fusion paths, the only representation of the $\psi_2e^{4i\phi/\sqrt{q_0}}$ quasi-hole that allows it to have the same operator content regardless of the path in which it is fused is $H_j^2H_k^2,$ where $j\neq k$, so this representation is the one we use.

Before closing this section, we remark that the representations in \cref{table:ident} can be understood also in terms of the wave functions in section \ref{sec:wavefcns}. For instance, identifying the $(\sigma_1,\frac{1}{3M+2})$ quasi-hole with $H_j$ for some $j\in\{a,b,c\}$ corresponds to how the wave function \cref{wavefcn sigma1} has zeros with the electrons in one group, and that group only. From \cref{anticomm}, we see that the choice of $j$ in the $H_j(w)$ operator selects one group of electrons with which factors $(z_i-w)$ appear. The freedom to select $j\in\{a,b,c\}$ arbitrarily then represents the way in which the zeros can be with any electron group. Similar arguments apply to the other quasi-holes: for instance, the $(\epsilon,\frac{3}{3M+2})$ quasi-hole wave function has one zero with one electron group and a double zero with another electron group; c.f. the wave function \cref{wavefcn eps} and the restriction $j\neq k$ in \cref{table:ident}. The representation of the Laughlin quasi-hole as $H_aH_bH_c$, meanwhile, reflects the way in which each electron subset in the wave function \cref{wavefcn Laughlin} has a zero with the quasi-hole.

\section{MPS description: auxiliary Hilbert space}
\label{sec:Hilbert}

In this section, we start our discussion of the MPS description for FQH states \cite{Zaletel2012MPS,estienne2013fractional}.
We follow the conventions of \cite{Kjall2018MPS}.
In particular, we discuss mapping the expectation value \cref{wavefcn} to a finite cylinder geometry, and introduce the quantum numbers characterising the auxiliary Hilbert space used for the MPS.

Even though FQH states live in a disc geometry experimentally, it turns out to be profitable to instead compute observables in a finite cylinder geometry through the mapping \cref{map} which turns the LLL single-particle orbital into 
\begin{equation}
\label{eq:single-particle-orbital}
    \phi_l(\tau,x)=\frac{1}{\sqrt{Ll_B\sqrt{\pi}}}e^{-\frac{i}{l_B^2}\tau_lx}e^{-\frac{1}{2l_B^2}(\tau-\tau_l)^2},
\end{equation}
with $\tau$ the length coordinate and $x$ the angular coordinate along the cylinder, as well as $\tau_l:=l\delta\tau=\frac{2\pi l_B^2}{L}l$. One can see that the orbitals are peaked around $\tau=\tau_l$, with an inter-peak distance of $\delta\tau=2\pi l_B^2/L$.

We denote the orbital occupation numbers for electrons of type $a, b, c$ by $m_{a,l},m_{b,l},m_{c,l}$, while the number of
quasi-holes of type $a, b, c$ inserted between orbitals $l$ and $l+1$ are denoted by $n_{a,l},n_{b,l},n_{c,l}$. We note that we will only
consider situations where we insert a quasi-hole (which can be composed of several quasi-holes of type $a, b, c$) in one location.
Since the electrons are fermionic (and we are effectively dealing with spin-less electrons), we demand $m_{a,j},m_{b,j},m_{c,j}\in\{0,1\}$ with at most one occupation number being nonzero, i.e. $m_{a,j}=m_{b,j}=m_{c,j}=0$ for empty orbitals and $m_{a,j}+m_{b,j}+m_{c,j}=1$ for occupied orbitals.

To represent the states using matrix product state techniques, we insert auxiliary Hilbert spaces along the bonds between the $\tau=\tau_l$ orbitals. On the level of the expectation value, this corresponds to inserting between each pair of adjacent operators in the expectation value a resolution of the identity as
\begin{align}\label{resid}
        \mathbf{1}= \sum_{\{Q,P,\mu\}}\ket{\{Q_j,P_j,\mu_j\}_{j=0}^2}\bra{\{Q_j,P_j,\mu_j\}_{j=0}^2},
\end{align}
where the sum is understood as being over all allowed values of the charge quantum numbers $Q_0,Q_1,Q_2$, momenta $P_0,P_1,P_2$ and partitions $\mu_0,\mu_1,\mu_2$, all of which will be defined shortly. We define the charge quantum numbers through ($i=0,1,2$)
\begin{equation}
\begin{cases}\label{piQ}
    \pi_{i,0}\ket{Q_0,Q_1,Q_2}=\frac{Q_i}{\sqrt{q_i}}\ket{Q_0,Q_1,Q_2},\\
    a_{i,n}\ket{Q_0,Q_1,Q_2}=0,\quad n>0
    \end{cases}
\end{equation}
where the $\pi_{i,0}$ operator is the logarithmic mode of either $\phi,\chi_1$ or $\chi_2$ (for $i=0,1$ and $2$, respectively), and the $a_{i,n}$ are the annihilation operators belonging to the same field (c.f. eq. \eqref{modexpbos}). We may then add or subtract charge through
\begin{align}\label{addcharge}
    &e^{i\frac{\alpha_0}{\sqrt{q_0}} \phi_0}e^{i\frac{\alpha_1}{\sqrt{q_1}} \chi_{1,0}}e^{i\frac{\alpha_2}{\sqrt{q_2}} \chi_{2,0}}\ket{Q_0,Q_1,Q_2}=\nonumber\\&\ket{Q_0+\alpha_0,Q_1+\alpha_1,Q_2+\alpha_2},
\end{align}
where the numbers $q_i$ describe the ``elementary charge" of the $\phi_0,\chi_1$ or $\chi_2$ field, and are related to the radii $R_i$ of the compactified bosons through $R_i=\sqrt{q_i}$. Here, $\phi_{0},\chi_{1,0},\chi_{2,0}$ are the zero modes of the fields $\phi,\chi_1$ and $\chi_2$. We remind the reader that the $Q_0$ quantum number represents the physical electric charge, while the charges $Q_1$ and $Q_2$ are topological and effectively encode the $\mathbb{Z}_3$ topological sectors. 
It follows from \cref{addcharge} that the charge eigenstates can be made explicit through the relation 
 
\begin{equation}
    \ket{Q_0,Q_1,Q_2}=e^{i\frac{Q_0}{\sqrt{q_0}}\phi_0}e^{i\frac{Q_1}{\sqrt{q_1}}\chi_{1,0}}e^{i\frac{Q_2}{\sqrt{q_2}}\chi_{2,0}}\ket{0}.
\end{equation}
\indent To define the momenta $P_0,P_1,P_2$, we must first define the partitions $\mu_0=(\mu_{0,1},\mu_{0,2}, \ldots ,\mu_{0,k}),\mu_1=(\mu_{1,1},\mu_{1,2}, \ldots ,\mu_{1,l})$ and $\mu_2=(\mu_{2,1},\mu_{2,2}, \ldots ,\mu_{2,m})$, all of which are sets of weakly decreasing positive integers. Together with the independent creation operators $a_{0,-j},a_{1,-j},a_{2,-j}$ for the fields $\phi,\chi_1$ and $\chi_2$, and the occupation numbers $r_{i,j}$ for mode number $j$ of field number $i$, we may define 
\begin{align}    &\ket{Q_0,Q_1,Q_2,P_0,P_1,P_2,\mu_0,\mu_1,\mu_2} \nonumber\\&= \frac{1}{\sqrt{z_{\mu_0}z_{\mu_1}z_{\mu_2}}}\prod_{j=1}^{\infty}a^{r_{0,j}}_{0,-j}a^{r_{1,j}}_{1,-j}a^{r_{2,j}}_{2,-j}\ket{Q_0,Q_1,Q_2},
\end{align}
where the normalizing factors are $z_{\mu_i}=\prod_{j=1}^{\infty}j^{r_{i,j}}(r_{i,j}!)$ for $i=0,1,2$.
Now, the partitions and momenta can be explained as follows: the momenta are 
\begin{equation}
    P_i=\sum_{j>0}jr_{i,j},\quad i=0,1,2. 
\end{equation}
The partitions $\mu_i$ are partitions of the momenta, i.e. $P_i=\sum_j\mu_{i,j}$, and the number of parts of $\mu_i$ which are equal to a certain $j$ is exactly $r_{i,j}$. Finally, we note for later use that 
\begin{align}\label{oscP}
    &\sum_{j>0}(a_{0,-j}a_{0,j}+a_{1,-j}a_{1,j}+a_{2,-j}a_{2,j})\ket{\{Q_l,P_l,\mu_l\}_{l=0}^2}=\nonumber\\&(P_0+P_1+P_2)\ket{\{Q_l,P_l,\mu_l\}_{l=0}^2}.
\end{align}

In the language of MPS, the occupation numbers $m_{a,j},m_{b,j},m_{c,j}$ (and the corresponding numbers for the quasi-hole occupations) are the physical, or free, indices. The charge quantum numbers $Q_i$, momenta $P_i$ and partitions of the momenta $\mu_i$ are contracted over, and so correspond to auxiliary degrees of freedom connecting the states at different orbitals along the cylinder. The physical indices are contracted when calculating actual expectation values.

\section{MPS description: matrix elements}
\label{sec:matrices}

The MPS method can be successfully used for quantum Hall states for two reasons. Firstly, there is no need to repeatedly evaluate the wave function, as is necessary in a Monte Carlo approach. Secondly, the information needed to describe the entanglement between neighbouring orbitals does not increase as rapidly as one would naïvely expect. Instead, in gapped phases with a finite correlation length, this information is bounded from above (see \cite{EisertCramerPlenio2010} for a review).
Consequently, it suffices to use matrices of moderate, \textit{finite} dimension.

To construct a FQH state using MPS, we note that the wave functions can be written schematically as
\begin{equation}
    \Psi = \sum_{\lambda} c_{\lambda}\mathrm{sl}_{\lambda},
\end{equation}
where $\mathrm{sl}_{\lambda}$ is a Slater determinant corresponding to the set of occupied single-particle orbitals $\lambda=(l_{N_e},l_{N_e-1}, \ldots ,l_1)$ with $0\leq l_1<l_2< \cdots <l_{N_e}\leq N_{\phi}$. The coefficients $c_{\lambda}$ can be computed (although the MPS method does not rely on doing so) as 
\begin{align}
\label{eq:c_lambda}
    c_{\lambda} &=
    \bigg(\prod_{j=1}^{N_e}\int_{-L/2}^{L/2}\frac{dx_j}{L}\bigg) \\\nonumber
    &\langle \mathcal{O}_{bg} V_{d_{N_e}}(\tau_{l_{N_e}},x_{N_e}) \cdots \mathcal{O}_{bg}V_{d_1}(\tau_{l_1},x_1)\rangle,
\end{align}
where $d_1, \ldots ,d_{N_e}\in\{a,b,c\}$ represent different choices for the different electron operators, and where the $\mathcal{O}_{bg}$ background charge operator creates the background charge associated with one orbital rather than that of the whole system. For the state without quasi-holes, charge neutrality (and a non-zero expectation value) requires that the number of $a,b$ and $c$ electrons are all equal. The expectation value above is assumed to be $\tau$-ordered, i.e. $\tau_{l_{N_e}}>\tau_{l_{N_e-1}}> \cdots >\tau_{l_1}.$ A way to place operators at the appropriate time coordinates is by using the following time evolution operator:
\begin{equation}
    U(\tau'-\tau)=e^{-(\tau'-\tau)H},\quad H=\frac{2\pi}{L}L_0,
\end{equation}
where $L_0=\frac{1}{2}(\pi_{0,0}^2+\pi_{1,0}^2+\pi_{2,0}^2)+\sum_{j>0}(a_{0,-j}a_{0,j}+a_{1,-j}a_{1,j}+a_{2,-j}a_{2,j})$. To represent a continuously distributed background charge between the orbitals at $\tau$ and $\tau'=\tau+\delta\tau$, we ``spread out" the effect of the charge operator $\mathcal{O}_{bg}=e^{-i\frac{3}{\sqrt{q_0}}\phi_0}$ over a $\tau$ interval of width $\delta\tau/N$, time evolve by $\delta\tau/N$, add an additional small amount of charge, and so on. The continuous charge distribution is achieved as a limiting case as $N\rightarrow \infty.$ In other words, we write (see \cite{estienne2013fractional,Kjall2018MPS} for more details)
\begin{align}
\label{eq:time-evolution}
    U(\tau'-\tau)&=\lim_{N\rightarrow \infty}\big(e^{-\frac{2\pi\delta\tau}{NL}L_0}e^{-i\frac{3}{N\sqrt{q_0}}\phi_0}\big)^N\nonumber\\
    &=e^{-\frac{2\pi\delta\tau}{L}L_0-i\frac{3}{\sqrt{q_0}}\phi_0}\nonumber\\
    &=e^{-\frac{2\pi\delta\tau}{L}(L_0+\frac{3}{2\sqrt{q_0}}\pi_{0,0}+\frac{3}{2q_0})}e^{-i\frac{3}{\sqrt{q_0}}\phi_0},
\end{align}
where the last equality follows from the Baker-Campbell-Hausdorff theorem with $[\pi_{r,0},\phi_{s,0}]=-i\delta_{rs}$. Defining $U'(\delta\tau):=e^{-\frac{2\pi\delta\tau}{L}(L_0+\frac{3}{2\sqrt{q_0}}\pi_{0,0}+\frac{3}{2q_0})}$, we see that the time evolution and spread-out background charge together make the expectation value, for some ``out-charge" $Q_{0,out}$, 
\begin{align}
\label{eq:correlator}
    \bra{Q_{0,out}} &\mathcal{O}_{bg} V_{d_{N_e}}(\tau_{l_{N_e}},x_{N_e}) \cdots \mathcal{O}_{bg}V_{d_1}(\tau_{l_1},x_1)\ket{0}=\nonumber\\
    \bra{Q_{0,out}} &U'(\tau_{N_{\phi}+1}-\tau_{l_{N_e}})\mathcal{O}_{bg} V_{d_{N_e}}(0,x_{N_e})\nonumber\\
    &U'(\tau_{l_{N_e}}-\tau_{l_{N_e-1}})\mathcal{O}_{bg} V_{d_{N_e-1}}(0,x_{N_e-1})\cdots \nonumber\\
    &U'(\tau_{l_2}-\tau_{l_1})\mathcal{O}_{bg}V_{d_1}(0,x_1)U'(\tau_{l_1}-0)\ket{0} \ ,
\end{align}
where the rightmost state, or ``in state", is the vacuum. We point out that there will be factors of $U'(\delta\tau)\mathcal{O}_{bg}$ inserted above even at the positions where there are no electrons or quasi-holes, i.e. at the empty orbitals.
These factors are implicitly understood in \cref{eq:c_lambda} (without the operator $U'(\delta\tau)$) and \cref{eq:correlator}.

Using the electron and quasi-hole operators, as well as the spread-out background charge operator, the resolution of the identity \eqref{resid} gives matrix elements for each site. 
We denote the matrices as $B^{[i]}$ where the values $i=0,a,b,c$ signify an empty orbital ($i=0$) or an orbital occupied by an electron of type $a$, $b$ or $c$. The matrix element for an empty orbital corresponds to the expectation value
\begin{align}\label{B0}
    B^{[0]}&=\bra{\{Q_j',P_j',\mu_j'\}_{j=0}^2}U'(\delta\tau)e^{-i\frac{3}{\sqrt{q_0}}\phi_0}\ket{\{Q_j,P_j,\mu_j\}_{j=0}^2}\nonumber\\&=e^{-\frac{2\pi\delta\tau}{L}\big(\frac{(Q_0')^2}{2q_0}+\frac{3Q_0'}{2q_0}+\frac{(Q_1')^2}{2q_1}+\frac{Q_2'^2}{2q_2}+P_0'+P_1'+P_2'\big)}\nonumber\\
    &\times\delta_{Q_0',Q_0-3}\delta_{Q_1',Q_1}\delta_{Q_2',Q_2}\delta_{P_0',P_0}\delta_{P_1',P_1}\delta_{P_2',P_2}\nonumber\\&\times\delta_{\mu_0',\mu_0}\delta_{\mu_1',\mu_1}\delta_{\mu_2',\mu_2},
\end{align}
where we used the definitions of $U'(\delta\tau)$ and $L_0$, as well as \cref{piQ,oscP}. From the form of
$U'(\delta\tau)$, \cref{eq:time-evolution}, it is clear that there should be a factor $e^{-\frac{2\pi\delta\tau}{L} \frac{3}{2q_0}}$,
which amounts to an uninteresting overall normalization. We drop this factor in all matrix elements.

For the matrices corresponding to occupied orbitals, we wish to trade the $z_j$ position dependence for orbital dependence. To do this, we recall three facts about generic vertex operators of a free boson field $V(z) = :e^{i\beta\phi(z)}:$. Firstly, one can perform the Fourier expansion
\begin{equation}\label{modexp}
    V(z)=\sum_{l\in\mathbb{Z}}z^lV_{-h-l},\quad V_{-h-l}= \frac{1}{2\pi i}\oint \frac{dz}{z}z^{-l}V(z),
\end{equation}
with $h$ the conformal dimension of the mode. Secondly, we have
\begin{equation}\label{vertexelem}
\bra{Q',P',\mu'}V(z)\ket{Q,P,\mu}=\delta_{Q',Q+\sqrt{q}\beta}z^{\frac{\beta Q}{\sqrt{q}}+P'-P}A_{\mu',\mu}^{\beta},
\end{equation}
where
\begin{widetext}
\begin{equation}
    A_{\mu',\mu}^{\beta}=\prod_{j=1}^{\infty}\sum_{s=0}^{s_j}\sum_{r=0}^{r_j}\delta_{r_j-r,s_j-s}\frac{(-1)^r}{\sqrt{r!s!}}\bigg(\frac{\beta}{\sqrt{j}}\bigg)^{r+s}\sqrt{{\binom{s_j}{s}}{\binom{r_j}{r}}}.
\end{equation}
Thirdly, from \cref{modexp,vertexelem} it follows that
\begin{align}
    \bra{Q',P',\mu'}V_{-h}\ket{Q,P,\mu}&= \frac{1}{2\pi i}\oint \frac{dz}{z}\delta_{Q',Q+\sqrt{q}\beta}z^{\frac{\beta Q}{\sqrt{q}}+P'-P}A_{\mu',\mu}^{\beta}\nonumber\\&=\delta_{Q',Q+\sqrt{q}\beta}\delta_{P',P-\frac{\beta Q}{\sqrt{q}}}A_{\mu',\mu}^{\beta}.
\end{align}
\end{widetext}
This information can be used to derive the MPS matrices for the different occupied orbitals. Since the different boson fields are independent, it follows from the above that 
\begin{align}\label{ba}
    B^{[a]}&=
    \int_{-L/2}^{L/2}\frac{dx}{L}\bra{\{Q_j',P_j',\mu_j'\}_{j=0}^2}\nonumber\\
    &\qquad\quad\qquad\quad U'(\delta\tau)e^{-i\frac{3}{\sqrt{q_0}}\phi_0}V_a\ket{\{Q_j,P_j,\mu_j\}_{j=0}^2}\nonumber\\&=e^{-\frac{2\pi\delta\tau}{L}\big(\frac{(Q_0')^2}{2q_0}+\frac{3Q_0'}{2q_0}+\frac{(Q_1')^2}{2q_1}+\frac{Q_2'^2}{2q_2}+P_0'+P_1'+P_2'\big)}\nonumber\\&\times\delta_{Q_0',Q_0+3M-1}\delta_{Q_1',Q_1+4}\delta_{Q_2',Q_2}\nonumber\\&\times\delta_{(P_0+P_1)',(P_0+P_1)-\frac{3M+2}{q_0}Q_0-\frac{4}{q_1}Q_1}\delta_{P_2',P_2}\nonumber\\&\times A^{\frac{3M+2}{\sqrt{q_0}}}_{\mu_0',\mu_0}A^{\frac{4}{\sqrt{q_1}}}_{\mu_1',\mu_1}\delta_{\mu_2',\mu_2},
\end{align}
where $P_2'=P_2,\mu_2'=\mu_2$ are both direct consequences of the fact that $V_a$ does not depend on $\chi_2$.
We remind the reader that the $e^{-i\frac{3}{\sqrt{q_0}}\phi_0}$ factor is the $\mathcal{O}_{bg}$ background charge operator of section \ref{sec:operators}, and is needed for charge neutrality. Similarly,
\begin{align}\label{bb}
    B^{[b]}&=
    \int_{-L/2}^{L/2}\frac{dx}{L}\bra{\{Q_j',P_j',\mu_j'\}_{j=0}^2}\nonumber\\
    &\qquad \qquad\qquad U'(\delta\tau)e^{-i\frac{3}{\sqrt{q_0}}\phi_0}V_b\ket{\{Q_j,P_j,\mu_j\}_{j=0}^2}\nonumber\\&=e^{-\frac{2\pi\delta\tau}{L}\big(\frac{(Q_0')^2}{2q_0}+\frac{3Q_0'}{2q_0}+\frac{(Q_1')^2}{2q_1}+\frac{Q_2'^2}{2q_2}+P_0'+P_1'+P_2'\big)}\nonumber\\
    &\times\delta_{Q_0',Q_0+3M-1}\delta_{Q_1',Q_1-2}\delta_{Q_2',Q_2+2}\nonumber\\&\times \delta_{(P_0+P_1+P_2)',(P_0+P_1+P_2)-\frac{3M+2}{q_0}Q_0+\frac{2}{q_1}Q_1-\frac{2}{q_2}Q_2}\nonumber\\&\times A^{\frac{3M+2}{\sqrt{q_0}}}_{\mu_0',\mu_0}A^{\frac{-2}{\sqrt{q_1}}}_{\mu_1',\mu_1}A^{\frac{2}{\sqrt{q_2}}}_{\mu_2',\mu_2},
\end{align}
and 
\begin{align}\label{bc}
    B^{[c]}&=
    \int_{-L/2}^{L/2}\frac{dx}{L}\bra{\{Q_j',P_j',\mu_j'\}_{j=0}^2}\nonumber\\
    &\qquad\qquad\qquad U'(\delta\tau)e^{-i\frac{3}{\sqrt{q_0}}\phi_0}V_c\ket{\{Q_j,P_j,\mu_j\}_{j=0}^2}\nonumber\\&=e^{-\frac{2\pi\delta\tau}{L}\big(\frac{(Q_0')^2}{2q_0}+\frac{3Q_0'}{2q_0}+\frac{(Q_1')^2}{2q_1}+\frac{Q_2'^2}{2q_2}+P_0'+P_1'+P_2'\big)}\nonumber\\&\times\delta_{Q_0',Q_0+3M-1}\delta_{Q_1',Q_1-2}\delta_{Q_2',Q_2-2}\nonumber\\&\times\delta_{(P_0+P_1+P_2)',(P_0+P_1+P_2)-\frac{3M+2}{q_0}Q_0+\frac{2}{q_1}Q_1+\frac{2}{q_2}Q_2}\nonumber\\&\times A^{\frac{3M+2}{\sqrt{q_0}}}_{\mu_0',\mu_0}A^{\frac{-2}{\sqrt{q_1}}}_{\mu_1',\mu_1}A^{\frac{-2}{\sqrt{q_2}}}_{\mu_2',\mu_2}.
\end{align}
Using similar reasoning, the elements of the quasi-hole matrices $H_{\tilde{l}}^{[a]},H_{\tilde{l}}^{[b]},H_{\tilde{l}}^{[c]}$ due to a quasi-hole insertion between orbitals $\tilde{l}-1$ and $\tilde{l}$ can be computed. We first state the matrix elements, and then explain where the different factors come from. If the quasi-hole is inserted at $\tau=\tau_{\alpha}$, the matrix elements become  
\begin{align}\label{ha}
    H_{\tilde{l}}^{[a]}&=(-1)^{\frac{Q_0+3\tilde{l}}{3(3M+2)}+\frac{Q_1}{6}}
    \nonumber\\&\times e^{-\frac{2\pi i x_{\alpha}}{L}\big(\frac{Q_0+3\tilde{l}}{q_0}+\frac{2Q_1}{q_1}+(P_0+P_1)'-(P_0+P_1)\big)}\nonumber\\
    &\times e^{\frac{2\pi}{L}(\tilde{l}\delta\tau-\tau_{\alpha})\big(\frac{Q_0^2}{2q_0}+\frac{3Q_0}{2q_0}+\frac{Q_1^2}{2q_1}+\frac{Q_2^2}{2q_2}+P_0+P_1+P_2\big)}
    \nonumber\\
    &
    \times e^{-\frac{2\pi}{L}(\tilde{l}\delta\tau-\tau_{\alpha})\big(\frac{(Q'_0)^2}{2q_0}+\frac{3Q'_0}{2q_0}+\frac{(Q'_1)^2}{2q_1}+\frac{(Q'_2)^2}{2q_2}+P_0'+P_1'+P_2'\big)}\nonumber\\
    &\times\delta_{Q_0',Q_0+1}\delta_{Q_1',Q_1+2}\delta_{Q_2',Q_2}\nonumber\\&\times\delta_{P_2',P_2}A_{\mu_0',\mu_0}^{\frac{1}{\sqrt{q_0}}}A_{\mu_1',\mu_1}^{\frac{2}{\sqrt{q_1}}}\delta_{\mu_2',\mu_2},
\end{align}
where $P_2$ trivially remains constant as was the case for $B^{[a]}$.
Here, $(\tau_{\alpha},x_{\alpha})$ are the coordinates of the quasi-hole. Furthermore, 
\begin{align}\label{hb}
    H_{\tilde{l}}^{[b]}&=(-1)^{\frac{Q_0+3\tilde{l}}{3(3M+2)}-\frac{Q_1}{12}+\frac{Q_2}{4}}\nonumber\\&\times e^{-\frac{2\pi ix_{\alpha}}{L}\big(\frac{Q_0+3\tilde{l}}{q_0}-\frac{Q_1}{q_1}+\frac{Q_2}{q_2}+(P_0+P_1+P_2)'-(P_0+P_1+P_2)\big)}\nonumber\\
    &\times e^{\frac{2\pi}{L}(\tilde{l}\delta\tau-\tau_{\alpha})\big(\frac{Q_0^2}{2q_0}+\frac{3Q_0}{2q_0}+\frac{Q_1^2}{2q_1}+\frac{Q_2^2}{2q_2}+P_0+P_1+P_2\big)}
    \nonumber\\
    &\times e^{-\frac{2\pi}{L}(\tilde{l}\delta\tau-\tau_{\alpha})\big(\frac{(Q_0')^2}{2q_0}+\frac{3Q'_0}{2q_0}+\frac{(Q'_1)^2}{2q_1}+\frac{(Q'_2)^2}{2q_2}+P_0'+P_1'+P_2'\big)}
    \nonumber\\&\times\delta_{Q_0',Q_0+1}\delta_{Q_1',Q_1-1}\delta_{Q_2',Q_2+1}\nonumber\\&\times A_{\mu_0',\mu_0}^{\frac{1}{\sqrt{q_0}}}A_{\mu_1',\mu_1}^{-\frac{1}{\sqrt{q_1}}}A_{\mu_2',\mu_2}^{\frac{1}{\sqrt{q_2}}},
\end{align}
and
\begin{align}\label{hc}
    H_{\tilde{l}}^{[c]}&=(-1)^{\frac{Q_0+3\tilde{l}}{3(3M+2)}-\frac{Q_1}{12}-\frac{Q_2}{4}}
    \nonumber\\&\times e^{-\frac{2\pi ix_{\alpha}}{L}\big(\frac{Q_0+3\tilde{l}}{q_0}-\frac{Q_1}{q_1}-\frac{Q_2}{q_2}+(P_0+P_1+P_2)'-(P_0+P_1+P_2)\big)}\nonumber\\
    &\times e^{\frac{2\pi}{L}(\tilde{l}\delta\tau-\tau_{\alpha})\big(\frac{Q_0^2}{2q_0}+\frac{3Q_0}{2q_0}+\frac{Q_1^2}{2q_1}+\frac{Q_2^2}{2q_2}+P_0+P_1+P_2\big)}\nonumber\\
   &\times e^{\frac{-2\pi}{L}(\tilde{l}\delta\tau-\tau_{\alpha})\big(\frac{(Q'_0)^2}{2q_0}+\frac{3Q'_0}{2q_0}+\frac{(Q'_1)^2}{2q_1}+\frac{(Q'_2)^2}{2q_2}+P_0'+P_1'+P_2'\big)}\nonumber\\ &\times\delta_{Q_0',Q_0+1}\delta_{Q_1',Q_1-1}\delta_{Q_2',Q_2-1}\nonumber\\&\times A_{\mu_0',\mu_0}^{\frac{1}{\sqrt{q_0}}}A_{\mu_1',\mu_1}^{-\frac{1}{\sqrt{q_1}}}A_{\mu_2',\mu_2}^{-\frac{1}{\sqrt{q_2}}}.
\end{align}
The Kronecker $\delta$'s for the charge quantum numbers and the factors $A_{\mu_0',\mu_0}^{\frac{1}{\sqrt{q_0}}}$ etc. originate in calculating the matrix elements using \cref{vertexelem}. The exponents of the type $e^{-\frac{2\pi i x_\alpha}{L} c}$ (where $c$ depends on the quantum numbers and $\tilde{l}$) is what remains of the coordinate dependence after setting $\tau_\alpha = 0$ (recall that the $\tau$ dependence is ``taken care of" by the free time evolution; see \cref{eq:correlator}).

To explain the $\tau_\alpha$ dependent exponential factors (see also \cite{Wu2015MPS}), we note that the matrix elements corresponding to the actual orbitals (either empty, or with an electron of arbitrary type) incorporate the effect of the free time evolution from the orbital to the next one. If we insert the matrix corresponding to the quasi-hole at position $\tau_\alpha$ in between the matrices corresponding to the orbitals $\tilde{l}-1$ and $\tilde{l}$ (such that $(\tilde{l}-1)\delta\tau<\tau_{\alpha}<\tilde{l}\delta\tau$), we need to time evolve \textit{backwards} from $\tau=\tilde{l}\delta\tau$ to $\tau=\tau_{\alpha}$, apply the quasi-hole operator with its $\tau$ coordinate set to $0$, and then time-evolve \textit{forward} again up to $\tau=\tilde{l}\delta\tau$ before applying the next operator. This time evolution produces the $\tau_\alpha$ dependent exponentials in the matrix elements \cref{ha,hb,hc}.

Finally, we need to explain the (charge quantum number dependent) signs. These signs have their origin in the factors $(z-w)$ that are present for an electron and a quasi-hole belonging to the same ``particle subset", i.e. from the anti-commutation between electrons and quasi-holes of the same type. Due to the Kronecker $\delta$'s for the charge quantum numbers in the matrix elements for the empty and occupied orbitals, one can at any given orbital deduce the number of matrices corresponding to an electron of a given type that were inserted already. Using this information gives rise to the signs present in the matrix elements \cref{ha,hb,hc}.

With all the matrix elements that we need for the MPS calculations in place, we would like to discuss the effect of the time evolution in more detail. That is, the combined effect of the exponentials of the form $e^{-\frac{2\pi\delta\tau}{L}( \cdot )}$ in the matrix elements \cref{B0,ba,bb,bc,ha,hb,hc}. We denote this factor (which via the quantum numbers depends on the orbital occupation numbers) by $\mathcal{U}$. In the absence of quasi-holes, this factor is given by
\begin{widetext}
\begin{equation}\label{normaliz}
\mathcal{U} =  \exp\bigg(-\frac{2\pi\delta\tau}{L} \sum_{j=0}^{N_{\phi}+1}\bigg[\frac{Q^2_{0,j}}{2q_0}+\frac{3Q_{0,j}}{2q_0}+\frac{Q_{1,j}^2}{2q_1}+\frac{Q_{2,j}^2}{2q_2}+(P_0+P_1+P_2)_j\bigg]\bigg) \ .
\end{equation}
\end{widetext}
This expression is modified slightly if a quasi-hole is present in the system, because the matrix elements in \cref{ha,hb,hc} affect the quantum numbers and contain additional exponential factors contributing to $\mathcal{U}$. We recall that $\delta\tau=2\pi l_B^2/L$
and that we set $l_B = 1$.

The exponential $\mathcal{U}$ of \cref{normaliz} can be seen to decay with growing charge and momentum quantum numbers, which leads to a natural truncation of the auxiliary Hilbert space. The states with high values of the $|Q|$ and $P$ numbers (and consequently many possible partitions $\mu$ of $P$, which increases the necessary dimensions of the auxiliary Hilbert spaces) are also the states where the exponential factor is the closest to zero. It should be mentioned, however, that the factor $\frac{2\pi\delta\tau}{L} = \bigl(\frac{2\pi l_B}{L}\bigr)^2$ is actually quite small (compared to one) for the circumferences usually considered, $L \sim 20-30 l_B$.

The exponent $\mathcal{U}$ in \cref{normaliz} can be computed by using the Kronecker $\delta$'s present in the matrix elements to write the quantum numbers in terms of the occupation numbers $m_{a,j}$, $m_{b,j}$ and $m_{c,j}$. We present the details of the calculation in \cref{app:exponent}. In order to be able to state the result in the presence of a quasi-hole, we need to mention that the wave function in that case can be written in the form
\begin{equation}\label{ws}
    \sum_{s=0}^{N_e}w^sP_s(\{z_j\})
\end{equation}    
for $N_e$ electrons. The parameter $s$ depends on the orbital occupation numbers. In \cref{app:exponent}, we show that
\begin{equation}\label{ssimeq}
    s\simeq -\sum_{j=0}^{N_{\phi}}j(m_{a,j}+m_{b,j}+m_{c,j}) \ ,
\end{equation}
where the symbol ``$\simeq$" means equality up to terms that do not depend on how the particles are distributed along the MPS cylinder, that is, terms that only depend on $N_e$ and $N_\phi$ (which only influence the overall normalisation of the state).
Using this notation, we obtain the result
\begin{widetext}
\begin{align}
\label{normQH}
\mathcal{U} \simeq
\exp\bigg(\frac{2\pi\delta\tau}{L}\sum_{j=0}^{N_{\phi}}\bigg[\frac{j^2}{2}(m_{a,j}+m_{b,j}+m_{c,j}) \bigg]
+\frac{2\pi}{L}\bigg(\tau_{\alpha}-\frac{M+2}{2}\delta\tau\bigg)s\bigg) \ .
\end{align}
\end{widetext}
Several remarks about this factor are in order. The first term in the exponent precisely corresponds to the factor that is necessary to incorporate the normalisation of the single particle orbitals on the cylinder, which are given in \cref{eq:single-particle-orbital} (see \cite{Kjall2018MPS} for a more detailed discussion). The second term in the exponential gives the $\tau_\alpha$ dependence. We note that there is a displacement by $-\frac{M+2}{2}\delta \tau$ for the location of the quasi-hole in \cref{normQH}. The quantity $\frac{M+2}{2}$ that appears is exactly the scaling dimension of the electron operators. This is to be expected, because the conformal mapping from the plane to the cylinder \cref{map} introduces a factor $z_j^h$ for the electron coordinates when calculating the correlation functions (see f.i. \cite{Belavin1984conformal,Francesco2012CFT}). This effectively results in a shift of the FQH droplet along the $\tau$ direction by $\frac{M+2}{2}\delta\tau$, in accordance with \cref{normQH}. When doing actual MPS calculations, one has to adjust the input parameter $\tau_\alpha$ for the location of the quasi-hole accordingly.

\section{Implementation}
\label{sec:implementation}
In this section, we provide some information on the actual implementation of the MPS formalism that
we use to analyse the $k=3$ Read-Rezayi states. We consider large but finite systems, with up to $N_e = 300$ electrons.
Because we consider finite systems, we can study edge effects, such as the edge spin. We report on these results in the
paper \cite{Fagerlund2024nonAbelian}.
We limit ourselves to situations where we insert a single quasi-hole at the center of the droplet.
For our purposes, a single quasi-hole suffices, and having only a single quasi-hole allows for an
efficient evaluation of the correlation matrix. This is needed when calculating the density profile, which
we use to obtain the charge and the spin of the quasi-hole.

In particular, we insert the matrix corresponding to the quasi-hole in the middle of our system.
We then bring the MPS to ``left canonical form" from the first orbital up to the quasi-hole matrix \cite{Schollwock2011mps}.
In addition, the matrices corresponding to orbitals ``after" the quasi-hole matrix are brought
to ``right canonical form", starting from the largest orbital. Using this mixed canonical form, one
avoids the high memory cost associated with also bringing all matrices, including the quasi-hole
matrix, to, say, the ``left canonical form". The reason for this is that the quasi-hole matrix is much
less sparse than the electron matrices.

The cutoff on the auxiliary Hilbert space that we use is directly associated with the angular momentum
of the droplet. The angular momentum of the first $j$ orbitals of the droplet,
$$
L_z (j) = \sum_{k=0}^{j-1}k(m_{a,k}+m_{b,k}+m_{c,k}) \ ,
$$
can be expressed in terms of the quantum numbers at orbital $j$, see
\cref{consP}. Both the angular momentum and $Q_0$ are good
quantum numbers (though the angular momentum only in the absence of
quasi-holes). We therefore implement a cutoff $P_{\rm max}$ as follows
\begin{equation}
P_0 + P_1 + P_2 + \frac{Q_1^2}{24} + \frac{Q_2^2}{8} \leq P_{\rm max} \ .
\end{equation}
We do not implement additional cutoffs on the charge quantum numbers.
The maximum cutoff we consider is $P_{\rm max} = 12$.
This already leads to rather large auxiliary dimensions (which are
bond dependent). For each of the quasi-hole types, we list the largest bond dimension (which occurs at
the insertion of the quasi-hole matrix), and the total number of auxiliary states used (that is, the
number of states in the union of all bond auxiliary spaces) for cutoff $12$, as shown in table \ref{tab:MPS}. 
\begin{table}[ht]
\begin{tabular}{c|c|c|c}
quasi-hole & $N_e$ & max & total\\
\hline
$(\sigma_1,1/5)$ & 300 & 158734 & 1007095 \\
$(\sigma_2,2/5)$ & 300 & 160496 & 1051656 \\
$(\psi_1,2/5)$ & 299 & 165182 & 796320 \\
$(\mathbf{1},3/5)$ & 300 & 160383 & 688409 \\
$(\epsilon,3/5)$ & 299 & 159445 & 951187 \\
$(\psi_2,4/5)$ & 298 & 177800 & 951154 \\
\end{tabular}
\caption{Some MPS details for the different quasi-hole states, for cutoff $P_{\rm max} = 12$.}
\label{tab:MPS}
\end{table}

For completeness, we give some details concerning the quantum numbers of the ``in" and ``out" states of the
MPS description in the presence of quasi-holes. The ``in" and ``out" momenta are always zero, $P_1 = P_2 = P_3 = 0$.
The same is true for the ``in" charges $Q_0 = Q_1 = Q_2 = 0$.
In the case of a droplet without quasi-holes (and $N_e \mod 3 = 0$), the ``out" charges are given by
$(Q_0, Q_1, Q_2 ) = (3(M+1), 0, 0)$.
In \cref{tab:quantum-numbers}, we specify the ``out" charges for the six different types of quasi-holes
that we consider. The out charge $Q_1$ is not always zero, because of the way the single quasi-holes states
are defined (i.e., in some cases, we need to send quasi-holes to the far end of the cylinder). In addition,
the number of electrons is not always a multiple of three, as indicated in the table. This means that
the value of $\Delta N_\phi$, for a given quasi-hole, as defined by
$N_\phi = (3 M + 2)/3 N_e - (M + 2) + \Delta N_\phi$, is not always an integer (but $N_\phi$ is, of course),
as indicated in the table.

\begin{table}[ht]
\begin{tabular}{c|c|c|c}
Quasi-hole & $N_e \mod 3$ & $\Delta N_\phi$ & Out charges\\
\hline
$(\sigma_1,1/5)$ & 0 & 1 & $(3(M+1)-2, 2, 0)$\\
$(\sigma_2,2/5)$ & 0 & 1 & $(3(M+1)-1, -2, 0)$\\
$(\psi_1,2/5)$ & 2 & $\frac{2}{3}$ & $(3(M+1), 0, 0)$\\
$(\mathbf{1},3/5)$ & 0 & 1 & $(3(M+1), 0, 0)$\\
$(\epsilon,3/5)$ & 2 & $\frac{5}{3}$ & $(3(M+1)-2, 2, 0)$\\
$(\psi_2,4/5)$ & 1 & $\frac{4}{3}$ & $(3(M+1), 0, 0)$
\end{tabular}
\caption{Details on the quantum numbers.}
\label{tab:quantum-numbers}
\end{table}

\section{Density and quasi-hole charges with MPS}
\label{sec:density}
In order to find out how the quasi-holes distort the electron fluid of the quantum Hall system, we wish to compute the  density 
\begin{equation}\label{density}
    \rho(r)=\int d^2r_2 \cdots dr^2_{N_e}\braket{\psi}{r,r_2, \ldots ,r_{N_e}}\braket{r,r_2, \ldots ,r_{N_e}}{\psi}.
\end{equation}
The density expression in \cref{density} can be rewritten in a form more suitable for MPS (see f.i. \cite{Kjall2018MPS})
\begin{align}\label{densityMPS}
    \rho(\tau,x)=\sum_{m,n}\bigg[\frac{1}{L\sqrt{\pi}l_B}&\bra{\psi}c_m^{\dagger}c_n\ket{\psi}e^{ix(m-n)\frac{2\pi}{L}}\nonumber\\
    \times &e^{-\frac{1}{2l_B^2}(\tau-\tau_m)^2}e^{-\frac{1}{2l_B^2}(\tau-\tau_n)^2}\bigg].
\end{align}
\indent Above, we represent (but do not explicitly compute) the wave function using the expectation value in \cref{wavefcn}, with the electron and quasi-hole insertions chosen to reproduce relevant Read-Rezayi wave functions from \cref{RR,wavefcn sigma1,wavefcn sigma2,wavefcn psi1,wavefcn psi2,wavefcn Laughlin,wavefcn eps}.
We then use the MPS representation of these wave functions to calculate the correlation matrix elements
$\bra{\psi}c_m^{\dagger}c_n\ket{\psi}$, from which one easily obtains the density $\rho(\tau,x)$.

From the density $\rho(\tau, x)$ for a state with a quasi-hole, one calculates the charge and spin of this quasi-hole.
We first discuss the density profiles and charges of the quasi-holes, but 
postpone the discussion of the spin to \cref{sec:compute spin}.

\subsection{Density profiles}\label{sec:density-profiles}

\begin{figure}
  \includegraphics[width=\linewidth]{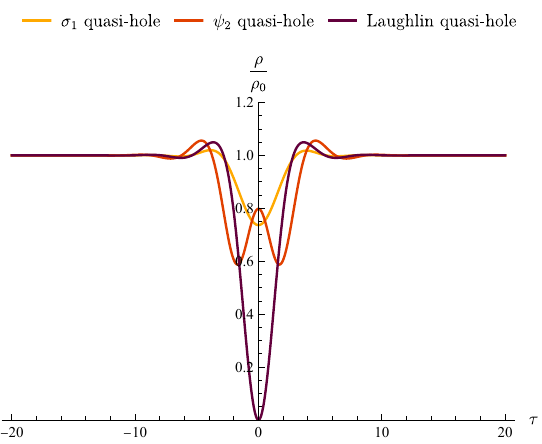}
  \caption{Scaled density profiles at the quasi-hole center $x=0$ for the $(\sigma_1,1/5)$, $(1,3/5)$ (or Laughlin) and $(\psi_2,4/5)$ quasi-holes.
  The circumference $L=20$, although there is no qualitative difference from $L=22$ or $L=24.$ The cutoff $P_{max}=12.$ We note the similarity between our Laughlin quasi-hole and the $\mathbb{Z}_3$ quasi-hole profile in fig. $2a$ of ref. \cite{Wu2014Braiding}.}
  \label{fig:prof1}
\end{figure}

\begin{figure}
  \includegraphics[width=\linewidth]{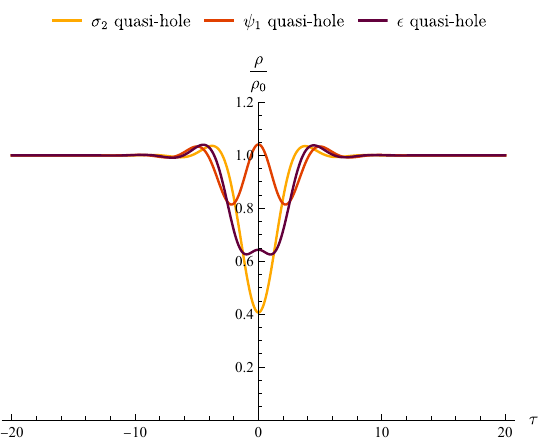}
  \caption{Scaled density profiles at the quasi-hole center $x=0$ for the $(\sigma_2,2/5),(\psi_1,2/5)$ and $(\epsilon,3/5)$ quasi-holes.
  The circumference and cutoff are $L=20,P_{max}=12.$ }
  \label{fig:prof2}
\end{figure}

In \cref{fig:prof1,fig:prof2}, we plot the (scaled) density profiles $\rho(\tau, 0)/\rho_0$, through the center of the quasi-holes,
where $\rho_0 = \nu/(2\pi)$ is the background density of the FQH fluid. The circumference of the cylinder is $L = 20$, and the cutoff used is $P_{\rm max} = 12$.

There are several features worth noticing. First of all, the quasi-holes are quite large. Deviations from the background density are
discernible roughly up to $\tau = 10$, which means that on cylinders with circumference $L \lesssim 20$, the quasi-hole would ``touch itself"
in the circumference direction $x$. In addition, the (radial) sizes of the six different quasi-holes we consider are comparable.
The shapes of the $(\sigma_1, 1/5)$, $(\sigma_2, 2/5)$ and $(1, 3/5)$ quasi-holes are similar.  The difference between them
lies in the ``depth" of the profiles.
The same is true for the $(\psi_1, 2/5)$ and $(\psi_2, 4/5)$ quasi-holes.


\subsection{Quasi-hole charges}\label{sec:compute charge}
An observable of interest of the quasi-holes is their charge.
The purpose of this section is to explain how the charges of the different quasi-holes can be computed from the density profiles
discussed above, and to give the results.

If the density with a quasi-hole in the system is given by $\rho_{qh}(\tau,x)$, the quasi-hole charge $Q_{qh}$ is given by
\begin{equation}\label{ch}
    Q_{qh}=\int d^2r (\rho_{qh}(\tau,x)-\rho_0).
\end{equation}
Throughout the paper, we assume that the quasi-hole is placed at the origin, i.e. at $\tau=x=0$. While the total quasi-hole charge is to be regarded as an input parameter for the MPS scheme, and can simply be read off from the $Q_{0}$ quantum numbers of the quasi-hole operators chosen, computing the charge is nonetheless useful as it provides a verification that the numerical implementation works as intended. It also allows for error assessment of the MPS scheme.

We use two different methods to calculate the quasi-hole charge as a function of the distance between the quasi-hole center and the outermost parts of the integration region.
Due to the finite circumference of the cylinder, if we express the integral in \cref{ch} in polar coordinates
$\tau=r\cos\theta,x=r\sin\theta$ as
\begin{equation}\label{Qinteg}
    Q_{qh}(r_{max}) =\int_0^{2\pi}\int_{0}^{r_{max}} \big(\rho_{qh}(\tau,x)-\rho_0\big)rdrd\theta,
\end{equation}
we need to restrict $r_{max} \leq L/2$.

To remedy this, we follow two different approaches. The simplest approach is to assume that the quasi-hole has rotational symmetry,
and integrate along $r$, setting $\theta = 0$ (or $\theta = \pi$). This simple approach also serves as a check on the convergence
in terms of $L$.
If the cylinder circumference $L$ is small, there is some self-interference of the quasi-holes
around the cylinder in the $x$ direction, leading to quasi-hole profiles that are not entirely rotationally symmetric.
This in turn can lead to deviations from the expected charge values in the large $r_{max}$ limit.
To be explicit, when calculating the quasi-hole charge by means of a one-dimensional integral, we calculate%
\footnote{The MPS formulation is not entirely symmetric, and it can happen that the dimensions of the matrices are
larger ``before" or ``after" the matrix corresponding to the quasi-hole. We perform the one-dimensional integrals
in the direction of the larger matrices.}
\begin{equation}
\label{QintegLine}
    Q_{qh}^{1d}(r_{max}) 
    = 2\pi\int_{0}^{r_{max}} \big(\rho_{qh}(\tau,0)-\rho_0\big)\tau d\tau.
\end{equation}
The expression for $\rho_{qh}(\tau,x)$ involves a sum over products of one-particle orbitals, weighted by the numerically obtained coefficients $\bra{\psi}c_m^{\dagger}c_n\ket{\psi}$; c.f. \cref{densityMPS}.
The $\tau$ integral over these products of one-particle orbitals can be done analytically at $x=0$,
resulting in an expression involving exponential functions and error functions\footnote{This expression is long and not particularly informative. Consequently, we will not reproduce it here.}.
We use this analytical expression when calculating $Q_{qh}^{1d}(r_{max})$.
By calculating the charge using \cref{QintegLine}, the integration region is, for $\tau > L/2$, larger than the
actual cylinder. This might seem odd at first, but we are interested in the charge (and more importantly, the spin in \cref{sec:compute spin} below),
in the limit of large circumferences.
Because in practice the MPS approach only converges for moderate circumferences, we do our best effort to generate convergent MPS data for as large a system as possible, and then calculate the charge and spin
in the way one would do this in the large circumference limit.

The second approach is to perform the two-dimensional integral, but taking the finite circumference into account.
That is, we integrate over $0\leq \theta < 2\pi$ for $0\leq r_{max} \leq L/2$ and over the largest possible range of $\theta$ for $r_{max}>L/2.$
How large this range is does, of course, depend on $r_{max}$. In this case, we are forced to assume that the profile is rotationally symmetric.
We note that in principle, one could perform the full two-dimensional integral up to $r_{max} = L/2$, and continue with
a one-dimensional integral. This approach would lead to a charge $Q_{qh} (r_{max})$ with a discontinuous first derivative
at $r_{max} = L/2$, so we do not consider this approach here.
The two-dimensional integral expression for the charge reads
\begin{equation}
\label{QintegDouble}
    Q_{qh}^{2d}(r_{max}) 
    = \int_{0}^{r_{max}} \int_{C_\theta(r)} w(r)\big(\rho_{qh}(\tau,x)-\rho_0\big)rdrd\theta \ ,
\end{equation}
where the range of $\theta$ is given by
\begin{align}
\label{eq:theta-range}
    C_\theta(r \leq L/2) =& [ 0, 2\pi ) \\
    C_\theta(r > L/2) =& [-\arcsin(\frac{L}{2r}),\arcsin(\frac{L}{2r})] \cup \nonumber \\
    & [\pi - \arcsin(\frac{L}{2r}), \pi + \arcsin(\frac{L}{2r})] \nonumber
\end{align}
and the ``weight factor" $w(r)$ is
\begin{align}
\label{eq:weight-factor}
    w(r \leq L/2) &= 1 \\
    w(r > L/2) &= \frac{\pi}{2 \arcsin(\frac{L}{2r})} \ .\nonumber
\end{align}
The range $C_\theta$ ensures that the integration region we are integrating over actually exists on the
cylinder with a finite size. Because we want to calculate the charge in the way one would do in the
large circumference limit, we introduce a (radius dependent) weight factor $w(r)$, which ``extends" the
integration region to $0 \leq \theta < 2\pi$, assuming rotational symmetry of the quasi-hole.

To compute the charge, the most accurate method is the two-dimensional integral in \cref{QintegDouble}, since the cylinder circumference is small enough to allow some self-interference of the quasi-holes around the cylinder in the $x$ direction.
However, a method analogous to \cref{QintegLine} turns out to give excellent accuracy for the spin computations, which we describe in section \ref{sec:compute spin}.
\begin{figure}[t]
    \centering
    \includegraphics[width=\linewidth]{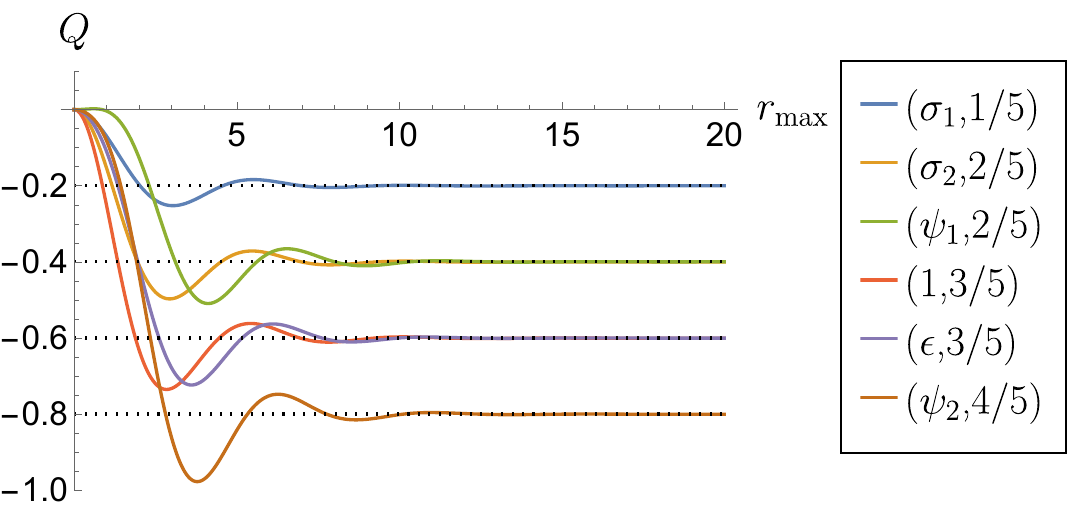}
    \caption[width=\linewidth]{Charges $Q_{qh}^{2d}$ of the different quasi-holes for circumference $L=20$ and cutoff $P_{max}=12$, computed using the double integral method in \cref{QintegDouble}. The dotted lines show the values expected from the CFT description of each quasi-hole, i.e. from the coefficient of $\phi(w)$ in the corresponding vertex operator.}
    \label{fig:Q}
\end{figure}
We plot the quasi-hole charges $Q^{2d}_{qh}(r_{max})$ in \cref{fig:Q}. We note that the values rapidly converge to the values one would expect from the CFT description, i.e. from the vertex operators used to represent the different quasi-holes (the expected value is reached at $r _{max}\approx L/2$).
Although this is not surprising, it is still valuable information: it serves as a verification that our MPS technique and its implementation faithfully reproduce the correct properties of the quasi-holes.

\section{Braiding phases and spins}\label{sec:spin}
In order to demonstrate that the Berry phase of the RR state vanishes in the ``minimal representation'', we first compute the values of the quasi-hole spins \textit{given that} the Berry phase vanishes. This is done in the current section. The spin values thus obtained will be seen in section \ref{sec:compute spin} to match values calculated using MPS data without imposing any assumptions about the Berry phase, showing that the Berry phase indeed does vanish.

The spins of the quasi-holes can be computed using the spin-statistics relation (SSR) derived on general ground for FQH systems in \cite{Nardin2023SSR}. This relation gives the braiding parameter $\kappa_{ab}^c$ of quasi-holes $a$ and $b$ in fusion channel $c$ in terms of the individual spins $J_a,J_b$ of the $a$ and $b$ quasi-holes and the spin $J_{ab}^c$ of the fusion product of $a$ and $b$ in fusion channel $c$ through\footnote{We point out that the letters $a,b$ and $c$ are generic labels, and have no relation to the indices $a,b$ and $c$ as used for the electron and quasi-hole operators. Which interpretation is intended should be clear from the context.}
\begin{equation}\label{ssr}
    \kappa_{ab}^c=J_a+J_b-J_{ab}^c \;\;\mathrm{mod}\;1.
\end{equation} Here, the spin is defined as
\begin{equation}\label{J}
    J=\int d^2r \bigg(\frac{r^2}{2l_B^2}-1\bigg)(\rho_{qh}(\tau,x)-\rho_0) \ ,
\end{equation}
where $\rho_{qh}(\tau,x)$ is the probability density of a system with a quasi-hole at the origin (using the coordinates
$\tau=r\cos\theta,x=r\sin\theta$) and $\rho_0$ is the background density when the system contains no quasi-holes \cite{Nardin2023SSR}.

To predict the individual spins, one can -- analogously to the calculation for the Moore-Read state in \cite{Nardin2023SSR} -- compute the braid parameters $\kappa_{ab}^c$ from the CFT description, using the ``minimal" description. 
We write the operators creating the quasi-holes as $A e^{i \alpha \phi_0}$, $B e^{i \beta \phi_0}$ and $C e^{i (\alpha + \beta) \phi_0}$,
where $A$, $B$ and $C$ correspond to operators in the $\mathbb{Z}_3$ parafermion theory, with scaling dimensions
$h_A$, $h_B$ and $h_C$ respectively. Then we have the following OPE
\begin{align}
    &A(w_1)e^{i\alpha\phi_0(w_1)}B(w_2)e^{i\beta\phi_0(w_2)}= \\ \nonumber
    &(w_1-w_2)^{h_C-h_A-h_B+\alpha\beta}c_{A,B}^C e^{i (\alpha + \beta) \phi_0 (w_2)}C(w_2) + \cdots \ ,
\end{align}
where $c_{A,B}^C$ is an OPE constant whose value we do not need.
Under the assumption that the braiding phase is contained in the monodromy, the braiding parameter may simply be read off as $\kappa_{ab}^c=h_C-h_A-h_B+\alpha\beta$. 

To be explicit, the scaling dimensions are given by
\begin{align}
h_{\mathbf{1}} &= 0 & h_{\psi_1} &=h_{\psi_2}=2/3 \\\nonumber
h_{\epsilon} & = 2/5 &h_{\sigma_1} &=h_{\sigma_2}=1/15
\end{align}
and the $U(1)$ chiral boson factor is given by $e^{i\frac{d}{\sqrt{q_0}}\phi}$, where $q_0=3(3M+2)$ and $d$ is given by the numerator of the charge of the quasi-hole, $q_{\rm qh} = d/(3M+2)$.

To calculate the spins of the quasi-holes using the SSR, we also need the spin of the Laughlin quasi-hole.
Generically, the spin of the Laughlin quasi-hole is give by $J_{\rm Laughlin} = -\frac{\nu}{2}+ \frac{\nu S}{2}$, see \cite{Comparin_2022} for details.
For the $k=3$ Read-Rezayi states we have $\nu = \frac{3}{3M+2}$ and $S = M+2$, resulting in
$J_{(1,\frac{3}{3M+2})}=\frac{3(M+1)}{2(3M+2)}$.

We now have all the information necessary to calculate the spins for the various quasi-holes.
The braid parameters are obtained from the OPEs. Considering different fusions of
quasi-holes, where one fusion should lead to the Laughlin quasi-hole, we obtain a linear system of
equations for the unknown spins, modulo one.

As a simple example, we consider the fusion of two $(\sigma_1,\frac{1}{3M+2})$
quasi-holes to a $(\sigma_2,\frac{2}{3M+2})$ quasi-hole, and the fusion of
$(\sigma_1,\frac{1}{3M+2})$ and $(\sigma_2,\frac{2}{3M+2})$ to the Laughlin quasi-hole.
The braid parameters are given by
$\kappa_{\sigma_1,\sigma_1}^{\sigma_2} = \frac{1-M}{5(3M+2)}$
and
$\kappa_{\sigma_1,\sigma_2}^{1}=\frac{2(1-M)}{5(3M+2)}$
(we drop the charge of the quasi-holes in the labels of the braid parameters, in order to avoid clutter).
The SSR relations then become (modulo $1$)
\begin{align}
\label{linsys}
\frac{1-M}{5(3M+2)} & =2J_{(\sigma_1,\frac{1}{3M+2})}-J_{(\sigma_2,\frac{2}{3M+2})},
\\ \nonumber
\frac{2(1-M)}{5(3M+2)} &=J_{(\sigma_1,\frac{1}{3M+2})}+J_{(\sigma_2,\frac{2}{3M+2})}-\frac{3(M+1)}{2(3M+2)} \ ,
\end{align}
resulting in
$J_{(\sigma_1,\frac{1}{3M+2})}=\frac{3M+7}{10(3M+2)}$ and $J_{(\sigma_2,\frac{2}{3M+2})}=\frac{2(2M+3)}{5(3M+2)}$. 

The method outlined and illustrated above can be carried out in much the same manner for all the different quasi-holes, to provide all the spin values of interest. The crucial assumption in doing so is that \textit{all the statistics is contained in the monodromy}. Under this assumption, we obtain the spin predictions in \cref{tab:spins}.
This assumption has been shown to hold for the Laughlin \cite{ArovasSchriefferWilczek1984} and Moore-Read \cite{,Bonderson2011Plasma,Herland2012Screening} states using the plasma analogy.
For the $k=3$ Read-Rezayi states this was shown by calculating the
braiding phase explicitly numerically \cite{Wu2014Braiding}.

We will verify the spin values in \cref{tab:spins} numerically by using the free boson MPS scheme of the current paper.
We present these numerical results in section \ref{sec:compute spin}, which in turn builds on the density profiles of section \ref{sec:density}. These numerical values will be seen to agree well with the predictions in \cref{tab:spins}, thus supporting the conclusion drawn in \cite{Wu2014Braiding} that the Berry phase of the $k=3$ Read-Rezayi state vanishes.
Unlike the approach in \cite{Wu2014Braiding}, we draw this conclusion without having to explicitly braid quasi-holes, resulting
in a simpler calculation.
Instead, all that is required is the SSR of \cite{Nardin2023SSR} and some local information regarding the quasi-hole spins.  
\begin{table*}
    \centering
    {
    \begin{tabular}{|c||c|c|c|c|c|c|}
        \hline
        Quasi-hole & $(\sigma_1,1/5)$ & $(\sigma_2,2/5)$ & $(\psi_1,2/5)$ & $(1,3/5)$ & $(\epsilon,3/5)$ & $(\psi_2,4/5)$\\
        \hline
        Spin prediction, general $M$ & $\frac{3M+7}{10(3M+2)}$ &$\frac{2(2M+3)}{5(3M+2)}$ & $-\frac{M}{3M+2}$ & $\frac{3(M+1)}{2(3M+2)}$ & $\frac{3M+7}{10(3M+2)}$ & $0$ \\
        \hline
        Spin prediction, $M=1$ & $1/5$ & $2/5$ & $-1/5$ & $3/5$ & $1/5$ & $0$ \\\hline
    \end{tabular}
    }
    \caption{Spin predictions from minimal representations of the quasi-holes. We include both the value for general $M$ and the value for $M=1$, which is the value the numerical calculations have been done for.}
    \label{tab:spins}
\end{table*}

\section{Quasi-hole spins}\label{sec:compute spin}
Analogously to how the charges of the quasi-holes are computed, there are two ways of calculating the quasi-hole spins \cref{J}.
We first state how we calculate the quasi-hole spin $J(r_{max})$ using a one-dimensional integral, assuming cylindrical symmetry
of the quasi-holes. In particular, we may compute it as
\begin{equation}
\label{JLine}
    J_{qh}^{1d}(r_{max})
    = 2\pi\int_{0}^{r_{max}} \bigg(\frac{\tau^2}{2l_B^2}-1\bigg)\big(\rho_{qh}(\tau,0)-\rho_0\big)\tau d\tau \ .
\end{equation}
We remind the reader that $\rho_{qh}$ is the density when a quasi-hole is centered at the point $\tau=x=0$, and that we use the coordinates $\tau=r\cos\theta,x=r\sin\theta$.
Just as was the case when calculating the quasi-hole charge using the one-dimensional integral, the analogous calculation for the spin can be carried out to a large extent
analytically. Again this involves a cumbersome expression which we omit.

The two-dimensional integral expression for the spin that we use is given by
\begin{align}
\label{JDouble}
    & J_{qh}^{2d}(r_{max}) = \\
    & \int_{0}^{r_{max}} \int_{C_\theta(r)} w(r)\bigg (\frac{r^2}{2l_B^2}-1\bigg) \big(\rho_{qh}(\tau,x)-\rho_0\big)rdrd\theta \ ,
    \nonumber
\end{align}
where $C_\theta(r)$ is given by \cref{eq:theta-range} and $w(r)$ by \cref{eq:weight-factor}.
The range $C_\theta(r)$ ensures that the integration region actually exists on the cylinder with finite circumference $L$,
while $w(r)$ ``extends" the integration region (assuming rotational symmetry), to $0 \leq \theta < 2\pi$ for $r > L/2$.
The reason for calculating the quasi-hole spin in this way, is that we are interested in the value one obtains in the large circumference limit.
On a cylinder with very large circumference, the range of $\theta$ would be $0\leq \theta < 2\pi$; we mimic this in eq.~\eqref{JDouble}
(and of course in the expression eq.~\eqref{JLine}, using a one-dimensional integral).

Unlike the charge computations in section \eqref{sec:compute charge}, the line integral method \cref{JLine} is more accurate than the double integral \cref{JDouble} (when comparing to the analytically expected value).
We discuss this behaviour below.

\begin{figure}
    \centering    \includegraphics[width=\linewidth]{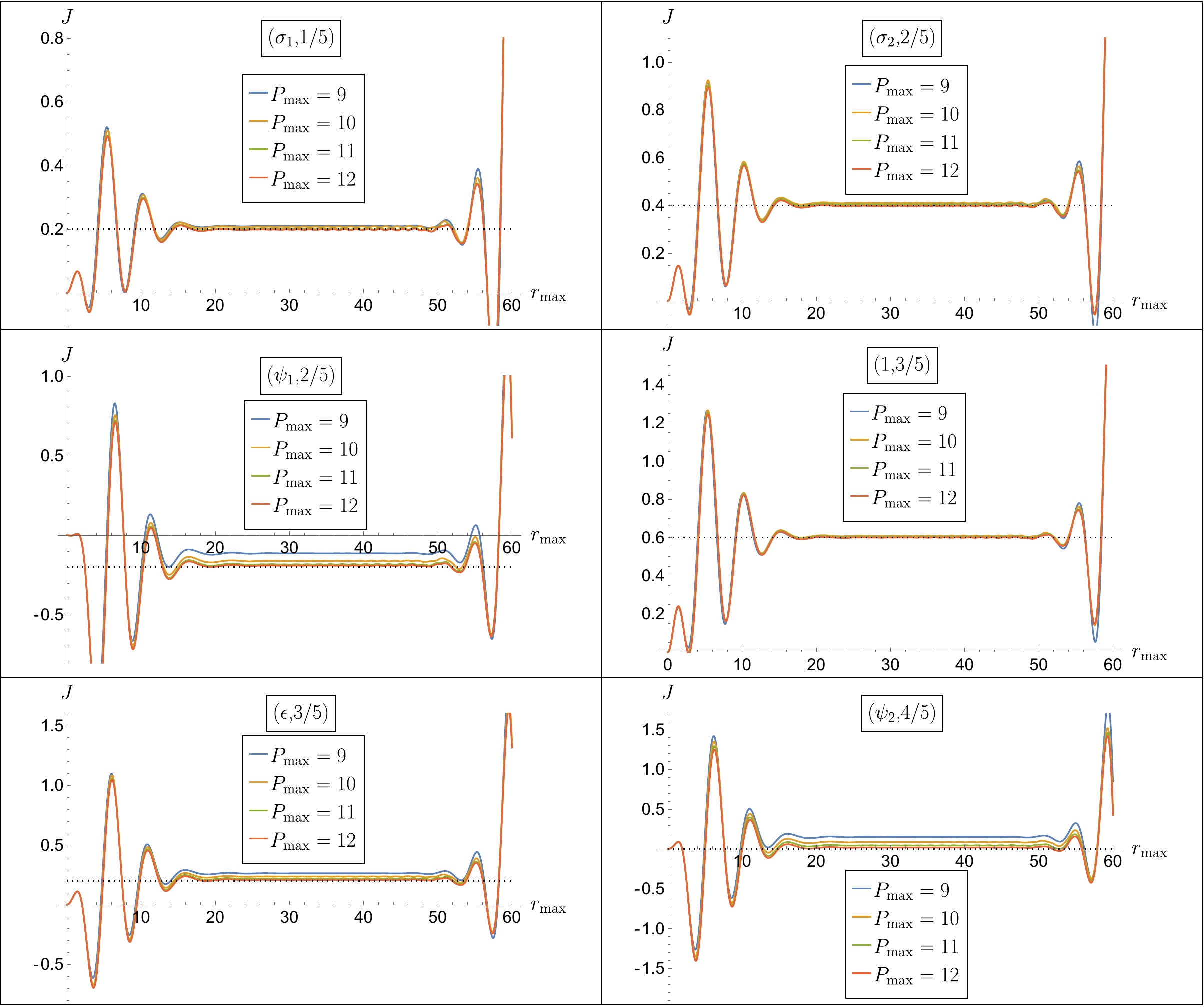}
    \caption{Spins $J$ as a function of the radius $r_{max}$ for different values of the cutoff $P_{max}$.
    All plots have been computed using the one-dimensional integral \cref{JLine}, and $L=20$ throughout.
    The dotted lines show the theoretically expected values when the Berry phase is $0$.
    For the $(\psi_2,4/5)$ quasi-hole, the dotted line coincides with the $r_{max}$ axis.}
    \label{fig:JvsrlineL20}
\end{figure}

We therefore focus here on the results obtained using the one-dimensional integral \cref{JLine}, which are shown in \cref{fig:JvsrlineL20} for cylinder circumference $L=20$ and MPS cutoffs $P_{max}\in\{10,11,12\}$. We note that the recurring behaviour is the same in all the figures: for small $r_{max}$, the integral oscillates due to large differences between the density of the quasi-hole state and the background density without quasi-holes. Then, there is a plateau, corresponding to small values of the integrand in the region where the quasi-hole barely perturbs the background density. Finally, there is a new region with oscillations corresponding to edge effects as the integration reaches the edges of the finite droplet.

\begin{figure}
  \centering\includegraphics[width=\linewidth]{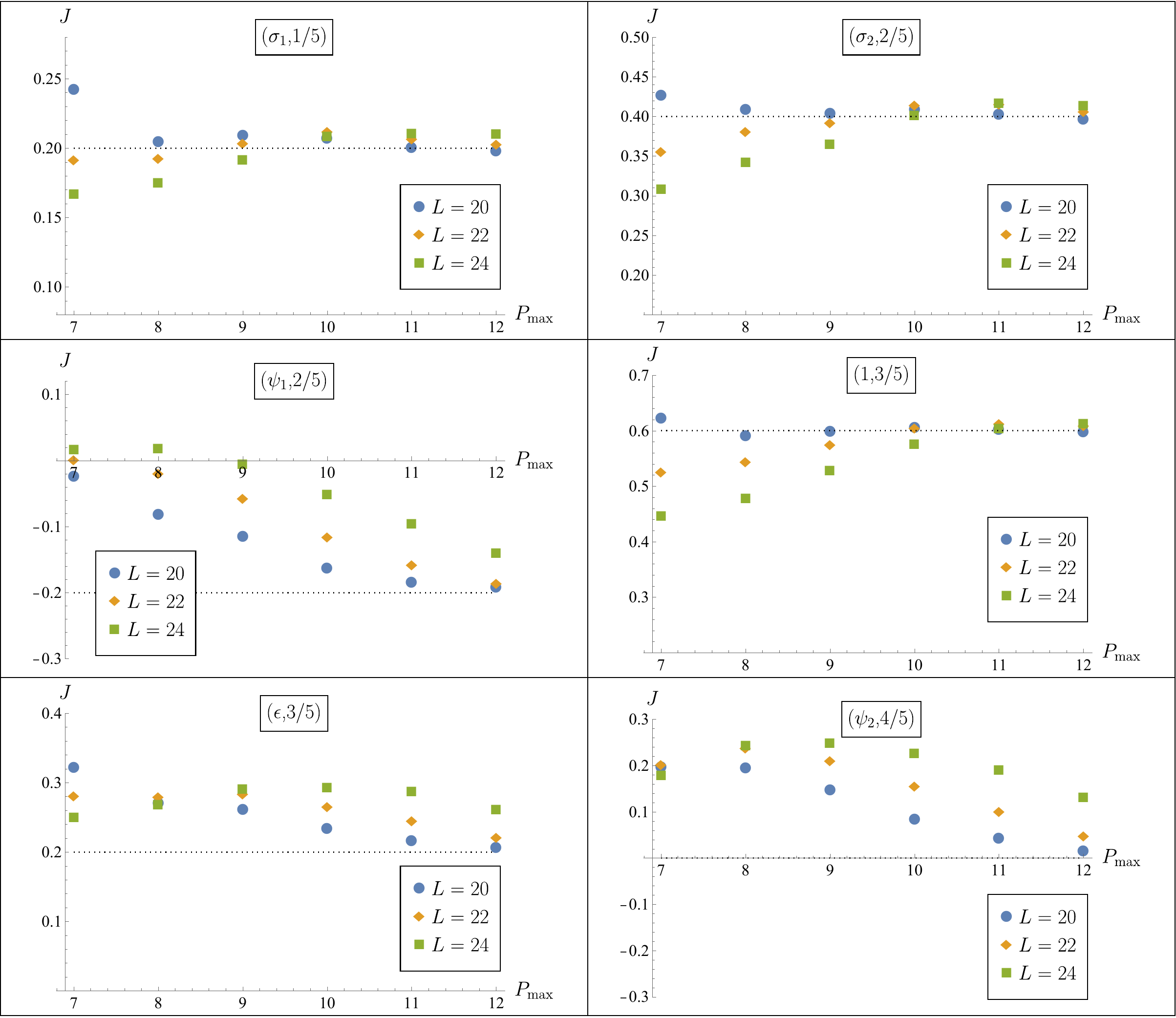}
  \caption[width=\linewidth]{Plateau-averaged values for the spins of various different quasi-hole types, against three different cylinder circumferences $L$. All calculations have been carried out using the spin expression \cref{JLine}. The dotted lines show the expected values assuming the Berry phase to be $0$. For the $(\psi_2,4/5)$ quasi-hole, the dotted line coincides with the $P_{max}$ axis.}
  \label{fig:JvsPline}
\end{figure}

To find a single spin prediction for each circumference $L\in\{20,22,24\}$ and cutoff $P_{max}\in\{7,8,9,10,11,12\}$, we average the spin values obtained above over the plateau for each $L$ and $P_{max}$. This procedure has three advantages. First of all, it allows us to eliminate the minor fluctuations that can be observed across the plateau, and hence minimize the effect of background effects due to e.g. the finite cylinder size. Secondly, comparing the values thus obtained shows how the MPS computation converges to a final prediction as the cutoff $P_{max}$ is increased. Thirdly, it allows us to compare different cylinder circumferences $L$, to see which cylinder size is optimal for each type of quasi-hole.\footnote{From figs. \ref{fig:prof1} and \ref{fig:prof2}, it is clear that the different quasi-holes have somewhat different sizes. Hence, one would expect different cylinder circumferences to be optimal for fitting the entire quasi-hole on the cylinder with negligible self-interaction effects around the backside of the cylinder, while simultaneously having the cylinder be small enough for fast convergence through the factor in \cref{normaliz}.}  
The results are plotted in \cref{fig:JvsPline}. 
\onecolumngrid

\newpage
\twocolumngrid
In fig. \ref{fig:JvsPline}, the values shown are obtained by using the spins $J$ as a function of the integration limit $r_{max}$, as seen in \cref{fig:JvsrlineL20}, for different circumferences $L=20,22,24$ and cutoffs $P_{max}=7,8,\ldots,12$. For each combination of $L$ and $P_{max}$, we numerically integrate $J(r_{max})$ over the intervals $r_{max}\in[30,40]$ for $L=20,$ $r_{max}\in[26,36]$ for $L=22$, and $r_{max}\in[26,30]$ for $L=24$, and divide the results by the appropriate interval widths. These intervals have been chosen since they are intervals where (for a fixed $L$ and cutoffs $P_{max}=10,11,12$) all quasi-hole types have a plateau, making it easier to compare the convergence behaviours for different quasi-holes. We note that in all figures, the values converge to the values predicted from the spin-statistics relation \cref{ssr}, although the precise pattern of the convergence (as a function of $L$ and $P_{max}$) differs between quasi-hole types. 

\section{Entanglement spectra}\label{sec:entanglement}
As a verification of our MPS description, we also perform computations of the entanglement entropy, and in particular, the entanglement spectrum as introduced in \cite{Li_2008}.
This allows us to verify that our free boson description indeed captures the $\mathbb{Z}_3$ parafermion structure of the states as it should.
The entanglement entropy of the $k=3$ RR state was calculated before in \cite{estienne2015entropies}. We therefore concentrate on
the entanglement spectrum, which contains more information, but mention that our calculations of the entanglement entropy are consistent with those presented in \cite{estienne2015entropies}.
Thus, we also reproduce the correct topological entropy \cite{KitaevPreskill06,LevinWen06}, but do not provide these details here.

\begin{figure}
    \centering
    \includegraphics[width=\linewidth]{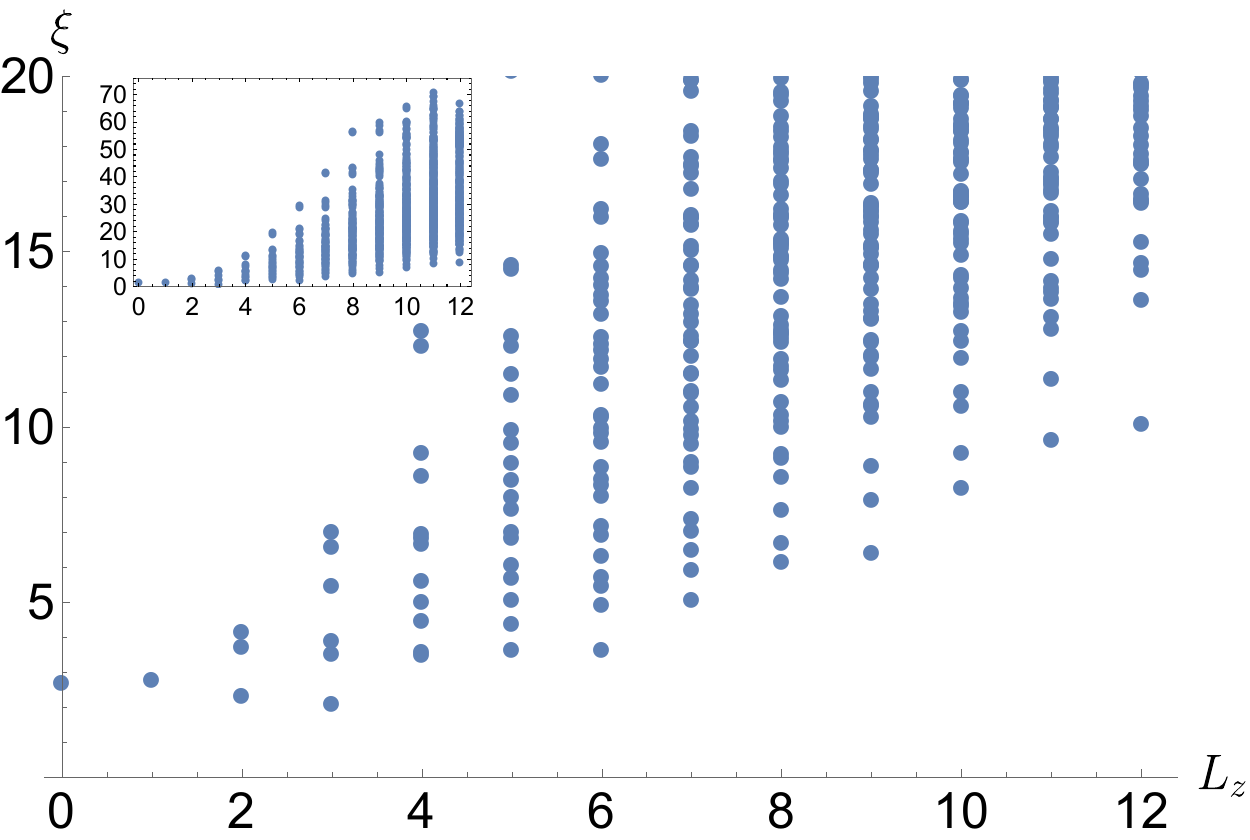}
    \caption{The low-lying part of the entanglement spectrum for a system with circumference $L=20$, $N_e = 120$ electrons and cutoff $P_{max}=12$.
    Both subsystems have $S_A = S_B = 99$ orbitals and contain $N_A = N_B = 60$ electrons. The inset shows the full entanglement spectrum with all the levels.}
    \label{fig:ent}
\end{figure}

To calculate the entanglement spectrum, we equally divide the system (without bulk quasi-holes)
in two parts A and B, and write A (B) in ``left (right) canonical form", using the conserved quantum numbers.
In \cref{fig:ent}, we plot the entanglement levels $e^{-\xi_i}$ against the angular momentum $L_z$ for the parameter choices $L=20, N_e = 120, P_{max}=12$.
Both A and B consist of $S_A = 99$ orbitals and the plot is for the case where both subsystems contain $N_A = N_B = 60$ electrons.
The ``in charges" $(Q_0,Q_1,Q_2) = (0,0,0)$, which means that we do not have any quasi-holes at the edges of the cylinder. 
Note that we have shifted $L_z$ such that its lowest value is zero.
There are a few remarks to be made about \cref{fig:ent}. We start by discussing the low-lying entanglement levels.

On average, the lowest level for each angular momentum $L_z$ increases with increasing $L_z$, but this increase is slow and irregularly spaced. Upon increasing $P_{max}$, one obtains a better approximation of the state one calculates. However, how important these extra contributions are varies irregularly with $P_{max}$.
This makes it difficult to predict what cutoff value is needed for a given accuracy when computing e.g. the spins of the quasi-holes. The spin of the $(\sigma_1,1/5)$ quasi-hole in the upper left panel of \cref{fig:JvsPline} illustrates this rather clearly for $L=20$. The irregular convergence of the spin to the analytically expected value is not easy to predict. For instance, the value moves away from the prediction as $P_{max}$ increments from $8$ to $9$.
Even if one could fit a curve through the points by e.g. the least-squares method, the predictive value of such a curve is limited. This is because only few data points are available and because the behaviour is so erratic. Regardless, we find that the MPS computations converge with increasing $P_{max}$. However, this requires quite high values
of $P_{max}$, even for the smaller circumferences like $L=20$ (we recall that the MPS description converges faster for smaller values of $L$, due to the exponential factor \cref{normQH} from the free time evolution).

We now turn our attention to the highest entanglement levels for a given $L_z$.
The most important feature to notice is that the highest level for a given $L_z$
increases rapidly as $L_z$ grows. From $L_z = 4$ onwards, the two highest entanglement levels are close together. This pair
is still present for $L_z = 8$ (see the inset of \cref{fig:ent}).
Continuing their expected location to $L_z = 9$, we find that these levels would have values $e^{-\xi} \gtrsim 70$,
roughly corresponding to machine precision. Indeed, these two expected states at level $L_z=9$ are ``missing" (c.f. the inset of \cref{fig:ent}). This is important when comparing the number of observed entanglement levels to the expected number, which we do next.

A way to characterise the structure of the quantum Hall liquid is to examine the number of states per angular momentum $L_z$ \cite{Li_2008}.
The data in \cref{fig:ent} gives the state counting
$(1,1,3,6,12,21,39,64,108,\ldots)$. This matches
the state counting of the vacuum sector of the $\mathbb{Z}_3$ parafermion CFT (see for
instance \cite{Ardonne_2005}) up to angular momentum $L_z=8$. At level $L_z = 9$, we obtain 171 entanglement levels, while
the expected number is 173. As we explained above, the ``discrepancy" is caused by the small values of the two highest entanglement levels,
implying that they can not be resolved using machine precision calculations.

\begin{figure}[t]
    \centering
    \includegraphics[width=\linewidth]{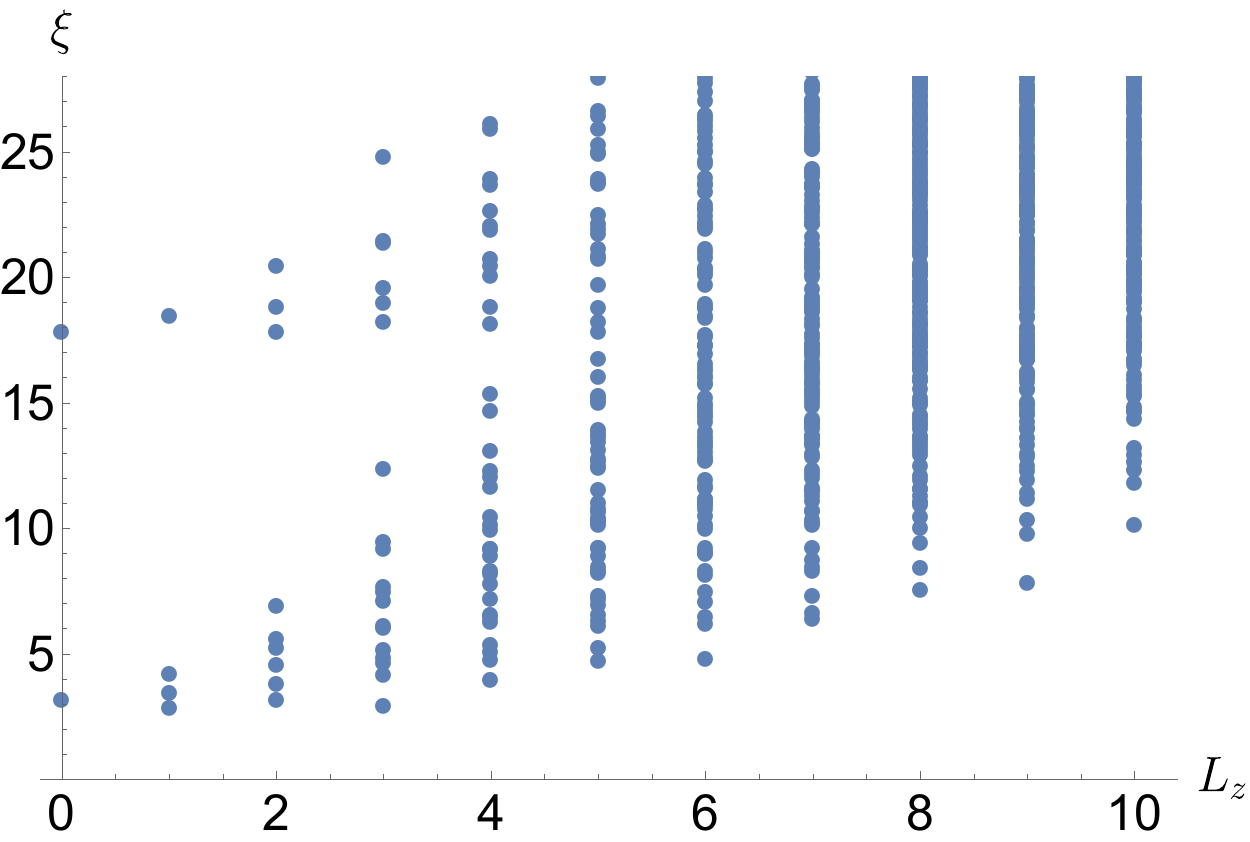}
    \caption{The low-lying part of the entanglement spectrum for a system with circumference $L=20$, $N_e = 120$ electrons and cutoff $P_{max}=12$,
    with the boundary charge $(3,0,2)$, corresponding to an $(\epsilon,3/5)$ quasihole.
    Both subsystems have $S_A = S_B = 99$ orbitals and contain $N_A = N_B = 60$ electrons.}
    \label{fig:ent-epsilon}
\end{figure}

We have also calculated the entanglement spectrum using the same parameters as above, but with a boundary charge $(Q_0,Q_1,Q_2) = (3,0,2)$,
which corresponds to an $(\epsilon,3/5)$ quasi-hole at each edge; the spectrum is given in \cref{fig:ent-epsilon}.
We clearly see that the entanglement spectrum consists of two branches at low $L_z$. These two branches correspond to the
two fusion channels of the fusion $\epsilon\times\epsilon = 1 + \epsilon$.
Indeed, the counting of the lower branch reproduces the state counting of the $\epsilon$ sector of the $\mathbb{Z}_3$ parafermion CFT,
namely $(1,3,6,13,24,\ldots)$.
The branch of higher entanglement levels follows the vacuum sector of the $\mathbb{Z}_3$ CFT.
The two branches can be distinguished up to $L_z = 4$. For higher $L_z$ values $5\leq L_z \leq 6$, the number of entanglement levels matches the sum
of state counting of the $\epsilon$ and vacuum sectors. For $L_z > 6$, the number of observed entanglement levels is lower than
the CFT state counting. The discrepancy is again due to the small singular values, which can not be distinguished from zero at machine precision. 

Finally, we mention (without showing the actual plots) that the entanglement spectra also give the expected state counting when $N_A < N_B$.
For instance, in the case $N_A = 59$, $N_B = 61$, and $(Q_0,Q_1,Q_2) = (0,0,0)$, one expects the state counting of the $\psi_2$ sector of the $\mathbb{Z}_3$
CFT (which equals that of the $\psi_1$ sector).
In the case of  $N_A = 59$, $N_B = 61$, and $(Q_0,Q_1,Q_2) = (3,0,2)$, we expect two branches, matching the $\sigma_1$ and $\psi_2$ sectors.
We indeed obtain entanglement spectra showing this behaviour.

The analysis of the entanglement spectra clearly demonstrates that the free boson MPS description captures the underlying $\mathbb{Z}_3$ structure of the
$k=3$ RR states.

\section{Dependence of observables on circumference, cutoff and integration method}\label{sec:error}
As is apparent from the charge computations, the MPS computation is very accurate for finding density profiles for quasi-holes. The main sources of error have to do with finite-circumference effects of the MPS cylinder. Firstly, the circumference $L$ should not be too large, in order to allow for faster decay of less important terms through the time evolution factor \cref{normQH}. On the other hand, having too small a value of $L$ makes the quasi-hole interfere with itself around the cylinder.
This self-interference causes errors when calculating the charge and spin of the quasi-hole, using
\cref{QintegDouble,QintegLine} and \cref{JDouble,JLine}.

Thus, there is an optimum value of $L$, which differs between different quasi-holes simply because some are (somewhat) larger than others: compare e.g. the $(\psi_2,4/5)$ and $(\mathbf{1},3/5)$ profiles in \cref{fig:prof1}. 

Increasing the cutoff $P_{max}$ generically leads to an increase in the numerical accuracy.
Although the obtained spin values for some combinations of quasi-hole and cylinder circumference $L$ (e.g. the $(\epsilon,\frac{3}{5})$ quasi-hole for $L=24$ in \cref{fig:JvsPline}) initially diverge from the expected spin value upon increasing $P_{max}$, all quasi-hole spins eventually converge towards the expected value for high enough $P_{max}$.
The rate of convergence depends on the circumference, although not in the same manner for all quasi-holes.
Typically, convergence is faster for a smaller circumference: see e.g. \cref{fig:JvsPline} and the data for the $(\psi_2,\frac{4}{5})$ quasi-hole with $L=20$ compared to $L=22$, or the same quasi-hole with $L=22$ compared to $L=24$, as expected.

The rate of convergence does not depend linearly on the circumference $L$, however. As can be seen clearly in \cref{fig:JvsPline} for the $(\psi_1,\frac{2}{5})$, $(\epsilon,\frac{3}{5})$ and $(\psi_2,\frac{4}{5})$ quasi-holes, it sometimes occurs that the difference between spin values for $L=20$ and $L=22$ for high $P_{max}$ is much smaller than that between $L=22$ and $L=24$, even though this is not the case for all quasi-hole types. Although we do not have a complete explanation for this phenomenon, we think this behaviour is related to the somewhat irregular behaviour of the lowest lying entanglement levels as described in the previous section.

To illustrate the effect of the chosen integration method, we give an example where the charges and spins are computed using the two different methods. First, there is the method where rotational symmetry of the system is assumed for all $r_{max}$ and the integration is carried out using a line integral (\cref{QintegLine} for the charge, \cref{JLine} for the spin). Secondly, there is the method where the symmetry assumption is only invoked once the integration region reaches around the cylinder, which occurs at $r_{max}= L/2$  (\cref{QintegDouble} for the charge, \cref{JDouble} for the spin). The results for the $(\sigma_1,1/5)$ quasi-hole are plotted in \cref{fig:compareMethods}. We see that the line integral scheme works better than the double integral for the spin, since self-interference around the finite-circumference cylinder means that the quasi-hole is not perfectly rotationally symmetric. By integrating along the line $r=\tau$ (or $r=-\tau$), we avoid the distortions in the $x$ direction, which affect the spin through the $r^2$ weight of the integrand, unlike if we use the double integral \cref{JDouble}. The charges, meanwhile, involve no such higher moment, and are therefore more accurately computed using the double integral in
\cref{QintegDouble} that takes into consideration the entire integration region, even the charge on the backside of the cylinder. In this case, the line integral \cref{QintegLine} gives a lower accuracy since the charge density is distorted: the $(\tau,x)=(\pm r,0)$ directions are not representative for the charge distribution in all directions around the quasi-hole. Thus, the double integral method performs better for the charge, as can be seen in \cref{fig:compareMethods}.  
\begin{onecolumngrid}

\begin{figure}
    \centering    \includegraphics[width=\linewidth]{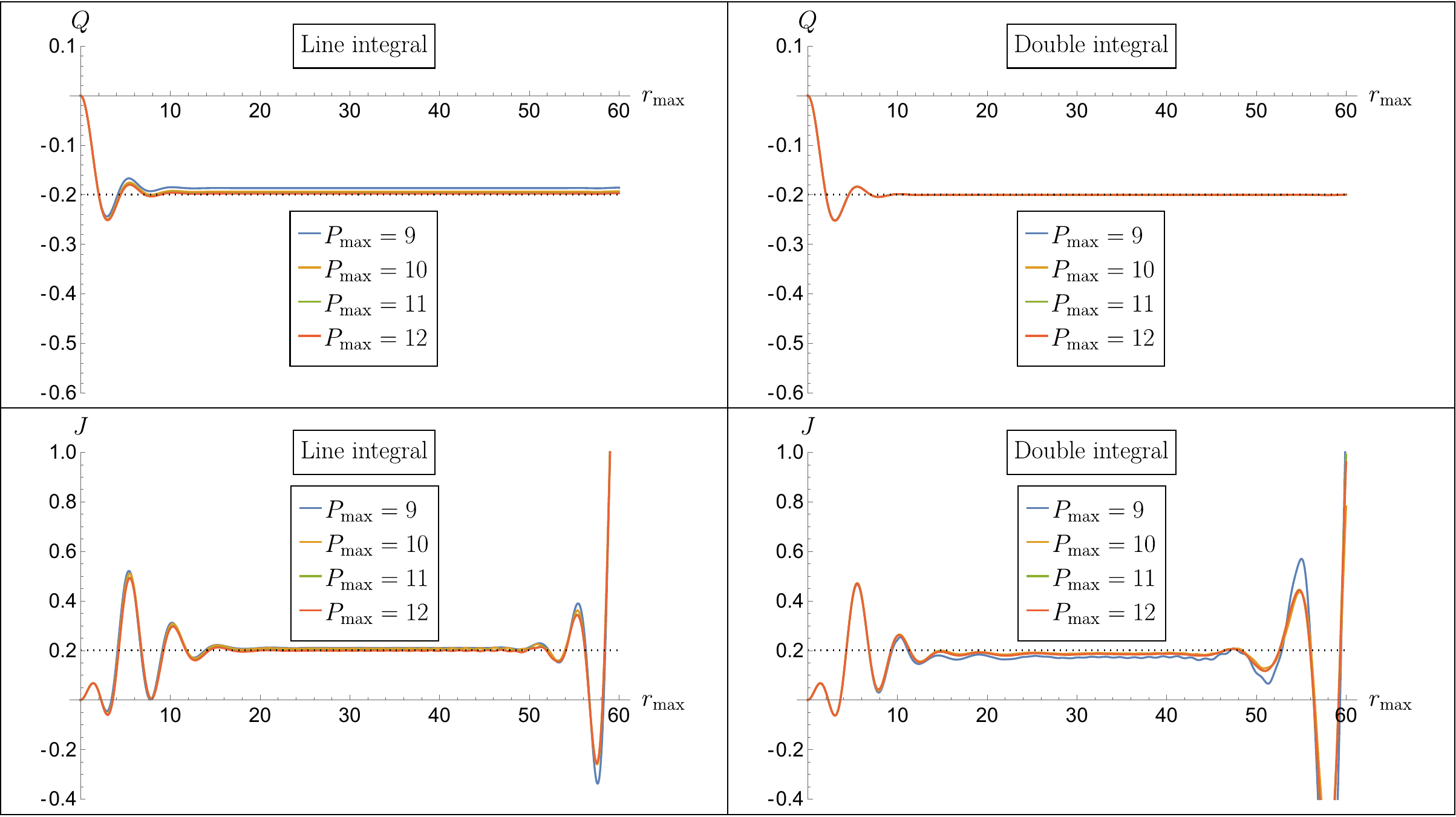}
    \caption[width=\linewidth]{Comparison between the line integral method \cref{QintegLine} and the double integral method \cref{QintegDouble} for the charge, and between the line integral \cref{JLine} and the double integral \cref{JDouble} for the spin. All results have been computed for the $(\sigma_1,1/5)$ quasi-hole. The circumference $L=20$ throughout. The dotted lines show the expected values.}
    \label{fig:compareMethods}
\end{figure}
    
\end{onecolumngrid}
\newpage
\twocolumngrid
\section{Discussion and conclusions}\label{sec:concl}
We have presented an MPS-based scheme for computing observables in the $k=3$ Read-Rezayi state. The observables we have focused on are the quasi-hole density profiles, charges, and spins, as well as the entanglement spectrum. The quasi-hole spins are of particular interest for two reasons. First of all, they are inherently interesting, and further emphasise the particle-like properties of Read-Rezayi quasi-holes. Secondly, the results agree well with the predictions from the SSR of ref. \cite{Nardin2023SSR}, under the assumption that the wave function has zero Berry phase and all its statistics information is contained in the monodromy. Therefore, we conclude that the statement ``holonomy equals monodromy" holds even for the $k=3$ Read-Rezayi state, which supports the conclusions drawn in \cite{Wu2014Braiding}. Interestingly, our method does not require explicit numerical braiding of quasi-holes to be performed, but relies solely on the spin-statistics theorem and the computed values of the spins, and is thus comparatively simple. Since the spins are local quantities, we may infer the non-local braiding behaviour and Berry phase from \textit{local} information, i.e. from the quasi-hole spins. This is not just surprising, but also practically useful since it makes drawing conclusions about Berry phases possible with lighter numerics than previously used, because there is no need to explicitly consider multi-anyon configurations. Therefore, our technique may render Berry phase computations for more complicated states, such as the Read-Rezayi states at larger values of $k$, more feasible in the future.

The relevance of the density profiles, charges and entanglement spectra is also considerable. The density profiles describe the Read-Rezayi quasi-holes in a way that is not readily apparent from the wave functions alone. The charges and entanglement spectra further characterise the properties of the Read-Rezayi state and its quasi-holes, and can also be used to assess the MPS technique introduced in the present work. Since the charge values can be read off from the CFT description, the charge computation can be used to evaluate the self-consistency of the numerics. The excellent accuracy with which the charges have been computed (see \cref{fig:Q}) is therefore a reassurance that our method is reliable. The entanglement spectra, in turn, can be used to verify that our free boson representation indeed captures the $\mathbb{Z}_3$ structure of the $k=3$ Read-Rezayi state, just like the more manifestly $\mathbb{Z}_3$-related representation of \cite{Wu2014Braiding,Wu2015MPS,estienne2015entropies}.

An advantage of the technique introduced in the present work is that it relies purely on free boson fields. The simplicity of the free boson CFT means that the analytical input to the MPS can be computed with relative ease, making it arguably more ``beginner-friendly" and approachable than the $\mathbb{Z}_3$ description that can be found in the literature (again, see refs. \cite{Wu2014Braiding,Wu2015MPS,estienne2015entropies}). Moreover, many interesting FQH states can be written in terms of several free chiral boson fields. Our results show that FQH states that can be written in terms of three free chiral bosons can be successfully analysed using MPS.

Although not explicitly shown in this paper, similar techniques have been used by the authors for the Moore-Read (i.e. $k=2$ Read-Rezayi) state, with two instead of three free boson fields. Even if the $k=3$ state is more involved, and requires an additional field, the two states can be handled on essentially the same footing, with no major conceptual differences. In principle, our method should be straightforwardly generalisable to higher Read-Rezayi states (i.e. to $k\geq 4$) by introducing additional boson fields.
The main obstacle in actually performing such calculations is that size of the auxiliary Hilbert space would become so large that meaningful calculations for, say, the quasi-hole spins would become impossible, without further optimisation of the code 
employed.

\section{Acknowledgements}
The authors wish to thank Leonardo Mazza and Alberto Nardin for interesting discussions, and for fruitful collaborations on a related paper. A.F. also wishes to thank Benoit Estienne for enjoyable discussions about MPS.
This research utilized the Sunrise HPC facility supported by the Technical Division at the Department of Physics, Stockholm University.

\bibliography{bibliography-long.bib}

\begin{thebibliography}{43}%
\makeatletter
\providecommand \@ifxundefined [1]{%
 \@ifx{#1\undefined}
}%
\providecommand \@ifnum [1]{%
 \ifnum #1\expandafter \@firstoftwo
 \else \expandafter \@secondoftwo
 \fi
}%
\providecommand \@ifx [1]{%
 \ifx #1\expandafter \@firstoftwo
 \else \expandafter \@secondoftwo
 \fi
}%
\providecommand \natexlab [1]{#1}%
\providecommand \enquote  [1]{``#1''}%
\providecommand \bibnamefont  [1]{#1}%
\providecommand \bibfnamefont [1]{#1}%
\providecommand \citenamefont [1]{#1}%
\providecommand \href@noop [0]{\@secondoftwo}%
\providecommand \href [0]{\begingroup \@sanitize@url \@href}%
\providecommand \@href[1]{\@@startlink{#1}\@@href}%
\providecommand \@@href[1]{\endgroup#1\@@endlink}%
\providecommand \@sanitize@url [0]{\catcode `\\12\catcode `\$12\catcode
  `\&12\catcode `\#12\catcode `\^12\catcode `\_12\catcode `\%12\relax}%
\providecommand \@@startlink[1]{}%
\providecommand \@@endlink[0]{}%
\providecommand \url  [0]{\begingroup\@sanitize@url \@url }%
\providecommand \@url [1]{\endgroup\@href {#1}{\urlprefix }}%
\providecommand \urlprefix  [0]{URL }%
\providecommand \Eprint [0]{\href }%
\providecommand \doibase [0]{https://doi.org/}%
\providecommand \selectlanguage [0]{\@gobble}%
\providecommand \bibinfo  [0]{\@secondoftwo}%
\providecommand \bibfield  [0]{\@secondoftwo}%
\providecommand \translation [1]{[#1]}%
\providecommand \BibitemOpen [0]{}%
\providecommand \bibitemStop [0]{}%
\providecommand \bibitemNoStop [0]{.\EOS\space}%
\providecommand \EOS [0]{\spacefactor3000\relax}%
\providecommand \BibitemShut  [1]{\csname bibitem#1\endcsname}%
\let\auto@bib@innerbib\@empty
\bibitem [{\citenamefont {Tsui}\ \emph {et~al.}(1982)\citenamefont {Tsui},
  \citenamefont {Stormer},\ and\ \citenamefont {Gossard}}]{Tsui1982FQHE}%
  \BibitemOpen
  \bibfield  {author} {\bibinfo {author} {\bibfnamefont {D.~C.}\ \bibnamefont
  {Tsui}}, \bibinfo {author} {\bibfnamefont {H.~L.}\ \bibnamefont {Stormer}},\
  and\ \bibinfo {author} {\bibfnamefont {A.~C.}\ \bibnamefont {Gossard}},\
  }\bibfield  {title} {\bibinfo {title} {{T}wo-{D}imensional {M}agnetotransport
  in the {E}xtreme {Q}uantum {L}imit},\ }\href
  {https://doi.org/10.1103/PhysRevLett.48.1559} {\bibfield  {journal} {\bibinfo
   {journal} {Phys. Rev. Lett.}\ }\textbf {\bibinfo {volume} {48}},\ \bibinfo
  {pages} {1559} (\bibinfo {year} {1982})}\BibitemShut {NoStop}%
\bibitem [{\citenamefont {Wen}(1995)}]{Wen1995Order}%
  \BibitemOpen
  \bibfield  {author} {\bibinfo {author} {\bibfnamefont {X.-G.}\ \bibnamefont
  {Wen}},\ }\bibfield  {title} {\bibinfo {title} {Topological orders and edge
  excitations in fractional quantum {H}all states},\ }\href
  {https://doi.org/10.1080/00018739500101566} {\bibfield  {journal} {\bibinfo
  {journal} {Advances in Physics}\ }\textbf {\bibinfo {volume} {44}},\ \bibinfo
  {pages} {405} (\bibinfo {year} {1995})},\ \Eprint
  {https://arxiv.org/abs/https://doi.org/10.1080/00018739500101566}
  {https://doi.org/10.1080/00018739500101566} \BibitemShut {NoStop}%
\bibitem [{\citenamefont {Nakamura}\ \emph {et~al.}(2020)\citenamefont
  {Nakamura}, \citenamefont {Liang},\ and\ \citenamefont {Gardner~\textit{et
  al}}}]{nakamura2020braiding}%
  \BibitemOpen
  \bibfield  {author} {\bibinfo {author} {\bibfnamefont {J.}~\bibnamefont
  {Nakamura}}, \bibinfo {author} {\bibfnamefont {S.}~\bibnamefont {Liang}},\
  and\ \bibinfo {author} {\bibfnamefont {G.}~\bibnamefont {Gardner~\textit{et
  al}}},\ }\bibfield  {title} {\bibinfo {title} {Direct observation of anyonic
  braiding statistics},\ }\href@noop {} {\bibfield  {journal} {\bibinfo
  {journal} {Nat. Phys.}\ }\textbf {\bibinfo {volume} {16}},\ \bibinfo {pages}
  {931–936} (\bibinfo {year} {2020})}\BibitemShut {NoStop}%
\bibitem [{\citenamefont {Read}\ and\ \citenamefont
  {Rezayi}(1999)}]{Read1999Parafermions}%
  \BibitemOpen
  \bibfield  {author} {\bibinfo {author} {\bibfnamefont {N.}~\bibnamefont
  {Read}}\ and\ \bibinfo {author} {\bibfnamefont {E.}~\bibnamefont {Rezayi}},\
  }\bibfield  {title} {\bibinfo {title} {Beyond paired quantum {H}all states:
  {P}arafermions and incompressible states in the first excited {L}andau
  level},\ }\href {https://doi.org/10.1103/PhysRevB.59.8084} {\bibfield
  {journal} {\bibinfo  {journal} {Phys. Rev. B}\ }\textbf {\bibinfo {volume}
  {59}},\ \bibinfo {pages} {8084} (\bibinfo {year} {1999})}\BibitemShut
  {NoStop}%
\bibitem [{\citenamefont {Laughlin}(1983)}]{Laughlin_1983}%
  \BibitemOpen
  \bibfield  {author} {\bibinfo {author} {\bibfnamefont {R.~B.}\ \bibnamefont
  {Laughlin}},\ }\bibfield  {title} {\bibinfo {title} {Anomalous {Q}uantum
  {H}all {E}ffect: {A}n {I}ncompressible {Q}uantum {F}luid with {F}ractionally
  {C}harged {E}xcitations},\ }\href
  {https://doi.org/10.1103/PhysRevLett.50.1395} {\bibfield  {journal} {\bibinfo
   {journal} {Phys. Rev. Lett.}\ }\textbf {\bibinfo {volume} {50}},\ \bibinfo
  {pages} {1395} (\bibinfo {year} {1983})}\BibitemShut {NoStop}%
\bibitem [{\citenamefont {Moore}\ and\ \citenamefont
  {Read}(1991)}]{Moore1991Nonabelions}%
  \BibitemOpen
  \bibfield  {author} {\bibinfo {author} {\bibfnamefont {G.}~\bibnamefont
  {Moore}}\ and\ \bibinfo {author} {\bibfnamefont {N.}~\bibnamefont {Read}},\
  }\bibfield  {title} {\bibinfo {title} {Nonabelions in the fractional quantum
  hall effect},\ }\href
  {https://doi.org/https://doi.org/10.1016/0550-3213(91)90407-O} {\bibfield
  {journal} {\bibinfo  {journal} {Nuclear Physics B}\ }\textbf {\bibinfo
  {volume} {360}},\ \bibinfo {pages} {362} (\bibinfo {year}
  {1991})}\BibitemShut {NoStop}%
\bibitem [{\citenamefont {Willett}\ \emph {et~al.}(1987)\citenamefont
  {Willett}, \citenamefont {Eisenstein}, \citenamefont {St\"ormer},
  \citenamefont {Tsui}, \citenamefont {Gossard},\ and\ \citenamefont
  {English}}]{Willett1987Observation}%
  \BibitemOpen
  \bibfield  {author} {\bibinfo {author} {\bibfnamefont {R.}~\bibnamefont
  {Willett}}, \bibinfo {author} {\bibfnamefont {J.~P.}\ \bibnamefont
  {Eisenstein}}, \bibinfo {author} {\bibfnamefont {H.~L.}\ \bibnamefont
  {St\"ormer}}, \bibinfo {author} {\bibfnamefont {D.~C.}\ \bibnamefont {Tsui}},
  \bibinfo {author} {\bibfnamefont {A.~C.}\ \bibnamefont {Gossard}},\ and\
  \bibinfo {author} {\bibfnamefont {J.~H.}\ \bibnamefont {English}},\
  }\bibfield  {title} {\bibinfo {title} {Observation of an even-denominator
  quantum number in the fractional quantum {H}all effect},\ }\href
  {https://doi.org/10.1103/PhysRevLett.59.1776} {\bibfield  {journal} {\bibinfo
   {journal} {Phys. Rev. Lett.}\ }\textbf {\bibinfo {volume} {59}},\ \bibinfo
  {pages} {1776} (\bibinfo {year} {1987})}\BibitemShut {NoStop}%
\bibitem [{\citenamefont {Greiter}\ \emph {et~al.}(1992)\citenamefont
  {Greiter}, \citenamefont {Wen},\ and\ \citenamefont
  {Wilczek}}]{Greiter1992paired}%
  \BibitemOpen
  \bibfield  {author} {\bibinfo {author} {\bibfnamefont {M.}~\bibnamefont
  {Greiter}}, \bibinfo {author} {\bibfnamefont {X.}~\bibnamefont {Wen}},\ and\
  \bibinfo {author} {\bibfnamefont {F.}~\bibnamefont {Wilczek}},\ }\bibfield
  {title} {\bibinfo {title} {Paired {H}all states},\ }\href
  {https://doi.org/https://doi.org/10.1016/0550-3213(92)90401-V} {\bibfield
  {journal} {\bibinfo  {journal} {Nuclear Physics B}\ }\textbf {\bibinfo
  {volume} {374}},\ \bibinfo {pages} {567} (\bibinfo {year}
  {1992})}\BibitemShut {NoStop}%
\bibitem [{\citenamefont {Xia}\ \emph {et~al.}(2004)\citenamefont {Xia},
  \citenamefont {Pan}, \citenamefont {Vicente}, \citenamefont {Adams},
  \citenamefont {Sullivan}, \citenamefont {Stormer}, \citenamefont {Tsui},
  \citenamefont {Pfeiffer}, \citenamefont {Baldwin},\ and\ \citenamefont
  {West}}]{Xia2004Correlation}%
  \BibitemOpen
  \bibfield  {author} {\bibinfo {author} {\bibfnamefont {J.~S.}\ \bibnamefont
  {Xia}}, \bibinfo {author} {\bibfnamefont {W.}~\bibnamefont {Pan}}, \bibinfo
  {author} {\bibfnamefont {C.~L.}\ \bibnamefont {Vicente}}, \bibinfo {author}
  {\bibfnamefont {E.~D.}\ \bibnamefont {Adams}}, \bibinfo {author}
  {\bibfnamefont {N.~S.}\ \bibnamefont {Sullivan}}, \bibinfo {author}
  {\bibfnamefont {H.~L.}\ \bibnamefont {Stormer}}, \bibinfo {author}
  {\bibfnamefont {D.~C.}\ \bibnamefont {Tsui}}, \bibinfo {author}
  {\bibfnamefont {L.~N.}\ \bibnamefont {Pfeiffer}}, \bibinfo {author}
  {\bibfnamefont {K.~W.}\ \bibnamefont {Baldwin}},\ and\ \bibinfo {author}
  {\bibfnamefont {K.~W.}\ \bibnamefont {West}},\ }\bibfield  {title} {\bibinfo
  {title} {Electron {C}orrelation in the {S}econd {L}andau {L}evel: {A}
  {C}ompetition {B}etween {M}any {N}early {D}egenerate {Q}uantum {P}hases},\
  }\href {https://doi.org/10.1103/PhysRevLett.93.176809} {\bibfield  {journal}
  {\bibinfo  {journal} {Phys. Rev. Lett.}\ }\textbf {\bibinfo {volume} {93}},\
  \bibinfo {pages} {176809} (\bibinfo {year} {2004})}\BibitemShut {NoStop}%
\bibitem [{\citenamefont {Orús}(2014)}]{Orus2014mps}%
  \BibitemOpen
  \bibfield  {author} {\bibinfo {author} {\bibfnamefont {R.}~\bibnamefont
  {Orús}},\ }\bibfield  {title} {\bibinfo {title} {A practical introduction to
  tensor networks: {M}atrix product states and projected entangled pair
  states},\ }\href {https://doi.org/https://doi.org/10.1016/j.aop.2014.06.013}
  {\bibfield  {journal} {\bibinfo  {journal} {Annals of Physics}\ }\textbf
  {\bibinfo {volume} {349}},\ \bibinfo {pages} {117} (\bibinfo {year}
  {2014})}\BibitemShut {NoStop}%
\bibitem [{\citenamefont {Schollwöck}(2011)}]{Schollwock2011mps}%
  \BibitemOpen
  \bibfield  {author} {\bibinfo {author} {\bibfnamefont {U.}~\bibnamefont
  {Schollwöck}},\ }\bibfield  {title} {\bibinfo {title} {The density-matrix
  renormalization group in the age of matrix product states},\ }\href
  {https://doi.org/https://doi.org/10.1016/j.aop.2010.09.012} {\bibfield
  {journal} {\bibinfo  {journal} {Annals of Physics}\ }\textbf {\bibinfo
  {volume} {326}},\ \bibinfo {pages} {96} (\bibinfo {year} {2011})},\ \bibinfo
  {note} {january 2011 Special Issue}\BibitemShut {NoStop}%
\bibitem [{\citenamefont {Zaletel}\ and\ \citenamefont
  {Mong}(2012)}]{Zaletel2012MPS}%
  \BibitemOpen
  \bibfield  {author} {\bibinfo {author} {\bibfnamefont {M.~P.}\ \bibnamefont
  {Zaletel}}\ and\ \bibinfo {author} {\bibfnamefont {R.~S.~K.}\ \bibnamefont
  {Mong}},\ }\bibfield  {title} {\bibinfo {title} {Exact matrix product states
  for quantum {H}all wave functions},\ }\href
  {https://doi.org/10.1103/PhysRevB.86.245305} {\bibfield  {journal} {\bibinfo
  {journal} {Phys. Rev. B}\ }\textbf {\bibinfo {volume} {86}},\ \bibinfo
  {pages} {245305} (\bibinfo {year} {2012})}\BibitemShut {NoStop}%
\bibitem [{\citenamefont {Metropolis}\ \emph {et~al.}(1953)\citenamefont
  {Metropolis}, \citenamefont {Rosenbluth}, \citenamefont {Rosenbluth},
  \citenamefont {Teller},\ and\ \citenamefont
  {Teller}}]{metropolis1953equation}%
  \BibitemOpen
  \bibfield  {author} {\bibinfo {author} {\bibfnamefont {N.}~\bibnamefont
  {Metropolis}}, \bibinfo {author} {\bibfnamefont {A.~W.}\ \bibnamefont
  {Rosenbluth}}, \bibinfo {author} {\bibfnamefont {M.~N.}\ \bibnamefont
  {Rosenbluth}}, \bibinfo {author} {\bibfnamefont {A.~H.}\ \bibnamefont
  {Teller}},\ and\ \bibinfo {author} {\bibfnamefont {E.}~\bibnamefont
  {Teller}},\ }\bibfield  {title} {\bibinfo {title} {Equation of state
  calculations by fast computing machines},\ }\href@noop {} {\bibfield
  {journal} {\bibinfo  {journal} {The journal of chemical physics}\ }\textbf
  {\bibinfo {volume} {21}},\ \bibinfo {pages} {1087} (\bibinfo {year}
  {1953})}\BibitemShut {NoStop}%
\bibitem [{\citenamefont {Estienne}\ \emph
  {et~al.}(2013{\natexlab{a}})\citenamefont {Estienne}, \citenamefont
  {Papi\ifmmode~\acute{c}\else \'{c}\fi{}}, \citenamefont {Regnault},\ and\
  \citenamefont {Bernevig}}]{Estienne_2013}%
  \BibitemOpen
  \bibfield  {author} {\bibinfo {author} {\bibfnamefont {B.}~\bibnamefont
  {Estienne}}, \bibinfo {author} {\bibfnamefont {Z.}~\bibnamefont
  {Papi\ifmmode~\acute{c}\else \'{c}\fi{}}}, \bibinfo {author} {\bibfnamefont
  {N.}~\bibnamefont {Regnault}},\ and\ \bibinfo {author} {\bibfnamefont
  {B.~A.}\ \bibnamefont {Bernevig}},\ }\bibfield  {title} {\bibinfo {title}
  {Matrix product states for trial quantum hall states},\ }\href
  {https://doi.org/10.1103/PhysRevB.87.161112} {\bibfield  {journal} {\bibinfo
  {journal} {Phys. Rev. B}\ }\textbf {\bibinfo {volume} {87}},\ \bibinfo
  {pages} {161112} (\bibinfo {year} {2013}{\natexlab{a}})}\BibitemShut
  {NoStop}%
\bibitem [{\citenamefont {Wu}\ \emph {et~al.}(2014)\citenamefont {Wu},
  \citenamefont {Estienne}, \citenamefont {Regnault},\ and\ \citenamefont
  {Bernevig}}]{Wu2014Braiding}%
  \BibitemOpen
  \bibfield  {author} {\bibinfo {author} {\bibfnamefont {Y.-L.}\ \bibnamefont
  {Wu}}, \bibinfo {author} {\bibfnamefont {B.}~\bibnamefont {Estienne}},
  \bibinfo {author} {\bibfnamefont {N.}~\bibnamefont {Regnault}},\ and\
  \bibinfo {author} {\bibfnamefont {B.~A.}\ \bibnamefont {Bernevig}},\
  }\bibfield  {title} {\bibinfo {title} {Braiding {N}on-{A}belian {Q}uasiholes
  in {F}ractional {Q}uantum {H}all {S}tates},\ }\href
  {https://doi.org/10.1103/PhysRevLett.113.116801} {\bibfield  {journal}
  {\bibinfo  {journal} {Phys. Rev. Lett.}\ }\textbf {\bibinfo {volume} {113}},\
  \bibinfo {pages} {116801} (\bibinfo {year} {2014})}\BibitemShut {NoStop}%
\bibitem [{\citenamefont {Wu}\ \emph {et~al.}(2015)\citenamefont {Wu},
  \citenamefont {Estienne}, \citenamefont {Regnault},\ and\ \citenamefont
  {Bernevig}}]{Wu2015MPS}%
  \BibitemOpen
  \bibfield  {author} {\bibinfo {author} {\bibfnamefont {Y.-L.}\ \bibnamefont
  {Wu}}, \bibinfo {author} {\bibfnamefont {B.}~\bibnamefont {Estienne}},
  \bibinfo {author} {\bibfnamefont {N.}~\bibnamefont {Regnault}},\ and\
  \bibinfo {author} {\bibfnamefont {B.~A.}\ \bibnamefont {Bernevig}},\
  }\bibfield  {title} {\bibinfo {title} {Matrix product state representation of
  non-{A}belian quasiholes},\ }\href
  {https://doi.org/10.1103/PhysRevB.92.045109} {\bibfield  {journal} {\bibinfo
  {journal} {Phys. Rev. B}\ }\textbf {\bibinfo {volume} {92}},\ \bibinfo
  {pages} {045109} (\bibinfo {year} {2015})}\BibitemShut {NoStop}%
\bibitem [{\citenamefont {Estienne}\ \emph {et~al.}(2015)\citenamefont
  {Estienne}, \citenamefont {Regnault},\ and\ \citenamefont
  {Bernevig}}]{estienne2015entropies}%
  \BibitemOpen
  \bibfield  {author} {\bibinfo {author} {\bibfnamefont {B.}~\bibnamefont
  {Estienne}}, \bibinfo {author} {\bibfnamefont {N.}~\bibnamefont {Regnault}},\
  and\ \bibinfo {author} {\bibfnamefont {B.~A.}\ \bibnamefont {Bernevig}},\
  }\bibfield  {title} {\bibinfo {title} {{C}orrelation {L}engths and
  {T}opological {E}ntanglement {E}ntropies of {U}nitary and {N}onunitary
  {F}ractional {Q}uantum {H}all {W}ave {F}unctions},\ }\href
  {https://doi.org/10.1103/PhysRevLett.114.186801} {\bibfield  {journal}
  {\bibinfo  {journal} {Phys. Rev. Lett.}\ }\textbf {\bibinfo {volume} {114}},\
  \bibinfo {pages} {186801} (\bibinfo {year} {2015})}\BibitemShut {NoStop}%
\bibitem [{\citenamefont {Herviou}\ and\ \citenamefont
  {Mila}(2024)}]{Herviou_2024}%
  \BibitemOpen
  \bibfield  {author} {\bibinfo {author} {\bibfnamefont {L.}~\bibnamefont
  {Herviou}}\ and\ \bibinfo {author} {\bibfnamefont {F.}~\bibnamefont {Mila}},\
  }\bibfield  {title} {\bibinfo {title} {Numerical investigation of the
  structure factors of the read-rezayi series},\ }\href
  {https://doi.org/10.1103/PhysRevB.110.045143} {\bibfield  {journal} {\bibinfo
   {journal} {Phys. Rev. B}\ }\textbf {\bibinfo {volume} {110}},\ \bibinfo
  {pages} {045143} (\bibinfo {year} {2024})}\BibitemShut {NoStop}%
\bibitem [{\citenamefont {Zamolodchikov}\ and\ \citenamefont
  {Fateev}(1985)}]{Fateev-Zamolodchikov1985}%
  \BibitemOpen
  \bibfield  {author} {\bibinfo {author} {\bibfnamefont {A.}~\bibnamefont
  {Zamolodchikov}}\ and\ \bibinfo {author} {\bibfnamefont {V.}~\bibnamefont
  {Fateev}},\ }\bibfield  {title} {\bibinfo {title} {Nonlocal (parafermion)
  currents in two-dimensional conformal quantum field theory and self-dual
  critical points in $z_n$-symmetric statistical systems},\ }\href@noop {}
  {\bibfield  {journal} {\bibinfo  {journal} {Sov. Phys.-JETP (Engl.
  Transl.);(United States)}\ }\textbf {\bibinfo {volume} {62}} (\bibinfo {year}
  {1985})}\BibitemShut {NoStop}%
\bibitem [{\citenamefont {Cappelli}\ \emph {et~al.}(2001)\citenamefont
  {Cappelli}, \citenamefont {Georgiev},\ and\ \citenamefont
  {Todorov}}]{Cappelli2001Parafermion}%
  \BibitemOpen
  \bibfield  {author} {\bibinfo {author} {\bibfnamefont {A.}~\bibnamefont
  {Cappelli}}, \bibinfo {author} {\bibfnamefont {L.~S.}\ \bibnamefont
  {Georgiev}},\ and\ \bibinfo {author} {\bibfnamefont {I.~T.}\ \bibnamefont
  {Todorov}},\ }\bibfield  {title} {\bibinfo {title} {Parafermion {H}all states
  from coset projections of abelian conformal theories},\ }\href
  {https://doi.org/https://doi.org/10.1016/S0550-3213(00)00774-4} {\bibfield
  {journal} {\bibinfo  {journal} {Nuclear Physics B}\ }\textbf {\bibinfo
  {volume} {599}},\ \bibinfo {pages} {499} (\bibinfo {year}
  {2001})}\BibitemShut {NoStop}%
\bibitem [{\citenamefont {Halperin}(1983)}]{Halperin_1983}%
  \BibitemOpen
  \bibfield  {author} {\bibinfo {author} {\bibfnamefont {B.~I.}\ \bibnamefont
  {Halperin}},\ }\bibfield  {title} {\bibinfo {title} {Theory of the quantized
  hall conductance},\ }\href
  {https://www.e-periodica.ch/digbib/view?pid=hpa-001:1983:56::1243#87}
  {\bibfield  {journal} {\bibinfo  {journal} {Helv. Phys. Acta}\ }\textbf
  {\bibinfo {volume} {56}},\ \bibinfo {pages} {75} (\bibinfo {year}
  {1983})}\BibitemShut {NoStop}%
\bibitem [{\citenamefont {Halperin}(1984)}]{Halperin_1984}%
  \BibitemOpen
  \bibfield  {author} {\bibinfo {author} {\bibfnamefont {B.~I.}\ \bibnamefont
  {Halperin}},\ }\bibfield  {title} {\bibinfo {title} {Statistics of
  quasiparticles and the hierarchy of fractional quantized hall states},\
  }\href {https://doi.org/10.1103/PhysRevLett.52.1583} {\bibfield  {journal}
  {\bibinfo  {journal} {Phys. Rev. Lett.}\ }\textbf {\bibinfo {volume} {52}},\
  \bibinfo {pages} {1583} (\bibinfo {year} {1984})}\BibitemShut {NoStop}%
\bibitem [{\citenamefont {Haldane}\ and\ \citenamefont
  {Rezayi}(1988)}]{Haldane_1988}%
  \BibitemOpen
  \bibfield  {author} {\bibinfo {author} {\bibfnamefont {F.~D.~M.}\
  \bibnamefont {Haldane}}\ and\ \bibinfo {author} {\bibfnamefont {E.~H.}\
  \bibnamefont {Rezayi}},\ }\bibfield  {title} {\bibinfo {title} {Spin-singlet
  wave function for the half-integral quantum hall effect},\ }\href
  {https://doi.org/10.1103/PhysRevLett.60.956} {\bibfield  {journal} {\bibinfo
  {journal} {Phys. Rev. Lett.}\ }\textbf {\bibinfo {volume} {60}},\ \bibinfo
  {pages} {956} (\bibinfo {year} {1988})}\BibitemShut {NoStop}%
\bibitem [{\citenamefont {Cr\'epel}\ \emph {et~al.}(2018)\citenamefont
  {Cr\'epel}, \citenamefont {Estienne}, \citenamefont {Bernevig}, \citenamefont
  {Lecheminant},\ and\ \citenamefont {Regnault}}]{Crepel_2018}%
  \BibitemOpen
  \bibfield  {author} {\bibinfo {author} {\bibfnamefont {V.}~\bibnamefont
  {Cr\'epel}}, \bibinfo {author} {\bibfnamefont {B.}~\bibnamefont {Estienne}},
  \bibinfo {author} {\bibfnamefont {B.~A.}\ \bibnamefont {Bernevig}}, \bibinfo
  {author} {\bibfnamefont {P.}~\bibnamefont {Lecheminant}},\ and\ \bibinfo
  {author} {\bibfnamefont {N.}~\bibnamefont {Regnault}},\ }\bibfield  {title}
  {\bibinfo {title} {Matrix product state description of halperin states},\
  }\href {https://doi.org/10.1103/PhysRevB.97.165136} {\bibfield  {journal}
  {\bibinfo  {journal} {Phys. Rev. B}\ }\textbf {\bibinfo {volume} {97}},\
  \bibinfo {pages} {165136} (\bibinfo {year} {2018})}\BibitemShut {NoStop}%
\bibitem [{\citenamefont {Cr\'epel}\ \emph {et~al.}(2019)\citenamefont
  {Cr\'epel}, \citenamefont {Regnault},\ and\ \citenamefont
  {Estienne}}]{Crepel_2019}%
  \BibitemOpen
  \bibfield  {author} {\bibinfo {author} {\bibfnamefont {V.}~\bibnamefont
  {Cr\'epel}}, \bibinfo {author} {\bibfnamefont {N.}~\bibnamefont {Regnault}},\
  and\ \bibinfo {author} {\bibfnamefont {B.}~\bibnamefont {Estienne}},\
  }\bibfield  {title} {\bibinfo {title} {Matrix product state description and
  gaplessness of the haldane-rezayi state},\ }\href
  {https://doi.org/10.1103/PhysRevB.100.125128} {\bibfield  {journal} {\bibinfo
   {journal} {Phys. Rev. B}\ }\textbf {\bibinfo {volume} {100}},\ \bibinfo
  {pages} {125128} (\bibinfo {year} {2019})}\BibitemShut {NoStop}%
\bibitem [{\citenamefont {Berry}(1984)}]{Berry1984phase}%
  \BibitemOpen
  \bibfield  {author} {\bibinfo {author} {\bibfnamefont {M.~V.}\ \bibnamefont
  {Berry}},\ }\bibfield  {title} {\bibinfo {title} {Quantal phase factors
  accompanying adiabatic changes},\ }\href
  {https://doi.org/10.1098/rspa.1984.0023} {\bibfield  {journal} {\bibinfo
  {journal} {Proceedings of the Royal Society of London. A. Mathematical and
  Physical Sciences}\ }\textbf {\bibinfo {volume} {392}},\ \bibinfo {pages}
  {45} (\bibinfo {year} {1984})}\BibitemShut {NoStop}%
\bibitem [{\citenamefont {Belavin}\ \emph {et~al.}(1984)\citenamefont
  {Belavin}, \citenamefont {Polyakov},\ and\ \citenamefont
  {Zamolodchikov}}]{Belavin1984conformal}%
  \BibitemOpen
  \bibfield  {author} {\bibinfo {author} {\bibfnamefont {A.}~\bibnamefont
  {Belavin}}, \bibinfo {author} {\bibfnamefont {A.}~\bibnamefont {Polyakov}},\
  and\ \bibinfo {author} {\bibfnamefont {A.}~\bibnamefont {Zamolodchikov}},\
  }\bibfield  {title} {\bibinfo {title} {Infinite conformal symmetry in
  two-dimensional quantum field theory},\ }\href
  {https://doi.org/https://doi.org/10.1016/0550-3213(84)90052-X} {\bibfield
  {journal} {\bibinfo  {journal} {Nuclear Physics B}\ }\textbf {\bibinfo
  {volume} {241}},\ \bibinfo {pages} {333} (\bibinfo {year}
  {1984})}\BibitemShut {NoStop}%
\bibitem [{\citenamefont {Bonderson}\ \emph {et~al.}(2011)\citenamefont
  {Bonderson}, \citenamefont {Gurarie},\ and\ \citenamefont
  {Nayak}}]{Bonderson2011Plasma}%
  \BibitemOpen
  \bibfield  {author} {\bibinfo {author} {\bibfnamefont {P.}~\bibnamefont
  {Bonderson}}, \bibinfo {author} {\bibfnamefont {V.}~\bibnamefont {Gurarie}},\
  and\ \bibinfo {author} {\bibfnamefont {C.}~\bibnamefont {Nayak}},\ }\bibfield
   {title} {\bibinfo {title} {Plasma analogy and non-{A}belian statistics for
  {I}sing-type quantum {H}all states},\ }\href
  {https://doi.org/10.1103/PhysRevB.83.075303} {\bibfield  {journal} {\bibinfo
  {journal} {Phys. Rev. B}\ }\textbf {\bibinfo {volume} {83}},\ \bibinfo
  {pages} {075303} (\bibinfo {year} {2011})}\BibitemShut {NoStop}%
\bibitem [{\citenamefont {Herland}\ \emph {et~al.}(2012)\citenamefont
  {Herland}, \citenamefont {Babaev}, \citenamefont {Bonderson}, \citenamefont
  {Gurarie}, \citenamefont {Nayak},\ and\ \citenamefont
  {Sudb\o{}}}]{Herland2012Screening}%
  \BibitemOpen
  \bibfield  {author} {\bibinfo {author} {\bibfnamefont {E.~V.}\ \bibnamefont
  {Herland}}, \bibinfo {author} {\bibfnamefont {E.}~\bibnamefont {Babaev}},
  \bibinfo {author} {\bibfnamefont {P.}~\bibnamefont {Bonderson}}, \bibinfo
  {author} {\bibfnamefont {V.}~\bibnamefont {Gurarie}}, \bibinfo {author}
  {\bibfnamefont {C.}~\bibnamefont {Nayak}},\ and\ \bibinfo {author}
  {\bibfnamefont {A.}~\bibnamefont {Sudb\o{}}},\ }\bibfield  {title} {\bibinfo
  {title} {Screening properties and phase transitions in unconventional plasmas
  for {I}sing-type quantum {H}all states},\ }\href
  {https://doi.org/10.1103/PhysRevB.85.024520} {\bibfield  {journal} {\bibinfo
  {journal} {Phys. Rev. B}\ }\textbf {\bibinfo {volume} {85}},\ \bibinfo
  {pages} {024520} (\bibinfo {year} {2012})}\BibitemShut {NoStop}%
\bibitem [{\citenamefont {Nardin}\ \emph {et~al.}(2023)\citenamefont {Nardin},
  \citenamefont {Ardonne},\ and\ \citenamefont {Mazza}}]{Nardin2023SSR}%
  \BibitemOpen
  \bibfield  {author} {\bibinfo {author} {\bibfnamefont {A.}~\bibnamefont
  {Nardin}}, \bibinfo {author} {\bibfnamefont {E.}~\bibnamefont {Ardonne}},\
  and\ \bibinfo {author} {\bibfnamefont {L.}~\bibnamefont {Mazza}},\ }\bibfield
   {title} {\bibinfo {title} {Spin-statistics relation for quantum {H}all
  states},\ }\href {https://doi.org/10.1103/PhysRevB.108.L041105} {\bibfield
  {journal} {\bibinfo  {journal} {Phys. Rev. B}\ }\textbf {\bibinfo {volume}
  {108}},\ \bibinfo {pages} {L041105} (\bibinfo {year} {2023})}\BibitemShut
  {NoStop}%
\bibitem [{\citenamefont {Fagerlund}\ \emph {et~al.}(2024)\citenamefont
  {Fagerlund}, \citenamefont {Nardin}, \citenamefont {Mazza},\ and\
  \citenamefont {Ardonne}}]{Fagerlund2024nonAbelian}%
  \BibitemOpen
  \bibfield  {author} {\bibinfo {author} {\bibfnamefont {A.}~\bibnamefont
  {Fagerlund}}, \bibinfo {author} {\bibfnamefont {A.}~\bibnamefont {Nardin}},
  \bibinfo {author} {\bibfnamefont {L.}~\bibnamefont {Mazza}},\ and\ \bibinfo
  {author} {\bibfnamefont {E.}~\bibnamefont {Ardonne}},\ }\bibfield  {title}
  {\bibinfo {title} {Spin fractionalization at the edge of quantum {H}all
  fluids induced by bulk quasiparticles},\ }\bibfield  {journal} {\bibinfo
  {journal} {arXiv:2412.14879}\ }\href
  {https://doi.org/10.48550/arXiv.2412.14879} {10.48550/arXiv.2412.14879}
  (\bibinfo {year} {2024})\BibitemShut {NoStop}%
\bibitem [{\citenamefont {Ardonne}\ and\ \citenamefont
  {Schoutens}(2007)}]{ARDONNE2007Registers}%
  \BibitemOpen
  \bibfield  {author} {\bibinfo {author} {\bibfnamefont {E.}~\bibnamefont
  {Ardonne}}\ and\ \bibinfo {author} {\bibfnamefont {K.}~\bibnamefont
  {Schoutens}},\ }\bibfield  {title} {\bibinfo {title} {Wavefunctions for
  topological quantum registers},\ }\href
  {https://doi.org/https://doi.org/10.1016/j.aop.2006.07.015} {\bibfield
  {journal} {\bibinfo  {journal} {Annals of Physics}\ }\textbf {\bibinfo
  {volume} {322}},\ \bibinfo {pages} {201} (\bibinfo {year} {2007})},\ \bibinfo
  {note} {january Special Issue 2007}\BibitemShut {NoStop}%
\bibitem [{\citenamefont {Kjäll}\ \emph {et~al.}(2018)\citenamefont {Kjäll},
  \citenamefont {Ardonne}, \citenamefont {Dwivedi}, \citenamefont {Hermanns},\
  and\ \citenamefont {Hansson}}]{Kjall2018MPS}%
  \BibitemOpen
  \bibfield  {author} {\bibinfo {author} {\bibfnamefont {J.}~\bibnamefont
  {Kjäll}}, \bibinfo {author} {\bibfnamefont {E.}~\bibnamefont {Ardonne}},
  \bibinfo {author} {\bibfnamefont {V.}~\bibnamefont {Dwivedi}}, \bibinfo
  {author} {\bibfnamefont {M.}~\bibnamefont {Hermanns}},\ and\ \bibinfo
  {author} {\bibfnamefont {T.~H.}\ \bibnamefont {Hansson}},\ }\bibfield
  {title} {\bibinfo {title} {Matrix product state representation of
  quasielectron wave functions},\ }\href
  {https://doi.org/10.1088/1742-5468/aab679} {\bibfield  {journal} {\bibinfo
  {journal} {Journal of Statistical Mechanics: Theory and Experiment}\ }\textbf
  {\bibinfo {volume} {2018}},\ \bibinfo {pages} {053101} (\bibinfo {year}
  {2018})}\BibitemShut {NoStop}%
\bibitem [{\citenamefont {Francesco}\ \emph {et~al.}(2012)\citenamefont
  {Francesco}, \citenamefont {Mathieu},\ and\ \citenamefont
  {S{\'e}n{\'e}chal}}]{Francesco2012CFT}%
  \BibitemOpen
  \bibfield  {author} {\bibinfo {author} {\bibfnamefont {P.}~\bibnamefont
  {Francesco}}, \bibinfo {author} {\bibfnamefont {P.}~\bibnamefont {Mathieu}},\
  and\ \bibinfo {author} {\bibfnamefont {D.}~\bibnamefont {S{\'e}n{\'e}chal}},\
  }\href@noop {} {\emph {\bibinfo {title} {Conformal field theory}}}\ (\bibinfo
   {publisher} {Springer Science \& Business Media},\ \bibinfo {year}
  {2012})\BibitemShut {NoStop}%
\bibitem [{\citenamefont {Wen}\ and\ \citenamefont {Zee}(1992)}]{WenZee1992}%
  \BibitemOpen
  \bibfield  {author} {\bibinfo {author} {\bibfnamefont {X.~G.}\ \bibnamefont
  {Wen}}\ and\ \bibinfo {author} {\bibfnamefont {A.}~\bibnamefont {Zee}},\
  }\bibfield  {title} {\bibinfo {title} {Shift and spin vector: New topological
  quantum numbers for the hall fluids},\ }\href
  {https://doi.org/10.1103/PhysRevLett.69.953} {\bibfield  {journal} {\bibinfo
  {journal} {Phys. Rev. Lett.}\ }\textbf {\bibinfo {volume} {69}},\ \bibinfo
  {pages} {953} (\bibinfo {year} {1992})}\BibitemShut {NoStop}%
\bibitem [{\citenamefont {Estienne}\ \emph
  {et~al.}(2013{\natexlab{b}})\citenamefont {Estienne}, \citenamefont
  {Regnault},\ and\ \citenamefont {Bernevig}}]{estienne2013fractional}%
  \BibitemOpen
  \bibfield  {author} {\bibinfo {author} {\bibfnamefont {B.}~\bibnamefont
  {Estienne}}, \bibinfo {author} {\bibfnamefont {N.}~\bibnamefont {Regnault}},\
  and\ \bibinfo {author} {\bibfnamefont {B.}~\bibnamefont {Bernevig}},\
  }\bibfield  {title} {\bibinfo {title} {Fractional quantum hall matrix product
  states for interacting conformal field theories},\ }\bibfield  {journal}
  {\bibinfo  {journal} {arXiv:1311.2936}\ }\href
  {https://doi.org/10.48550/arXiv.1311.2936} {10.48550/arXiv.1311.2936}
  (\bibinfo {year} {2013}{\natexlab{b}})\BibitemShut {NoStop}%
\bibitem [{\citenamefont {Eisert}\ \emph {et~al.}(2010)\citenamefont {Eisert},
  \citenamefont {Cramer},\ and\ \citenamefont
  {Plenio}}]{EisertCramerPlenio2010}%
  \BibitemOpen
  \bibfield  {author} {\bibinfo {author} {\bibfnamefont {J.}~\bibnamefont
  {Eisert}}, \bibinfo {author} {\bibfnamefont {M.}~\bibnamefont {Cramer}},\
  and\ \bibinfo {author} {\bibfnamefont {M.~B.}\ \bibnamefont {Plenio}},\
  }\bibfield  {title} {\bibinfo {title} {Colloquium: Area laws for the
  entanglement entropy},\ }\href {https://doi.org/10.1103/RevModPhys.82.277}
  {\bibfield  {journal} {\bibinfo  {journal} {Rev. Mod. Phys.}\ }\textbf
  {\bibinfo {volume} {82}},\ \bibinfo {pages} {277} (\bibinfo {year}
  {2010})}\BibitemShut {NoStop}%
\bibitem [{\citenamefont {Comparin}\ \emph {et~al.}(2022)\citenamefont
  {Comparin}, \citenamefont {Opler}, \citenamefont {Macaluso}, \citenamefont
  {Biella}, \citenamefont {Polychronakos},\ and\ \citenamefont
  {Mazza}}]{Comparin_2022}%
  \BibitemOpen
  \bibfield  {author} {\bibinfo {author} {\bibfnamefont {T.}~\bibnamefont
  {Comparin}}, \bibinfo {author} {\bibfnamefont {A.}~\bibnamefont {Opler}},
  \bibinfo {author} {\bibfnamefont {E.}~\bibnamefont {Macaluso}}, \bibinfo
  {author} {\bibfnamefont {A.}~\bibnamefont {Biella}}, \bibinfo {author}
  {\bibfnamefont {A.~P.}\ \bibnamefont {Polychronakos}},\ and\ \bibinfo
  {author} {\bibfnamefont {L.}~\bibnamefont {Mazza}},\ }\bibfield  {title}
  {\bibinfo {title} {Measurable fractional spin for quantum {H}all
  quasiparticles on the disk},\ }\href
  {https://doi.org/10.1103/PhysRevB.105.085125} {\bibfield  {journal} {\bibinfo
   {journal} {Phys. Rev. B}\ }\textbf {\bibinfo {volume} {105}},\ \bibinfo
  {pages} {085125} (\bibinfo {year} {2022})}\BibitemShut {NoStop}%
\bibitem [{\citenamefont {Arovas}\ \emph {et~al.}(1984)\citenamefont {Arovas},
  \citenamefont {Schrieffer},\ and\ \citenamefont
  {Wilczek}}]{ArovasSchriefferWilczek1984}%
  \BibitemOpen
  \bibfield  {author} {\bibinfo {author} {\bibfnamefont {D.}~\bibnamefont
  {Arovas}}, \bibinfo {author} {\bibfnamefont {J.~R.}\ \bibnamefont
  {Schrieffer}},\ and\ \bibinfo {author} {\bibfnamefont {F.}~\bibnamefont
  {Wilczek}},\ }\bibfield  {title} {\bibinfo {title} {Fractional statistics and
  the quantum hall effect},\ }\href
  {https://doi.org/10.1103/PhysRevLett.53.722} {\bibfield  {journal} {\bibinfo
  {journal} {Phys. Rev. Lett.}\ }\textbf {\bibinfo {volume} {53}},\ \bibinfo
  {pages} {722} (\bibinfo {year} {1984})}\BibitemShut {NoStop}%
\bibitem [{\citenamefont {Li}\ and\ \citenamefont {Haldane}(2008)}]{Li_2008}%
  \BibitemOpen
  \bibfield  {author} {\bibinfo {author} {\bibfnamefont {H.}~\bibnamefont
  {Li}}\ and\ \bibinfo {author} {\bibfnamefont {F.~D.~M.}\ \bibnamefont
  {Haldane}},\ }\bibfield  {title} {\bibinfo {title} {{E}ntanglement {S}pectrum
  as a {G}eneralization of {E}ntanglement {E}ntropy: {I}dentification of
  {T}opological {O}rder in {N}on-{A}belian {F}ractional {Q}uantum {H}all
  {E}ffect {S}tates},\ }\href {https://doi.org/10.1103/PhysRevLett.101.010504}
  {\bibfield  {journal} {\bibinfo  {journal} {Phys. Rev. Lett.}\ }\textbf
  {\bibinfo {volume} {101}},\ \bibinfo {pages} {010504} (\bibinfo {year}
  {2008})}\BibitemShut {NoStop}%
\bibitem [{\citenamefont {Kitaev}\ and\ \citenamefont
  {Preskill}(2006)}]{KitaevPreskill06}%
  \BibitemOpen
  \bibfield  {author} {\bibinfo {author} {\bibfnamefont {A.}~\bibnamefont
  {Kitaev}}\ and\ \bibinfo {author} {\bibfnamefont {J.}~\bibnamefont
  {Preskill}},\ }\bibfield  {title} {\bibinfo {title} {Topological entanglement
  entropy},\ }\href {https://doi.org/10.1103/PhysRevLett.96.110404} {\bibfield
  {journal} {\bibinfo  {journal} {Phys. Rev. Lett.}\ }\textbf {\bibinfo
  {volume} {96}},\ \bibinfo {pages} {110404} (\bibinfo {year}
  {2006})}\BibitemShut {NoStop}%
\bibitem [{\citenamefont {Levin}\ and\ \citenamefont {Wen}(2006)}]{LevinWen06}%
  \BibitemOpen
  \bibfield  {author} {\bibinfo {author} {\bibfnamefont {M.}~\bibnamefont
  {Levin}}\ and\ \bibinfo {author} {\bibfnamefont {X.-G.}\ \bibnamefont
  {Wen}},\ }\bibfield  {title} {\bibinfo {title} {Detecting topological order
  in a ground state wave function},\ }\href
  {https://doi.org/10.1103/PhysRevLett.96.110405} {\bibfield  {journal}
  {\bibinfo  {journal} {Phys. Rev. Lett.}\ }\textbf {\bibinfo {volume} {96}},\
  \bibinfo {pages} {110405} (\bibinfo {year} {2006})}\BibitemShut {NoStop}%
\bibitem [{\citenamefont {Ardonne}\ \emph {et~al.}(2004)\citenamefont
  {Ardonne}, \citenamefont {Kedem},\ and\ \citenamefont
  {Stone}}]{Ardonne_2005}%
  \BibitemOpen
  \bibfield  {author} {\bibinfo {author} {\bibfnamefont {E.}~\bibnamefont
  {Ardonne}}, \bibinfo {author} {\bibfnamefont {R.}~\bibnamefont {Kedem}},\
  and\ \bibinfo {author} {\bibfnamefont {M.}~\bibnamefont {Stone}},\ }\bibfield
   {title} {\bibinfo {title} {Filling the {B}ose sea: symmetric quantum {H}all
  edge states and affine characters},\ }\href
  {https://doi.org/10.1088/0305-4470/38/3/006} {\bibfield  {journal} {\bibinfo
  {journal} {Journal of Physics A: Mathematical and General}\ }\textbf
  {\bibinfo {volume} {38}},\ \bibinfo {pages} {617} (\bibinfo {year}
  {2004})}\BibitemShut {NoStop}%
\end{thebibliography}%

\appendix
\section{The time evolution factor}\label{app:exponent}

In this appendix, we demonstrate that the exponential \cref{normaliz}, repeated here for convenience,
\begin{widetext}
\begin{equation}
\label{eq:normaliz-app}
\mathcal{U} =
\exp\bigg(-\alpha \sum_{j=0}^{N_{\phi}+1}\bigg[\frac{Q^2_{0,j}}{2q_0}+\frac{3Q_{0,j}}{2q_0}+\frac{Q_{1,j}^2}{2q_1}+\frac{Q_{2,j}^2}{2q_2}+(P_0+P_1+P_2)_j\bigg]\bigg) \ ,
\end{equation}
\end{widetext}
together with the additional contributions due to the presence of a quasi-hole, equals the expression \cref{normQH}.
We introduced the parameter $\alpha = \frac{2\pi\delta\tau}{L}$.
The quantum numbers are denoted as $Q_{i,j}$ and $P_{i,j}$, where $i=0,1,2$ indicates the boson field, while
$j = 0, 1, \ldots, N_\phi+1$ denotes the auxiliary Hilbert space.
There are $N_\phi+1$ orbitals, indexed by $0,1,\ldots,N_\phi$, with orbital occupation numbers $m_{a,j}$, $m_{b,j}$ and $m_{c,j}$
taking values in $\{0,1\}$ and which satisfy $m_{a,j} + m_{b,j} + m_{c,j} \in \{0,1\}$ and
$\sum_{j=0}^{N_\phi} m_{a,j} + m_{b,j} + m_{c,j} = N_e$.
The $j^{th}$ term in \cref{eq:normaliz-app} comes from the auxiliary Hilbert space between orbitals $j-1$ and $j$. 
For convenience, we write $\mathcal{U} = e^{\mathcal{A}}$, and focus on $\mathcal{A}$.

To begin with, we note that the matrix elements \cref{ba,bb,bc} imply that the charge quantum numbers
at the auxiliary Hilbert space with index $j$ can be written as
\begin{align}
\label{Qrels}
Q_{0,j} &=(3M+2)\sum_{k=0}^{j-1}(m_{a,k}+m_{b,k}+m_{c,k})-3j \nonumber\\
Q_{1,j} &=2\sum_{k=0}^{j-1}(2m_{a,k}-m_{b,k}-m_{c,k}) \\
Q_{2,j} &=2\sum_{k=0}^{j-1}(m_{b,k}-m_{c,k}) \ . \nonumber
\end{align}
We recall that we have set $Q_{i,0}=0$ for $i=0,1,2$.
As for the momenta, the matrix elements \cref{ba,bb,bc} and the charge expressions \cref{Qrels} lead to 
 \begin{align}\label{termsofQ}
     &(P_0+P_1+P_2)_{j}=\nonumber\\
     &-\frac{3M+2}{q_0}\sum_{k=0}^{j-1}(m_{a,k}+m_{b,k}+m_{c,k})\nonumber\\&\times\bigg[(3M+2)\sum_{l=0}^{k-1}(m_{a,l}+m_{b,l}+m_{c,l})-3k\bigg]\nonumber\\&-\frac{4}{q_1}\sum_{k=0}^{j-1}\sum_{l=0}^{k-1}(2m_{a,k}-m_{b,k}-m_{c,k})(2m_{a,l}-m_{b,l}-m_{c,l})\nonumber\\
     &-\frac{4}{q_2}\sum_{k=0}^{j-1}\sum_{l=0}^{k-1}(m_{b,k}-m_{c,k})(m_{b,l}-m_{c,l}).
 \end{align}
This can in turn can be rewritten using the results \cref{Qrels} for the $Q_{i,j}$ quantum numbers and that there may at most be one electron per orbital, i.e. $m_{a,j},m_{b,j},m_{c,j}\in\{0,1\}$ and $m_{a,j}+m_{b,j}+m_{c,j}\in\{0,1\}$, as follows:
\begin{align}\label{consP}
    (P_0+P_1+P_2)_{j}&=\sum_{k=0}^{j-1}k(m_{a,k}+m_{b,k}+m_{c,k})\nonumber\\
    &-\frac{Q_{0,j}^2}{2q_0}+\frac{Q_{0,j}}{q_0}\bigg(\frac{3M+6}{2}-3j\bigg)\nonumber\\
    &-\frac{Q_{1,j}^2}{2q_1}-\frac{Q_{2,j}^2}{2q_2}-\frac{9j^2}{2q_0}+\frac{(q_0+12)j}{2q_0} \ .
\end{align}
Here, we have used that $q_0=3(3M+2),q_1=12,q_2=4$ for minor simplifications. By inserting \cref{consP} in \cref{eq:normaliz-app} and cancelling terms, we find that
$\mathcal{A}$ is given by
\begin{widetext}
\begin{equation}\label{simplify}
 \mathcal{A} \simeq -\alpha \sum_{j=1}^{N_{\phi}+1}\sum_{k=0}^{j-1}k(m_{a,k}+m_{b,k}+m_{c,k}) + \frac{\alpha}{2q_0}\sum_{j=1}^{N_{\phi}+1}\big(6j-3M-9\big)Q_{0,j} \ ,
\end{equation}
where we recall that $\simeq$ means ``up to terms that do not depend on the precise distribution of electrons over the orbitals".
We may express the above in terms of the occupation numbers $m_{a,j},m_{b,j},m_{c,j}$ by inserting $Q_{0,j}$ from \cref{Qrels}.
By making use of the relations (where we set $n=N_{\phi}+1$)
\begin{align}
    \sum_{j=1}^{n}\sum_{k=0}^{j-1}(m_{a,k}+m_{b,k}+m_{c,k}) &= \sum_{j=0}^{N_{\phi}}(n-j)(m_{a,j}+m_{b,j}+m_{c,j}) \\
    \sum_{j=1}^{n}\sum_{k=0}^{j-1}j(m_{a,k}+m_{b,k}+m_{c,k}) &=
    \frac{1}{2}\sum_{j=0}^{N_{\phi}}\big(n(n+1)-j(j+1)\big)(m_{a,j}+m_{b,j}+m_{c,j}) \\
    \sum_{j=1}^{n}\sum_{k=0}^{j-1}k(m_{a,k}+m_{b,k}+m_{c,k}) &= \sum_{j=0}^{N_{\phi}}j(n-j)(m_{a,j}+m_{b,j}+m_{c,j}) \ ,
\end{align}
\end{widetext}
the expression \cref{simplify} becomes
\begin{align}\label{newnorm}
\mathcal{A} \simeq \alpha \sum_{j=0}^{N_{\phi}}\bigg[&\frac{j^2}{2}(m_{a,j}+m_{b,j}+m_{c,j}) \\
\nonumber
&+j\bigg(\frac{M+2}{2}-n\bigg)(m_{a,j}+m_{b,j}+m_{c,j}) \biggr] \ .
\end{align}
We note that in the absence of quasi-holes, the expression $\sum_{j=0}^{N_\phi} j (m_{a,j}+m_{b,j}+m_{c,j})$
does not depend on how the electrons are distributed over the orbitals. This is not true in the presence of quasi-holes. We therefore need to keep this term, because it will be modified below when we consider quasi-holes.

With a single quasi-hole inserted, the exponent $\mathcal{A}$ changes.
In the main text, we already stated that in the presence of a single quasi-hole, the wave function can be written as in \cref{ws}
\begin{equation}
    \sum_{s=0}^{N_e}w^sP_s(\{z_j\}) \ .
\end{equation}
We want to obtain an expression for $s$ in terms of the orbital occupation numbers.
By counting powers of the electron coordinates, one sees that the total degree of the wave functions \cref{wavefcn eps,wavefcn Laughlin,wavefcn psi1,wavefcn psi2,wavefcn sigma1,wavefcn sigma2} in the electron coordinates is
$M N_e(N_e-1)/2 + N_e (N_e-3)/2 + \gamma N_e/2 + \delta$ where $\gamma$ is directly proportional to the quasi-hole charge
and $\delta$ also depends on the type of quasi-hole, as given in \cref{tab:gamma-delta} (although the specific values of $\gamma$ and $\delta$ are not necessary for the current argument).
\begin{table}[h]
\begin{tabular}{c | c | c}
quasi-hole & $\gamma$ & $\delta$ \\
\hline
$(\sigma_1, 1/5)$ & $1$ & $0$ \\
$(\sigma_2, 2/5)$ & $2$ & $0$ \\
$(\psi_1, 2/5)$ & $2$ & $-2/3$ \\
$(\mathbf{1}, 3/5)$ & $3$ & $0$ \\
$(\epsilon, 3/5)$ & $3$ & $-1/3$ \\
$(\psi_2, 4/5)$ & $4$ & $-2/3$ \\
\end{tabular}
\caption{The values of the parameters $\gamma$ and $\delta$ for the various quasi-holes.}
\label{tab:gamma-delta}
\end{table}
The total degree also equals $\sum_{j=0}^{N_{\phi}}j(m_{a,j}+m_{b,j}+m_{c,j})$.
Therefore, we obtain that
\begin{equation}\label{eq:ssimeq-app}
    s\simeq -\sum_{j=0}^{N_{\phi}}j(m_{a,j}+m_{b,j}+m_{c,j}).
\end{equation}

We now assume that a quasi-hole is inserted between orbitals $\tilde{l}-1$ and $\tilde{l}$.
The quasi-hole may be of either type $a,b$ or $c$.
For the auxiliary Hilbert spaces \textit{before} the quasi-hole, i.e. the terms with indices $j=1,2, \ldots ,\tilde{l}$ in \cref{eq:normaliz-app}, the contribution to $\mathcal{A}$ is just as before
\begin{equation}\label{before}
-\alpha 
\sum_{j=0}^{\tilde{l}}\bigg[\frac{Q^2_{0,j}}{2q_0}+\frac{3Q_{0,j}}{2q_0}+\frac{Q_{1,j}^2}{2q_1}+\frac{Q_{2,j}^2}{2q_2}+(P_0+P_1+P_2)_j\bigg].
\end{equation}
If the quasi-hole is of type $a$, the quasi-hole operator matrix elements \cref{ha} make the contributions \textit{after} the quasi-hole (orbital index $\tilde{l},\tilde{l}+1, \ldots ,N_{\phi}$ and auxiliary space index $\tilde{l}+1,\tilde{l}+2, \ldots ,N_{\phi}+1$) 
\begin{align}\label{deltaa}
    -\alpha
    \sum_{j=\tilde{l}+1}^{N_{\phi}+1}\bigg[&\frac{(Q_{0,j}+1)^2}{2q_0}+\frac{3(Q_{0,j}+1)}{2q_0} \\
    \nonumber
    &+\frac{(Q_{1,j}+2)^2}{2q_1}+\frac{Q_{2,j}^2}{2q_2} \\
    \nonumber
    &+P_{0,j}+P_{1,j}+P_{2,j}+\Delta(P_{0,j}+P_{1,j}+P_{2,j})\bigg] \ ,
\end{align}
since the charge of the quasi-hole changes the charges in all subsequent Hilbert spaces by a constant shift. Above, the quantum numbers $Q_{i,j},P_{i,j}$ are the values without quasi-holes, and the charge deviations are explicitly written out. The $\Delta(P_{0,j}+P_{1,j}+P_{2,j})$ term is a deviation from what $P_{0,j}+P_{1,j}+P_{2,j}$ would have been without quasi-holes. 
A way to understand the appearance of the extra charge terms and the term $\Delta(P_{0,j}+P_{1,j}+P_{2,j})$ is as follows: 
the charges are shifted because the quasi-hole carries $Q_0,Q_1$ and (in principle, although not for the quasi-hole of type $a$ we are considering here) $Q_2$ charge. The momenta depend on the charges through \cref{consP}, and shifting the charges therefore also typically shifts the sum of the momenta.

To actually compute the momentum shift, we note that
a quasi-hole operator of type $a$ has, due to \cref{ha}, matrix elements proportional to $e^{-\frac{2\pi i x_{\alpha}}{L}\big(\frac{Q_{0}+3\tilde{l}}{q_0}+\frac{2Q_1}{q_1}+(P_0+P_1)'-(P_0+P_1)\big)}$, where $x_{\alpha}$ is the coordinate of the quasi-hole along the cylinder circumference, $\tilde{l}$ is an orbital index, and primed (unprimed) momenta are the values immediately after (before) the quasi-hole insertion. However, from \cref{ws}, we know that the quasi-hole operator should give a factor $w^s$ on the plane. The latter can be mapped to the cylinder using \cref{map}, where the $x_{\alpha}$ dependence is contained in a factor of $e^{-i\frac{2\pi}{L}x_{\alpha}s}$. Setting the two exponentials proportional to one another implies that the sum of the momenta immediately after the $a$-type quasi-hole operator, and by the matrix elements \cref{ba,bb,bc} also the sum of the momenta on all subsequent auxiliary spaces, is shifted by an amount
\begin{align}\label{deltaPa}
    \Delta(P_0+P_1+P_2)_{j}=&-\frac{Q_{0,j}}{q_0}-\frac{2Q_{1,j}}{q_1}+s,
\end{align}
up to an additive constant which does not depend on the quantum numbers or $s$. Here, we have used the factor of $\delta_{P_2',P_2}$ in \cref{ha} to conclude that the above holds for the total shift $\Delta(P_0+P_1+P_2)_{\tilde{l}}=(P_0+P_1+P_2)_{\tilde{l}}'-(P_0+P_1+P_2)_{\tilde{l}}$ immediately after the quasi-hole and not just for $(P_0+P_1)_{\tilde{l}}'-(P_0+P_1)_{\tilde{l}}$.
Inserting \cref{deltaPa} in \cref{deltaa} and omitting unimportant overall constants then gives the following contribution to $\mathcal{A}$:
\begin{equation}\label{after}
-\alpha
\sum_{j=\tilde{l}+1}^{N_{\phi}+1}\bigg[\frac{Q_{0,j}^2}{2q_0}+\frac{3Q_{0,j}}{2q_0}+\frac{Q_{1,j}^2}{2q_1}+\frac{Q_{2,j}^2}{2q_2}+P_{0,j}+P_{1,j}+P_{2,j}+s\bigg].
\end{equation}
If the quasi-hole is instead of type $b,$ the contribution after the quasi-hole is, via \cref{hb},
\begin{align}
    -\alpha
    \sum_{j=\tilde{l}+1}^{N_{\phi}+1}\bigg[&\frac{(Q_{0,j}+1)^2}{2q_0}+\frac{3(Q_{0,j}+1)}{2q_0}
    \nonumber\\
    &+\frac{(Q_{1,j}-1)^2}{2q_1}+\frac{(Q_{2,j}+1)^2}{2q_2}\nonumber\\&
    +P_{0,j}+P_{1,j}+P_{2,j}+\Delta(P_{0,j}+P_{1,j}+P_{2,j})\bigg],
\end{align}
but here, the momentum shift instead becomes
\begin{align}\label{deltaPb}
    \Delta(P_0+P_1+P_2)_{j}=&-\frac{Q_{0,j}}{q_0}
    +\frac{Q_{1,j}}{q_1}-\frac{Q_{2,j}}{q_2}+s,
\end{align}
up to unimportant constants. Thus, we again arrive at \cref{after}.
Finally, for type $c$ we have, from \cref{hc}, 
\begin{align}
    -\alpha
    \sum_{j=\tilde{l}+1}^{N_{\phi}+1}\bigg[&\frac{(Q_{0,j}+1)^2}{2q_0}+\frac{3(Q_{0,j}+1)}{2q_0}
    \nonumber\\
    &+\frac{(Q_{1,j}-1)^2}{2q_1}+\frac{(Q_{2,j}-1)^2}{2q_2}
    \nonumber\\&+P_{0,j}+P_{1,j}+P_{2,j}+\Delta(P_{0,j}+P_{1,j}+P_{2,j})\bigg].
\end{align}
In this final case, the sum of the momenta changes as
\begin{align}\label{deltaPc}
    \Delta(P_0+P_1+P_2)_{j}=&-\frac{Q_{0,j}}{q_0} +\frac{Q_{1,j}}{q_1}+\frac{Q_{2,j}}{q_2}+s,
\end{align}
which again implies \cref{after}.
All in all, we regardless of the quasi-hole operator type find that, up to unimportant constants, the terms in the exponent $\mathcal{A}$ from orbitals after the quasi-hole are given by \cref{after}, 
 where the shift parameter $s$ comes from the exponent of the quasi-hole coordinate on the plane and is independent of $j$; c.f. \cref{ssimeq} and the surrounding discussion. Before moving on to the final parts of our derivation, we note that the coefficients in \cref{deltaPa,deltaPb,deltaPc} can be read off from the quasi-hole operators in \cref{qhops} as long as one remembers to change the signs.

The final contribution to the exponent in \cref{eq:normaliz-app} comes from the time evolution factors of \cref{ha,hb,hc}. This part is due to the quasi-hole itself, rather than its influence on the subsequent auxiliary Hilbert spaces. 
As before, we assume that the quasi-hole operator is inserted between orbitals $\tilde{l}-1$ and $\tilde{l}$, with $\tau=\tau_{\alpha}$.
Then, the quasi-hole time evolution factors give an extra contribution (c.f. the matrix elements  \cref{ha,hb,hc} and surrounding discussion)
\begin{align}\label{at0}
       &-\frac{2\pi}{L}(\tilde{l}\delta\tau-\tau_{\alpha})\bigg[\frac{(Q_0')^2-(Q_0)^2}{2q_0}+\frac{3(Q_0'-Q_0)}{2q_0}\nonumber\\&\quad\quad\quad\quad\quad\quad\quad+\frac{(Q_1')^2-(Q_1)^2}{2q_1}+\frac{(Q_2')^2-(Q_2)^2}{2q_2}\nonumber\\
    &\qquad\qquad\quad\quad\quad+P'_0+P'_1+P'_2-(P_0+P_1+P_2)\bigg],
\end{align}
where the ``primed" values are those immediately after the quasi-hole, and the ``unprimed" are those immediately before it. To simplify the above, we note that the parameter $s\simeq -\sum_{j=0}^{N_{\phi}}j(m_{a,j}+m_{b,j}+m_{c,j})$ is introduced as the power of the quasi-hole coordinate $w$ in the wave function; see eq. \eqref{ws}. If the quasi-hole is of type $a$, the matrix element \cref{ha} just as before implies that $P_0'+P_1'+P_2'-(P_0+P_1+P_2)=\Delta(P_0+P_1+P_2)$ is given by \cref{deltaPa}, up to unimportant constants. 
With the charges being affected as $Q_0'=Q_0+1,Q_1'=Q_1+2,Q_2'=Q_2$, \cref{at0} becomes 
\begin{align}\label{at}
    &-\frac{2\pi}{L}(\tilde{l}\delta\tau-\tau_{\alpha})\bigg[\frac{(Q_0+1)^2-(Q_0)^2}{2q_0}+\frac{3(Q_0+1-Q_0)}{2q_0}\nonumber\\&\quad\quad\quad\quad\quad\quad\quad+\frac{(Q_1+2)^2-(Q_1)^2}{2q_1}+\frac{(Q_2)^2-(Q_2)^2}{2q_2}\nonumber\\
    &\qquad\qquad\quad\quad\quad+s-\frac{Q_0
    }{q_0}-\frac{2Q_1}{q_1}\bigg]\nonumber\\
    &\simeq -\frac{2\pi}{L}(\tilde{l}\delta\tau-\tau_{\alpha})s,
\end{align}
where the step leading to the last line involves omitting unimportant state-independent constants. One can show in a completely analogous manner that if the quasi-hole operator is of type $b$ or $c$, the result is the same. 

Adding up all the contributions from \cref{before,after,at}, and using the previous result \cref{newnorm}, that $s\simeq -\sum_{j=0}^{N_{\phi}}j(m_{a,j}+m_{b,j}+m_{c,j})$ from \cref{ssimeq}, and that the parameter $\alpha=2\pi\delta\tau/L$, we arrive at the following expression for $\mathcal{U}$:
\begin{align}\label{normQHApp}
     \mathcal{U} \simeq
     \exp\bigg\{&\frac{2\pi\delta\tau}{L}\sum_{j=0}^{N_{\phi}}\bigg[\frac{j^2}{2}(m_{a,j}+m_{b,j}+m_{c,j}) \bigg]
     + \\ \nonumber
     &\frac{2\pi}{L}\bigg(\tau_{\alpha}-\frac{M+2}{2}\delta\tau\bigg)s\bigg\},
\end{align}
up to unimportant constants.
We note that the contributions proportional to $s n$ cancel each other. The same is true for the contributions proportional to $s \tilde{l}$.
Thus, we have derived \cref{normQH}.

\end{document}